\documentclass{article}
\usepackage[german, english]{babel}
\usepackage{a4wide}
\usepackage{amsmath}
\usepackage{amssymb}
\usepackage{ifthen}
\usepackage{epsfig}
\usepackage{setspace}
\usepackage{subfigure}
\usepackage{graphicx}

\newcounter{fig}

\newcommand{\bea}{\begin{eqnarray}}
\newcommand{\eea}{\end{eqnarray}}
\newcommand{\be}{\begin{equation}}
\newcommand{\ee}{\end{equation}}

\def\vecphi{{\pmb{\varphi}}}

\def\vecphi{{\pmb{\phi}}}

\def\bfph1{{\pmb{\phi}}^{(I)}}
\def\bfph2{{\pmb{\phi}^{(II)}}}

\begin{document}

\title{Spontaneous symmetry breaking in dual-core baby-Skyrmion systems}
\date{~}
\author{Boris A. Malomed$^{1}$, Yakov Shnir$^{2,3}$ and Gleb Zhilin$^{4}$ \\
$^{1}$Department of Physical Electronics, School of Electrical Engineering,\\
Faculty of Engineering, Tel Aviv University,\\
Tel Aviv 69978, Israel\\
$^{2}$Bogoliubov Laboratory for Theoretical Physics,\\
Joint Institute for Nuclear Research,\\
Dubna 141980, Moscow Region, Russia\\
$^{3}$Institute of Physics\\
Carl von Ossietzky University Oldenburg\\
Oldenburg D-26111, Germany\\
$^{4}$Department of Theoretical Physics and Astrophysics,\\
Belarusian State University, Minsk 220004, Belarus}
\maketitle

\begin{abstract}
We introduce a system composed of two (2+1)-dimensional baby-skyrmion models
(BSMs) set on parallel planes and linearly coupled by tunneling of fields.
This system can be realized in a dual-layer ferromagnetic medium. Unlike
dual-core models previously studied\ in nonlinear optics and BEC, here the
symmetry-breaking bifurcation (SBB) in solitons (baby skyrmions) occurs with
the increase of the inter-core coupling ($\kappa $), rather than with its
decrease, due to the fact that, even in the uncoupled system, neither core
may be empty. Prior to the onset of the symmetry breaking between the two
components of the solitons, they gradually separate in the opposite
directions, due to the increase of $\kappa $, which is explained in an
analytical form by means of an effective interaction potential. Such
evolution scenarios are produced for originally symmetric states with
topological charges in the two cores, $B^{(1)}=B^{(2)}=1,2,3,4$. The
evolution of mixed states, of the $\left( B^{(1)},B^{(2)}\right) =\left(
1,2\right) $ and $\left( 2,4\right) $ types, with the variation of $\kappa $
is studied too.
\end{abstract}


\section{Introduction}

Diverse models of the field theory support topological solitons. Many such
models have been intensively studied over last decades in a wide variety of
physical contexts. Perhaps one of the simplest examples is a modified
version of the non-linear $O(3)$ $\sigma $-model in $\left( 2+1\right) $
dimensions (the so-called \textquotedblleft baby-skyrmion model", BSM) \cite%
{BB,Bsk}. This is a low-dimensional simplified theory which emulates the
conventional Skyrme model in $\left( 3+1\right) $ dimensions \cite{Skyrme}
in many respects, and finds direct physical realizations. In particular,
hexagonal lattices of two-dimensional skyrmions were observed in a thin
ferromagnetic layer \cite{CondMatt}, and in a metallic itinerant-electron
magnet, where the skyrmion lattice was detected by results of neutron
scattering \cite{neutron}, and through a specific contribution to the
topological quantum Hall effect \cite{SkHall}.

According to the Derrick's theorem \cite{Derrick}, to support the solitons'
stability, the Lagrangian of the BSM should include, apart from the usual $%
O(3)$ sigma-model's kinetic term, also an interaction term quartic in
derivatives \footnote{%
Recently, some modification of the BSM with the Dzyaloshinskii-Moriya
interaction term was suggested to model noncentrosymmetric ferromagnetic
planar structures \cite{Bogdanov}.}, and the potential one, which does not
contain derivatives. Although the structure of the potential term is largely
arbitrary, its particular choice determines different ways of symmetry
breaking \cite{Ward,Hen,JSS}.

In various branches of physics, complex systems, involving several coupled
scalar field, arise (see, e.g., Refs. \cite%
{Rajaraman:1978kd,Bazeia:1995en,Malomed,Halavanau:2012dv}). Properties of
solitons in these extended models may be quite different from a
straightforward extension of the single-component counterpart. In
particular, the extended models can support non-topological solitons \cite%
{Rajaraman:1978kd,Halavanau:2012dv}.

Currently, there is a number of experimentally relevant multicomponent
systems, including those in non-conventional superconductivity models (see a
topical issue on this topic \cite{SC}), where a few superconducting bands
and a set of corresponding Josephson coupling constants between the
condensates appear. The systems of the latter type give rise to effective
chiral $\mathbb{C}P^{2}$ planar skyrmions, which were constructed in the
three-component Ginzburg-Landau model with broken time-reversal symmetry
\cite{Garaud:2012pn}. These configurations are actually bound states of
triplets of vortices, the system being symmetric with respect to the
dihedral group. It is also relevant to mention a recent experimental
observation of a bound state of two magnetic baby-skyrmions, and their
current-driven motion, in layered manganites \cite{bi}.

In this work we introduce a system composed of two replicas of the usual $%
\left( 2+1\right) $-dimensional BSM, set in two parallel planes
(\textquotedblleft cores"), which are linearly coupled by hopping
(tunneling) between them. The fields in the two cores are also referred to
below as \textquotedblleft sectors" of the coupled system. Depending on the
coupling strength, this dual system may manifest various symmetries and
symmetry-breaking scenarios. A system of this type may be realized, in
particular, in a bilayer ferromagnetic film, with the BSM implemented in
each layer.

Linearly coupled dual-core systems with nonlinearities in each core, alias
\textit{nonlinear couplers} \cite{Jensen}, where studied in detail in
relatively simple models of nonlinear optics and Bose-Einstein condensates
(BEC), represented by a single complex field in each core, which obey the
respective nonlinear Schr\"{o}dinger (NLS) or Gross-Pitaevskii (GP)
equations. If the intrinsic nonlinearity in the cores is strong enough, the
power exchange between them is effected by the intensity of the guided
waves, which is used for the design of all-optical switching devices \cite%
{switch}. In addition to the simplest dual-core system with the cubic
nonlinearity and single component in each core, realizations of nonlinear
couplers have been studied in many other settings, including the bimodal
propagation of light \cite{Trillo}, twin-core Bragg gratings \cite{Bragg},
\cite{Sukhorukov}, lossy optical couplers embedded into a gain medium \cite%
{Barash}, systems with saturable \cite{satur}, quadratic \cite{chi2,Shapira}%
, cubic-quintic (CQ) \cite{Albuch}, and nonlocal \cite{Nonlocal}
nonlinearities, dual-core traps for BEC \cite{BEC1}, parallel coupled arrays
of discrete waveguides \cite{Herring}, ($2+1$)-dimensional couplers for
spatiotemporal \textquotedblleft light bullets" in planar dual-core
waveguides \cite{Dror}, and $\mathcal{PT}$-symmetric nonlinear couplers,
both ($1+1$)-dimensional \cite{PT1} and ($2+1$)-dimensional \cite{PT2}. Note
that the ($2+1$)-dimensional dual-core waveguides have the same geometric
dimension as the dual-core BSM introduced in the present work.

A fundamental property of nonlinear couplers is the symmetry-breaking
bifurcation (SBB), which destabilizes obvious symmetric states in the system
and gives rise to asymmetric ones. The SBB was analyzed for spatially
uniform states \cite{Snyder} and solitons in twin-core waveguides \cite%
{Wabnitz}, as well as for gap solitons in Bragg gratings \cite{Bragg} with
the cubic nonlinearity (these results were originally reviewed in Ref. \cite%
{Wabnitz2}, and later in Ref. \cite{Progress}). The SBB was studied too for
solitons supported by the quadratic \cite{chi2} and CQ \cite{Albuch}
nonlinearities.

In all the above-mentioned systems, the study of the spontaneous symmetry
breaking followed the pattern which assumed that, in the limit case of the
uncoupled system, one core would carry a usual single-component soliton,
while the other one is empty. Then, with the increase of the coupling
constant, $\kappa $, the originally empty core is filled by the field
tunneling from the mate core. Eventually, only symmetric states, featuring
identical soliton components in both cores, exist above a critical value of $%
\kappa $, while below that value symmetric states are unstable, being
replaced by stable asymmetric ones. The principal difference of the
situation in the dual-core BSM considered below is that neither core is
supposed to be empty, due to the boundary conditions imposed on the fields
at infinity. Therefore, it gives rise to an altogether different scenario of
the symmetry breaking, with identical configurations in the uncoupled cores,
which start to separate in opposite (lateral) directions, and eventually
undergo an SBB,\ with the \emph{increase} (rather than decrease)\ of $\kappa
$; however, at critical value $\kappa =\kappa _{\mathrm{cr}}$ given below by
Eq. (\ref{cr}), the coupling cannot stabilize the baby skyrmion in the
system.

The rest of the paper is structured as follows. In the next section we
briefly review the model which support non-trivial soliton configurations of
the dual-core BSM. Numerical results are presented in Section 3, where we
consider various patterns of the evolution of the coupled configurations in
two different cases, \textit{viz}., the \textquotedblleft old" BSM \cite{Bsk}
and the \textquotedblleft new" double-vacuum model \cite{Weidig:1998ii}. For
the sake of compactness, we restrict the analysis to configurations with
topological charge $\leq 4$ in each sector. Conclusions and remarks are
formulated in Section 4.

\section{The model}

As said above, we consider a set of two coupled replicas of the nonlinear
modified $O(3)$ $\sigma $-model with the Skyrme term in $\left( 2+1\right) $
dimensions (i.e., the BSM), which is based on the following Lagrangian
density:

\begin{gather}
L=\sum_{a=1,2}L_{a}+L_{\mathrm{coupling}},  \notag \\
L_{a}=\frac{1}{2}\partial _{\mu }\mathbf{\phi }^{(a)}\cdot \partial ^{\mu }%
\mathbf{\phi }^{(a)}-\frac{1}{4}\left( \partial _{\mu }\mathbf{\phi }%
^{(a)}\times \partial _{\nu }\mathbf{\phi }^{(a)}\right) ^{2}-U\left(
\mathbf{\phi }^{(a)}\right) ,  \label{Lag} \\
L_{\mathrm{coupling}}=\kappa \phi _{3}^{(1)}\phi _{1}^{(2)},  \notag
\end{gather}%
where $\mathbf{\phi }^{(a)}=\left( \phi _{1}^{(a)},\phi _{2}^{(a)},\phi
_{3}^{(a)}\right) $, $a=1,2,$ are two vectorial triplets of scalar fields
which are subject to constraint
\begin{equation}
|\mathbf{\phi }^{(1,2)}|^{2}=1.  \label{unity}
\end{equation}%
Note that rescaling of the two-component model does not allow us to absorb
all the constants into rescaled parameters of the potentials. Here we focus
on three most essential coefficients of the model, \textit{viz}., two mass
parameters $\mu _{1,2}$, which are defined below, and inter-core coupling
constant $\kappa $.

The \textquotedblleft skew" form of the coupling potential, $L_{\mathrm{%
coupling}}$, which is defined in Eq. (\ref{Lag}), is the one which gives
rise to the symmetry breaking, see below. If, instead, a \textquotedblleft
straight" form is taken, with $L_{\mathrm{coupling}}=\kappa \phi
_{3}^{(1)}\phi _{3}^{(2)}$, it will not give rise to any symmetry breaking.

We consider fields $\mathbf{\phi }^{(a)}$ as maps, $\mathbf{\phi }^{(a)}:%
\mathbb{R}^{2}\rightarrow S^{2}$, which are characterized by two integers
(topological charges), $B^{(a)}=\pi _{2}(S^{2})$. Explicitly, they are given
as integrals of vectorial products,
\begin{equation}
B^{(a)}=\frac{1}{4\pi }\int\!\!\int \mathbf{\phi }^{(a)}\cdot \left(
\partial _{1}\mathbf{\phi }^{(a)}\times \partial _{2}\mathbf{\phi }%
^{(a)}\right) ~dxdy,  \label{charge}
\end{equation}%
thus the two-component configuration possesses topological charges in both
cores, that will be referred to as $(B^{(1)},B^{(2)})$. Note that the
symmetry of the configuration with respect to the reflections in the
internal space, $\phi _{2}^{(a)}\rightarrow -\phi _{2}^{(a)}$, which inverts
of the sign of the topological charge of the corresponding constituent,
allows us to restrict the consideration to positive values of $B^{(1,2)}$.

In the decoupled limit, $\kappa \rightarrow 0$, each component approaches
the vacuum value at the infinitely remote spatial boundary in the respective
two-dimensional plane. This boundary value is commonly taken as $\mathbf{%
\phi }_{\infty }^{(1,2)}=(0,0,1)$, thus an $O(3)$-symmetry-breaking
potential, $U\left( \mathbf{\phi }^{(a)}\right) $, which vanishes at the
boundary, stabilizes the configuration. In the dual-core system, we can
relax the restriction on the asymptotic value of one component, say $\mathbf{%
\phi }_{\infty }^{(2)}$. For the fields taking values on the unit sphere, as
per Eq. (\ref{unity}), we assume
\begin{equation}
\mathbf{\phi }_{\infty }^{(1)}=(0,0,1),  \label{1}
\end{equation}%
which provides the one-point compactification of the spatial boundary.

The explicit form of potential $U\left( \mathbf{\phi }^{(a)}\right) $ is
largely arbitrary. There are a few familiar examples, such as the BSM with
the so-called \textquotedblleft old" potential \cite{Bsk},%
\begin{equation}
U\left( \mathbf{\phi }^{(a)}\right) =\mu _{a}^{2}\left[ 1-\phi _{3}^{(a)}%
\right] \,,  \label{pot-old}
\end{equation}%
which corresponds to the unique vacuum of component $\mathbf{\phi }^{(a)}$,
or the double-vacuum model \cite{Weidig:1998ii}, with
\begin{equation}
U\left( \mathbf{\phi }^{(a)}\right) =\mu _{a}^{2}\left[ 1-(\phi
_{3}^{(a)})^{2}\right] .  \label{double}
\end{equation}%
Both potentials are invariant with respect to iso-rotations about the third
component $\phi _{3}^{(a)}$, hence at $\kappa =0$ the symmetry of the model
is broken\footnote{%
In the double vacuum model, there is an additional reflection symmetry, $%
Z_{2}\times Z_{2}$.} to $SO(2)\times SO(2)$. On the other hand, the vacuum
structure strongly depends on the value of the inter-core coupling, $\kappa $%
, as the coupling term in (\ref{Lag}) violates the rotational invariance,
and it can completely break the symmetry.

Note that the structure of the potential term is important for the
stability. For example, iso-rotations of skyrmions with topological
charge $B$ in the model with the \textquotedblleft old" potential
may break them into $B $ skyrmions with charge $1$ \cite{SH}.

The violation of the rotational invariance in the BSM has recently drawn a
great deal of attention. It was demonstrated that the effect strongly
depends on the particular choice of the above-mentioned potential \cite%
{Ward,Hen,JSS}. Here we consider another symmetry-breaking
mechanism, introduced by the linear coupling between the cores
carrying the two sectors of systems. As mentioned above, this
mechanism was previously studied in detail in systems of linearly
coupled NLS/GP equations.

First we consider the model with the double vacuum potentials in each
sector, hence the total potential is
\begin{equation}
U\left( \mathbf{\phi }^{(1)},\mathbf{\phi }^{(2)}\right) =\mu _{1}^{2}\left[
1-\left( \phi _{3}^{(1)}\right) ^{2}\right] +\mu _{2}^{2}\left[ 1-\left(
\phi _{3}^{(2)}\right) ^{2}\right] +\kappa \phi _{3}^{(1)}\phi _{1}^{(2)}\,.
\label{pot}
\end{equation}%
Evidently, the coupling between the sectors may stabilize the configuration
when both mass parameters $\mu _{1},\mu _{2}$ are zero.

\section{Numerical results}

\subsection{States with the equal topological charges in the two sectors}

In this section results produced by numerical solutions are presented for
the dual-core BSM. The solutions were chiefly constructed on an equidistant
grid in polar coordinates $\left( \rho ,\theta \right) $, employing the
compactified radial coordinate,
\begin{equation}
\xi =\rho /(1+\rho )\in \lbrack 0,1],  \label{xi}
\end{equation}%
and $\theta \in \lbrack 0,2\pi ]$, i.e. $x=\rho \sin \theta,~ y=\rho \cos
\theta$. To find minima of the functional corresponding to Lagrangian
density (\ref{Lag}), we have implemented a simple forward-differencing
scheme on a square lattice with spacing $\Delta x=0.01$. Typically, the
grids of size $120\times 120$ were used, the relative errors of the final
solutions being $\lesssim 10^{-4}$. To check our results for the
correctness, we evaluated the values of the topological charges of the
components by direct integration of expressions (\ref{charge}).

Initial configurations were taken as per the straightforward hedgehog
ansatz,
\begin{equation}
\phi _{1}^{(a)}=\sin \left( f(\rho )\right) \cos (B^{(a)}\theta ),\quad \phi
_{2}^{(a)}=\sin \left( f(\rho )\right) \sin (B^{(a)}\theta ),\quad \phi
_{3}^{(a)}=\cos \left( f(\rho )\right) ,  \label{rot-inv-ans}
\end{equation}%
where the input profile function is $f(\rho )=4\arctan \left( e^{-\rho
}\right) $. Evidently, this corresponds to the configuration with
topological charge $B^{(a)}$ and standard boundary conditions (b.c.) imposed
on the profile function, $f(\rho )$, in both sectors of the dual-core system
(\ref{Lag}). Note that ansatz (\ref{rot-inv-ans}) is rotationally invariant,
because the spatial $SO(2)$ rotation about the $z$-axis is equivalent to the
$O(2)$ iso-rotation about component $\phi _{3}^{(a)}$. However, in our
calculations we do not adopt any \textit{a priori }assumptions about spatial
symmetries of components of the field configuration, $\mathbf{\phi }^{(a)}$.

As said above [see Eq. (\ref{1})], for component $\mathbf{\phi }^{(1)}$ b.c.
is chosen as
\begin{equation}
\phi _{1}^{(1)}\biggl.\biggr|_{\rho \rightarrow \infty }\!\!\!\rightarrow
0\,,~~\phi _{2}^{(1)}\biggl.\biggr|_{\rho \rightarrow \infty
}\!\!\!\rightarrow 0\,,~~\phi _{3}^{(1)}\biggl.\biggr|_{\rho \rightarrow
\infty }\!\!\!\rightarrow 1\,,  \label{infty1}
\end{equation}%
while the second component is subject to b.c.
\begin{equation}
\partial _{\rho }\phi _{1}^{(2)}\biggl.\biggr|_{\rho \rightarrow \infty
}\!\!\!\rightarrow 0\,,~~\phi _{2}^{(2)}\biggl.\biggr|_{\rho \rightarrow
\infty }\!\!\!\rightarrow 0\,,~~\partial _{\rho }\phi _{3}^{(2)}\biggl.%
\biggr|_{\rho \rightarrow \infty }\!\!\!\rightarrow 0\,.  \label{infty2}
\end{equation}%
Then, the total potential energy of system (\ref{pot}) can be minimized.
Indeed, b.c. (\ref{infty1}), (\ref{infty2}), along with the restriction of
the fields to the unit sphere [Eq. (\ref{unity})], demonstrates that, as $%
\rho \rightarrow \infty $, the potential tends to $U\left( \mathbf{\phi }%
^{(1)},\mathbf{\phi }^{(2)}\right) _{\infty }=\mu _{2}^{2}\left( \phi
_{1}^{(2)}\right) ^{2}+\kappa \phi _{1}^{(2)}$, thus the originally unknown
asymptotic values of the components $\phi _{1}^{(2)}$ and $\phi _{3}^{(2)}$
depend on mass parameter $\mu _{2}$ and coupling constant $\kappa $.

Furthermore, the total asymptotic potential possess a minimum if ${\phi
_{1}^{(2)}}\left. {}\right\vert _{\infty }=-\kappa /\left( 2\mu
_{2}^{2}\right) \in \lbrack -1,1]$. We fix the scales by setting $\mu _{2}=1$%
, hence there is a critical value of the coupling,
\begin{equation}
\kappa _{\mathrm{cr}}=2 \, .  \label{cr}
\end{equation}

\begin{figure}[th]
\refstepcounter{fig}
\par
\begin{center}
\hspace{-9.5 cm} \setlength{\unitlength}{1cm} %
\includegraphics[height=7.9cm,angle=0,bb=00 00 00 799]{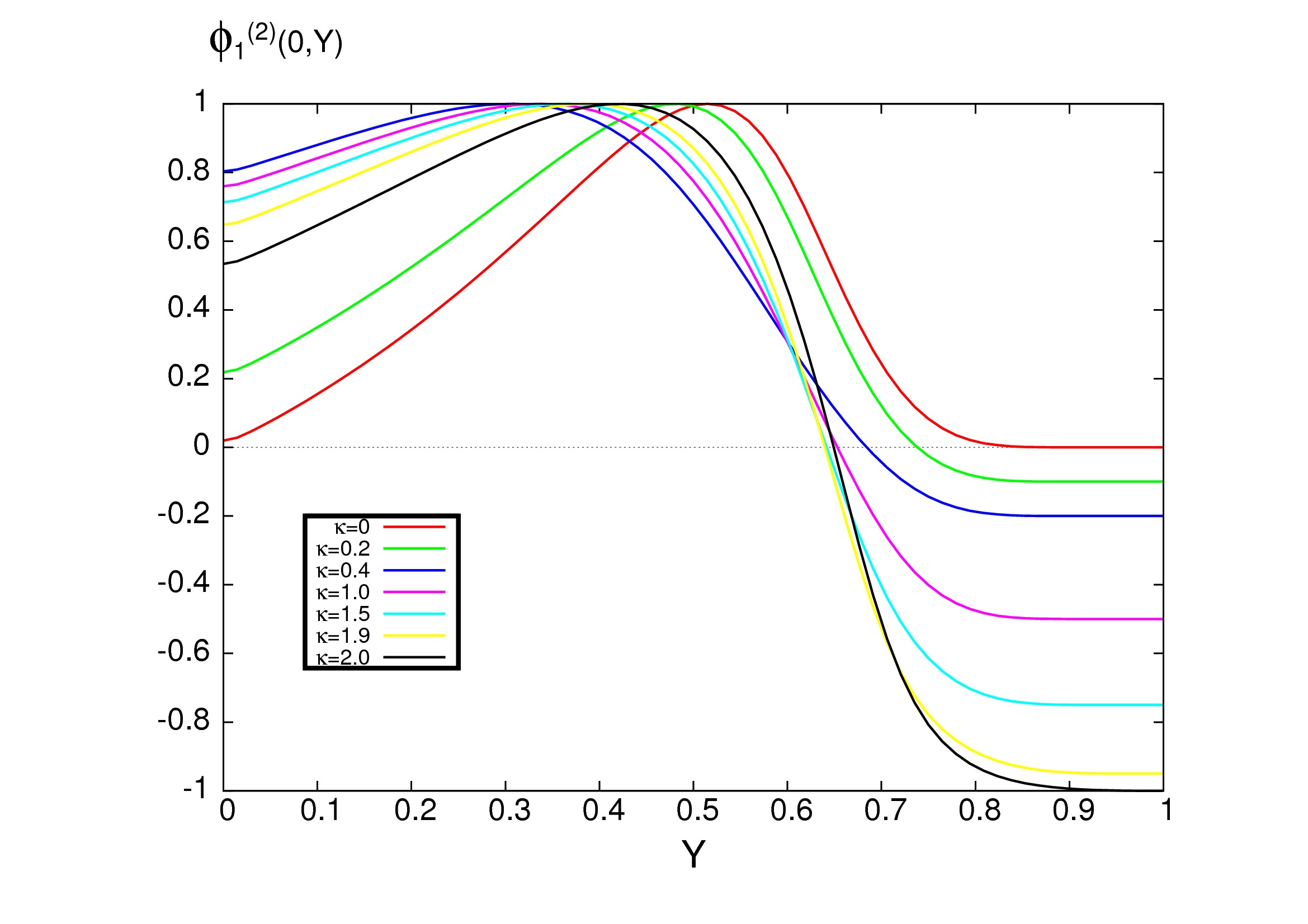}\hspace{%
7.8 cm} \includegraphics[height=7.9cm,angle=0,bb=00 00 00 799]{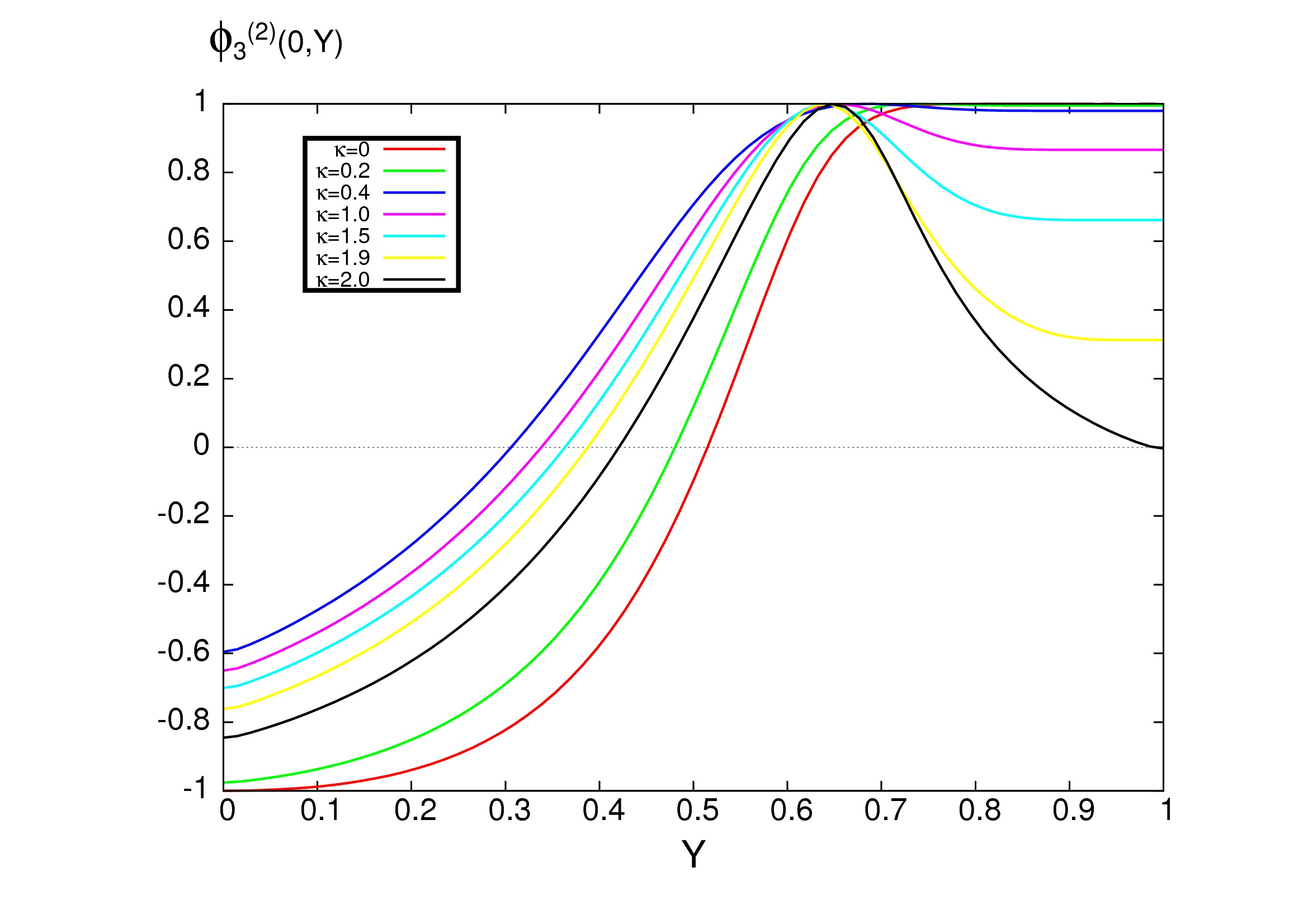}
%
\end{center}
\caption{(Color online) Field components $\protect\phi _{1}^{(2)}$ and $%
\protect\phi _{3}^{(2)}$ of the $(1,1)$ configuration along the positive $y$
axis compactified onto the unit interval, $\ Y=y/(1+y)\in $ $[0,1]$, cf. Eq.
(\protect\ref{xi}), in the model with potential (\protect\ref{pot}) at $%
\protect\kappa =0;0.2;0.4;1.0;1.5;1.9;2.0$. }
\label{f-19}
\end{figure}

Indeed, it is seen in Fig.~\ref{f-19}, which displays components $\phi
_{1}^{(2)}$ and $\phi _{3}^{(2)}$ of the $\left( B^{(1)},B^{(2)}\right)
=(1,1)$ configurations along the compactified $y$ axis, for some set of
values of coupling $\kappa $, that, as the coupling approaches the critical
value (\ref{cr}), the asymptotic behavior of component $\phi _{3}^{(2)}$
becomes different, ceasing to decay exponentially. Thus, in this limit the
total potential cannot stabilize the corresponding baby skyrmion in the
second sector, where the mode becomes unstable with respect to radiation of
scalar radiation waves.

As the coupling increases above the critical value (\ref{cr}), the
configuration again gets stabilized. Eventually, at large $\kappa $ the
fields approach the vacuum values at a finite distance from the center, and
their asymptotic values cease to vary.

\begin{figure}[tbh]
\refstepcounter{fig} \setlength{\unitlength}{1cm} \centering
\vspace{3.3 cm} \hspace{-0.0 cm} \includegraphics[height=5.2cm,angle=0,bb=00
00 325 725]{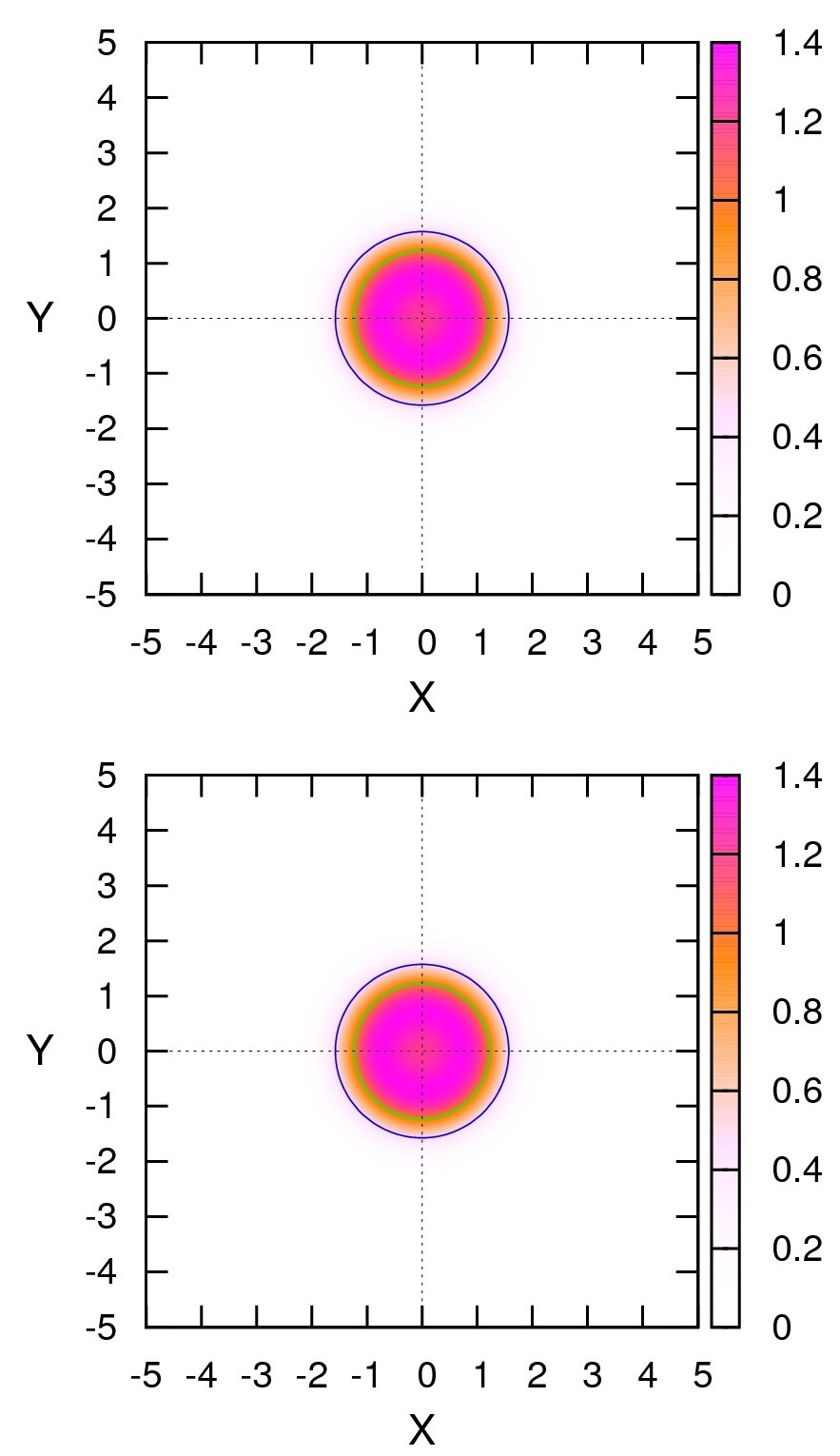}\hspace{0.2 cm} %
\includegraphics[height=5.2cm,angle=0,bb=00 00 325
725]{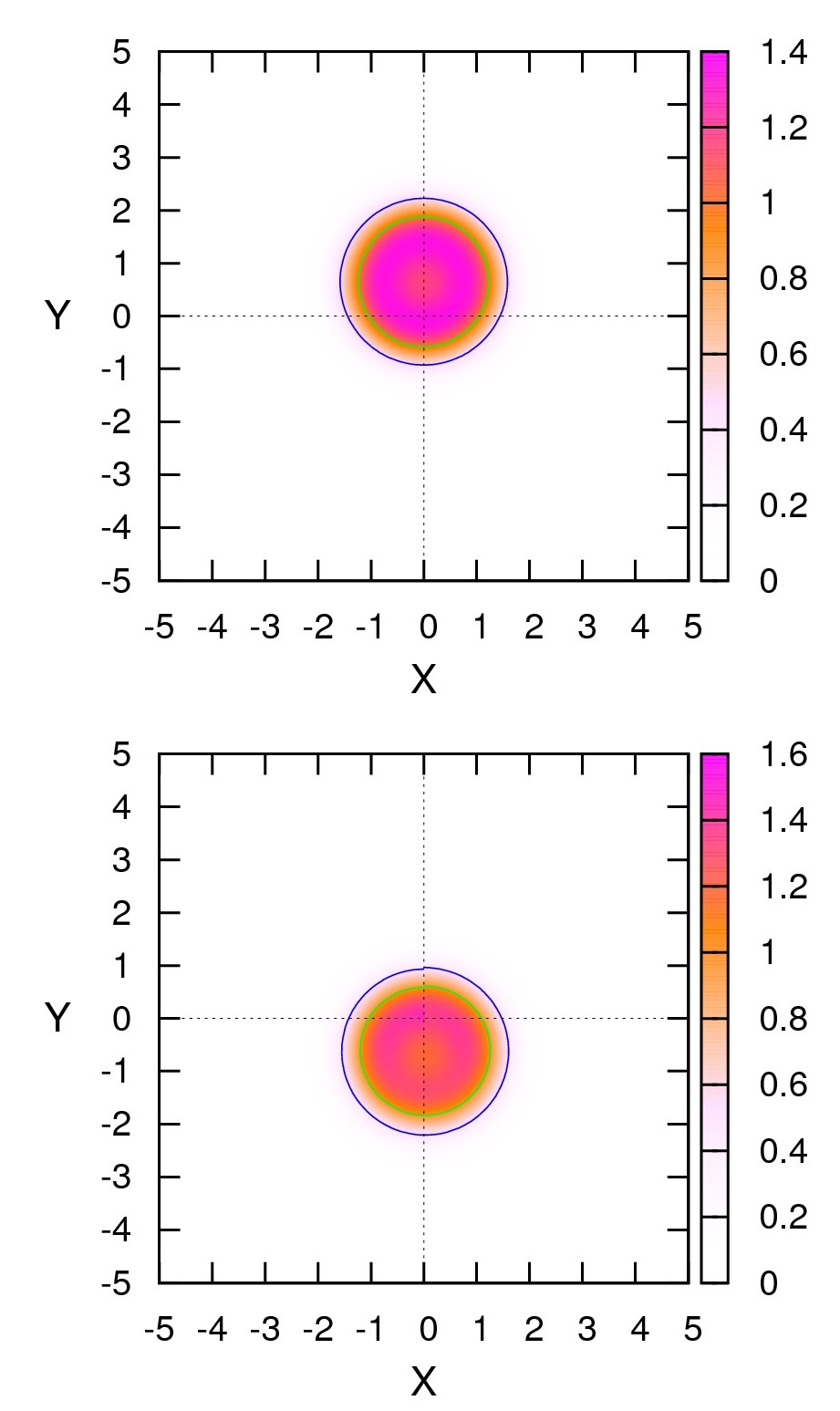}\hspace{0.2 cm} %
\includegraphics[height=5.2cm,angle=0,bb=00 00 325
725]{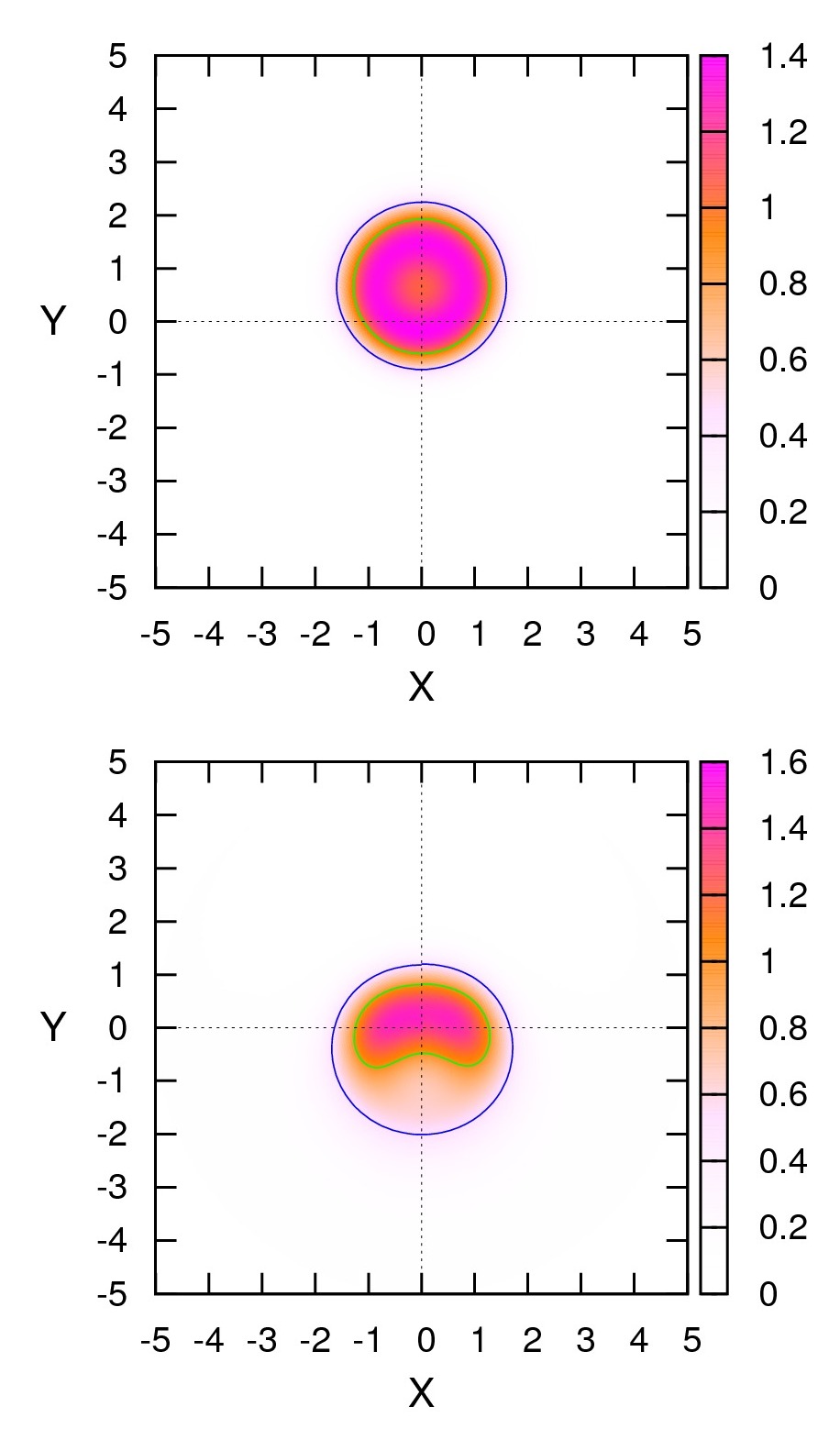}\hspace{0.2 cm} %
\includegraphics[height=5.2cm,angle=0,bb=00 00 325
725]{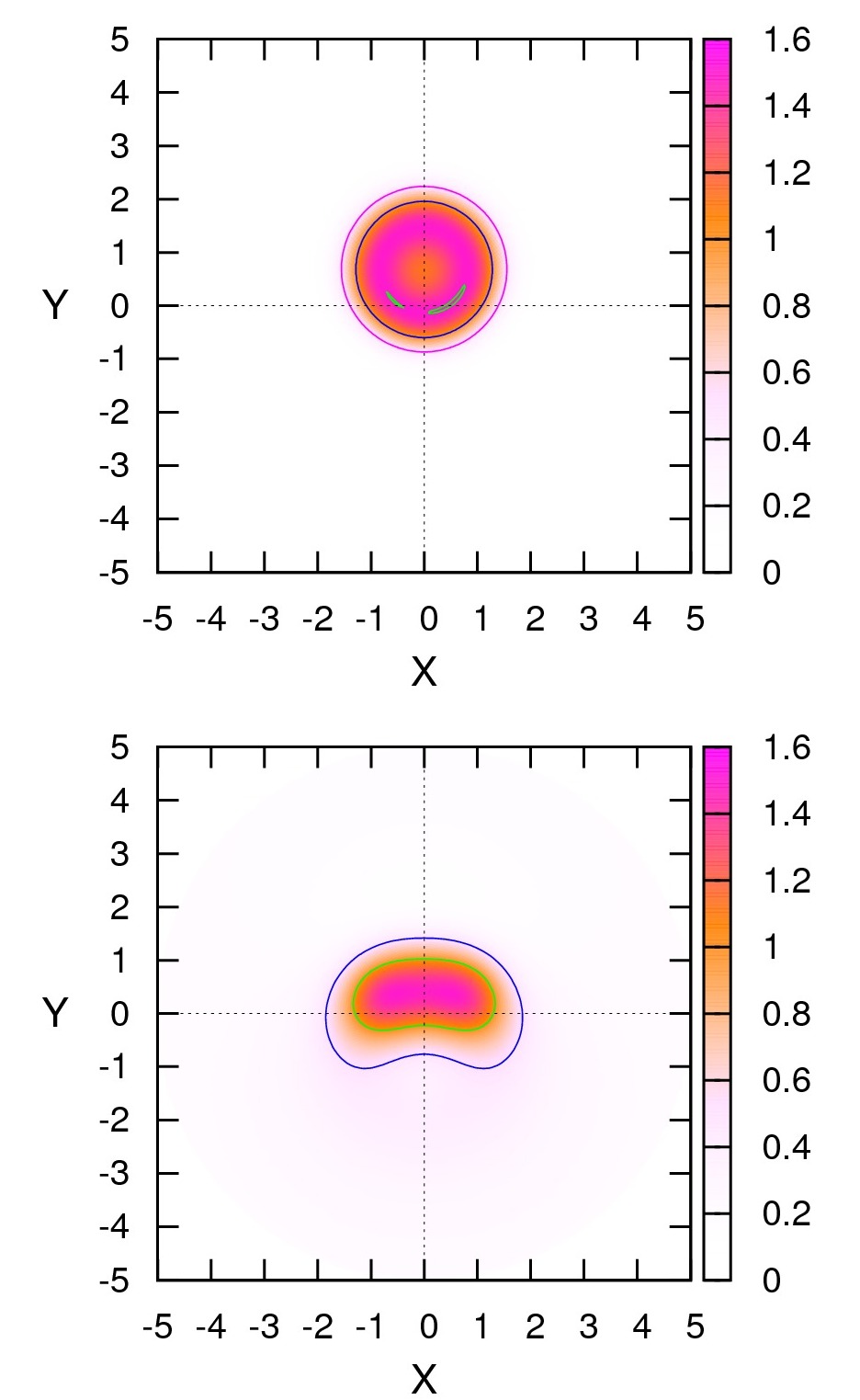}\hspace{0.2 cm} %
\includegraphics[height=5.2cm,angle=0,bb=00 00 325
725]{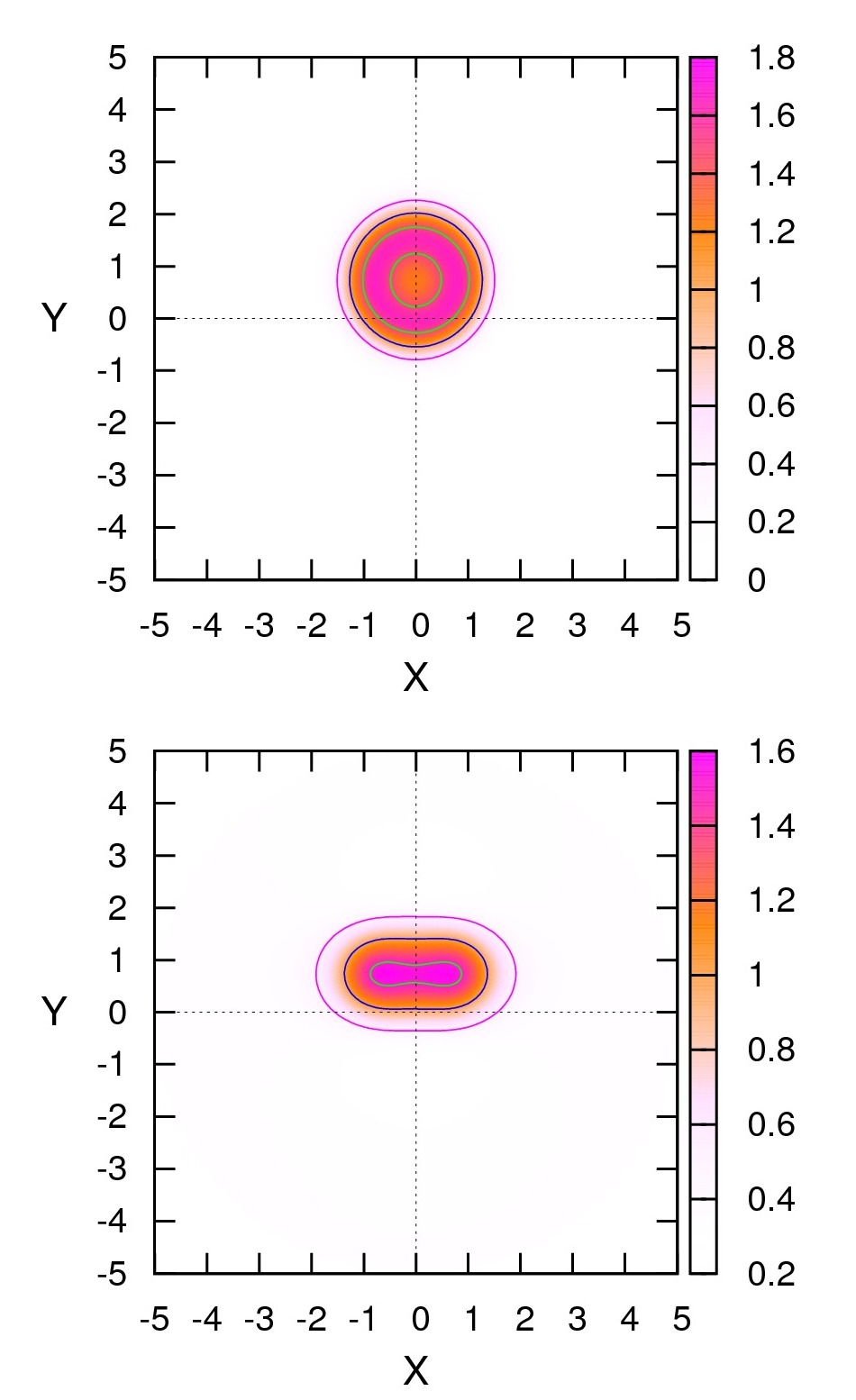}
\caption{(Color online) The upper and lower rows display contour plots of
the energy density in the two sectors of the $(1,1)$ configuration in the
model with potential (\protect\ref{pot}) at $\protect\kappa %
=0,0.1,1.0,1.5,2.0$ (from left to right).}
\label{f-1}
\end{figure}

\begin{figure}[tbh]
\refstepcounter{fig} \setlength{\unitlength}{1cm} \centering
\vspace{3.3 cm} \hspace{0.0 cm} \includegraphics[height=5.2cm,angle=0,bb=00
00 325 725]{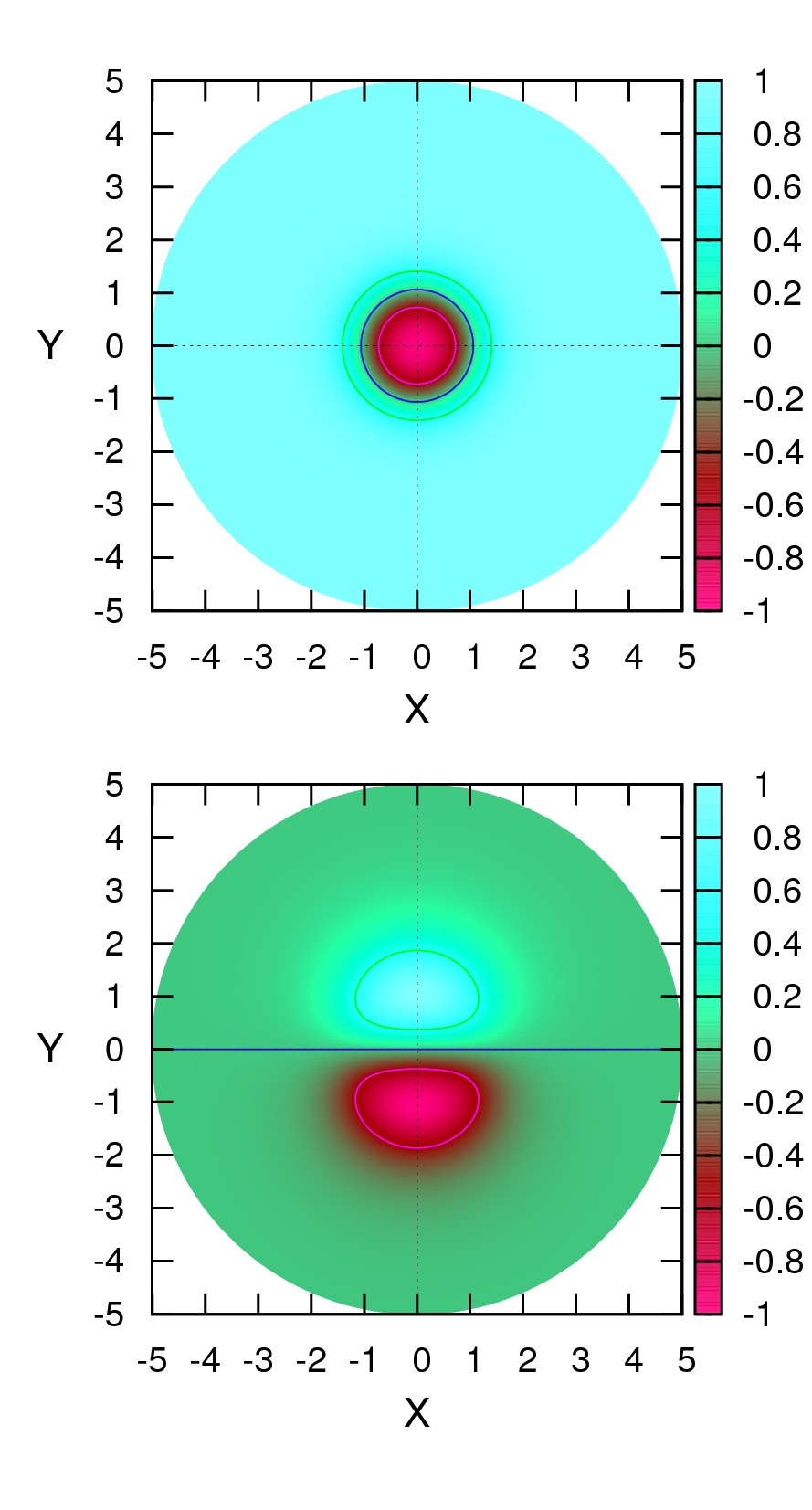}\hspace{0.2 cm} %
\includegraphics[height=5.2cm,angle=0,bb=00 00 325
725]{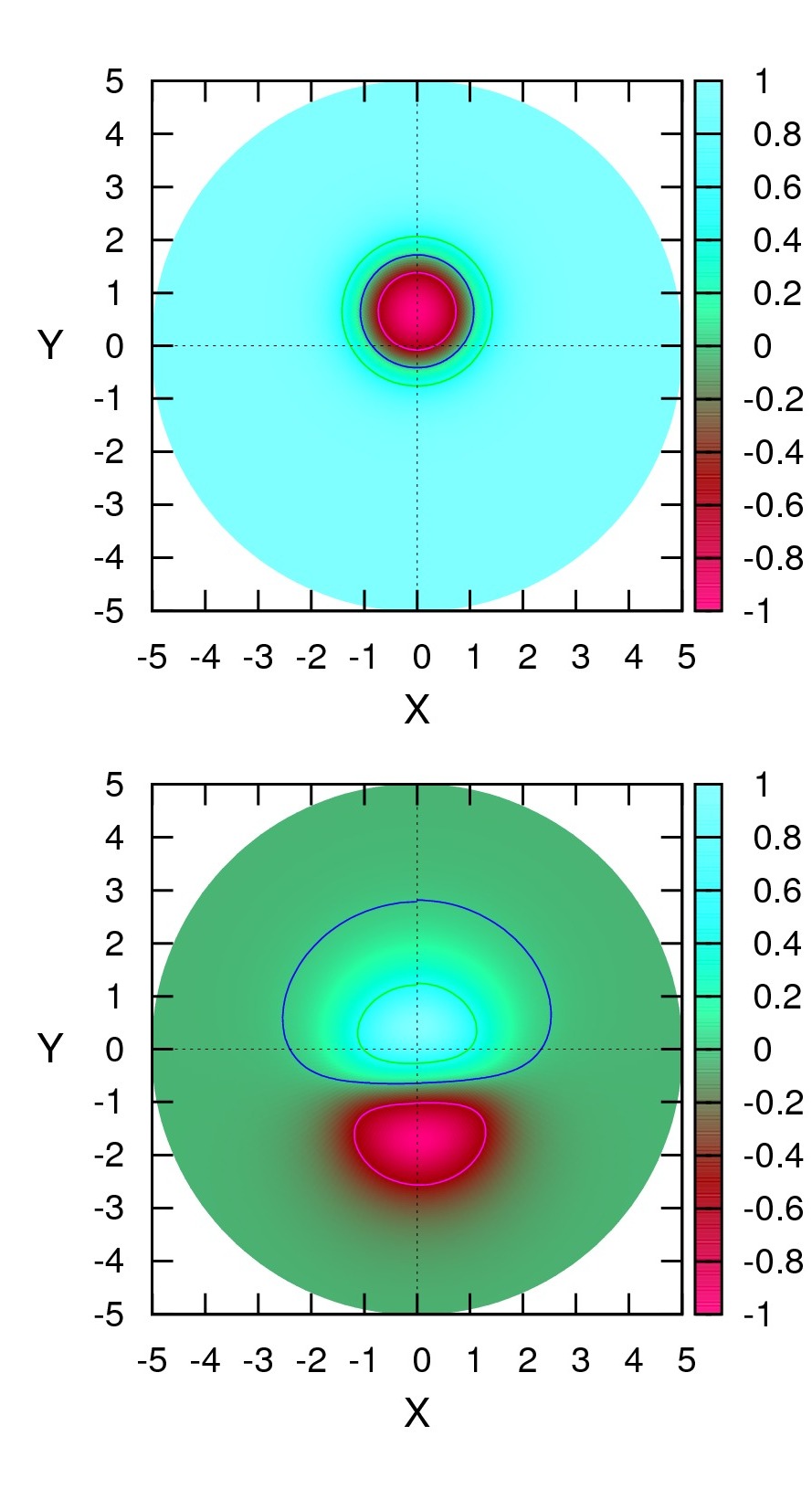}\hspace{0.2 cm} %
\includegraphics[height=5.2cm,angle=0,bb=00 00 325
725]{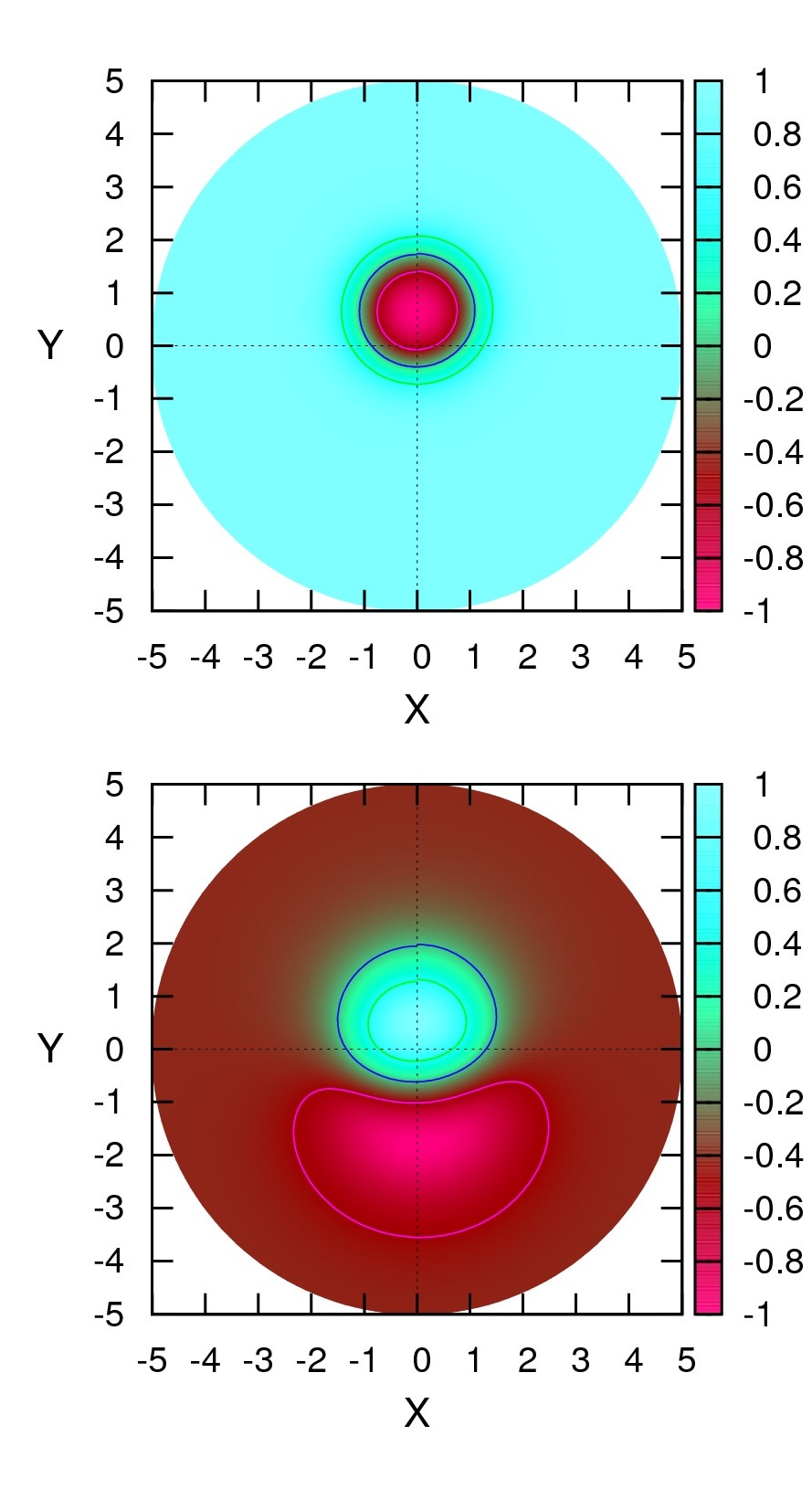}\hspace{0.2 cm} %
\includegraphics[height=5.2cm,angle=0,bb=00 00 325
725]{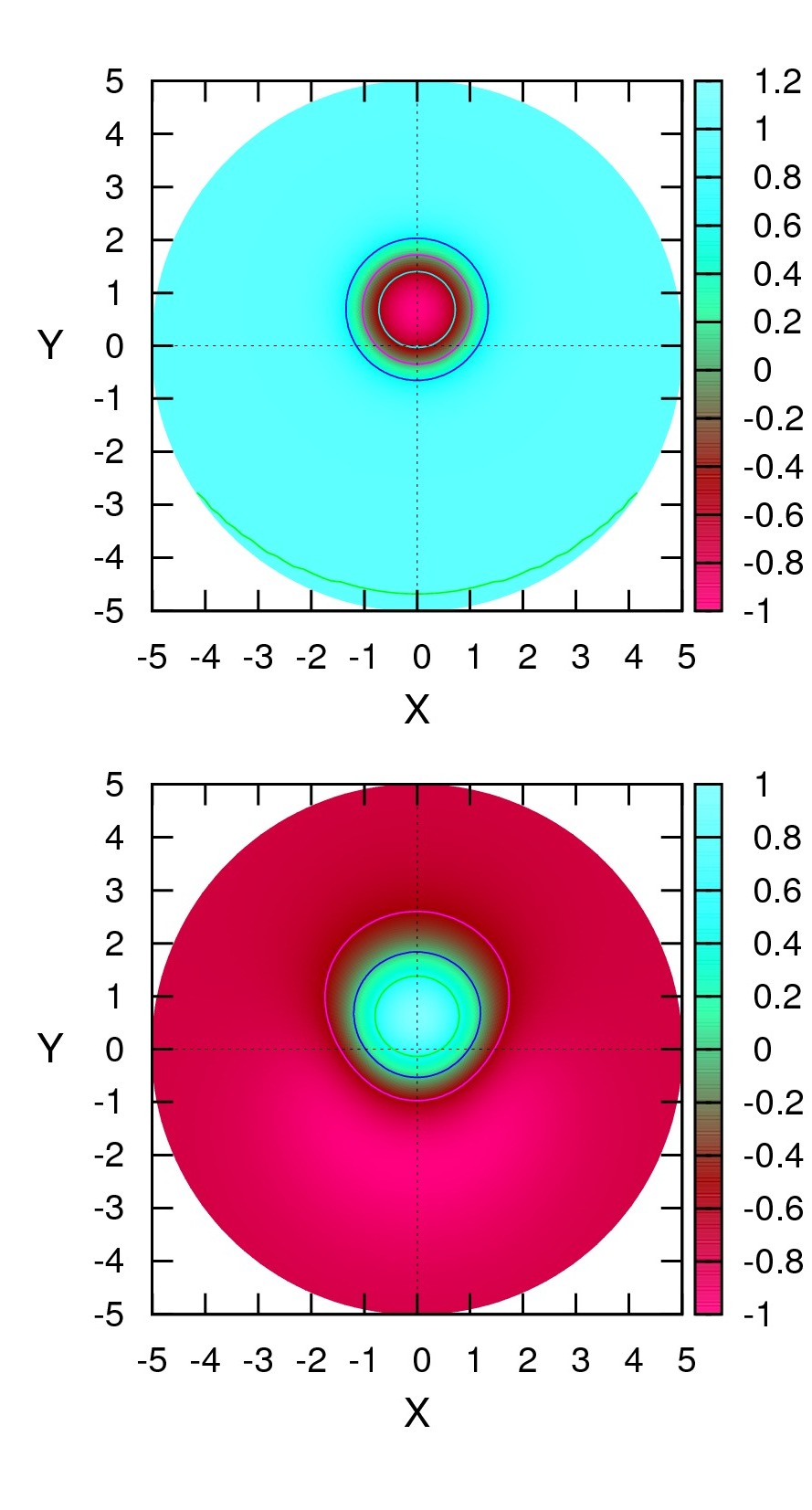}\hspace{0.2 cm} %
\includegraphics[height=5.2cm,angle=0,bb=00 00 325
725]{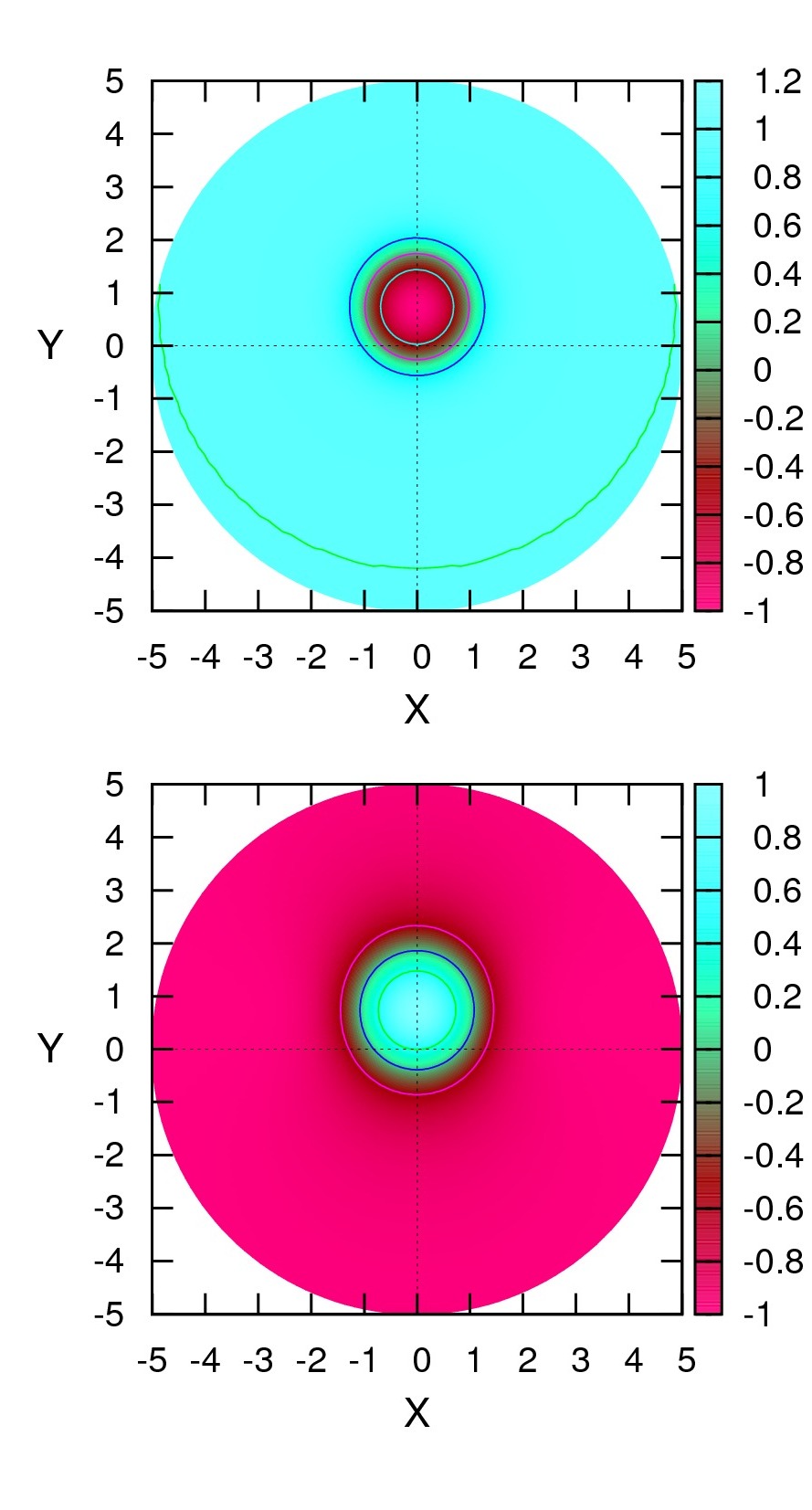}
\caption{(Color online) Contour plots of coupled components $\protect\phi %
_{3}^{(1)}$ and $\protect\phi _{1}^{(2)}$ (the upper and lower rows,
respectively) of the $(1,1)$ configuration in the model with potential (%
\protect\ref{pot}) at $\protect\kappa =0,0.1,0.8,1.5,2.0$ (from left to
right).}
\label{f-4}
\end{figure}

\begin{figure}[th]
\refstepcounter{fig} \setlength{\unitlength}{1cm} \centering
\vspace{3.3 cm} \hspace{-0.0 cm}
\includegraphics[height=5.4cm,angle=0,bb=00
00 325 725]{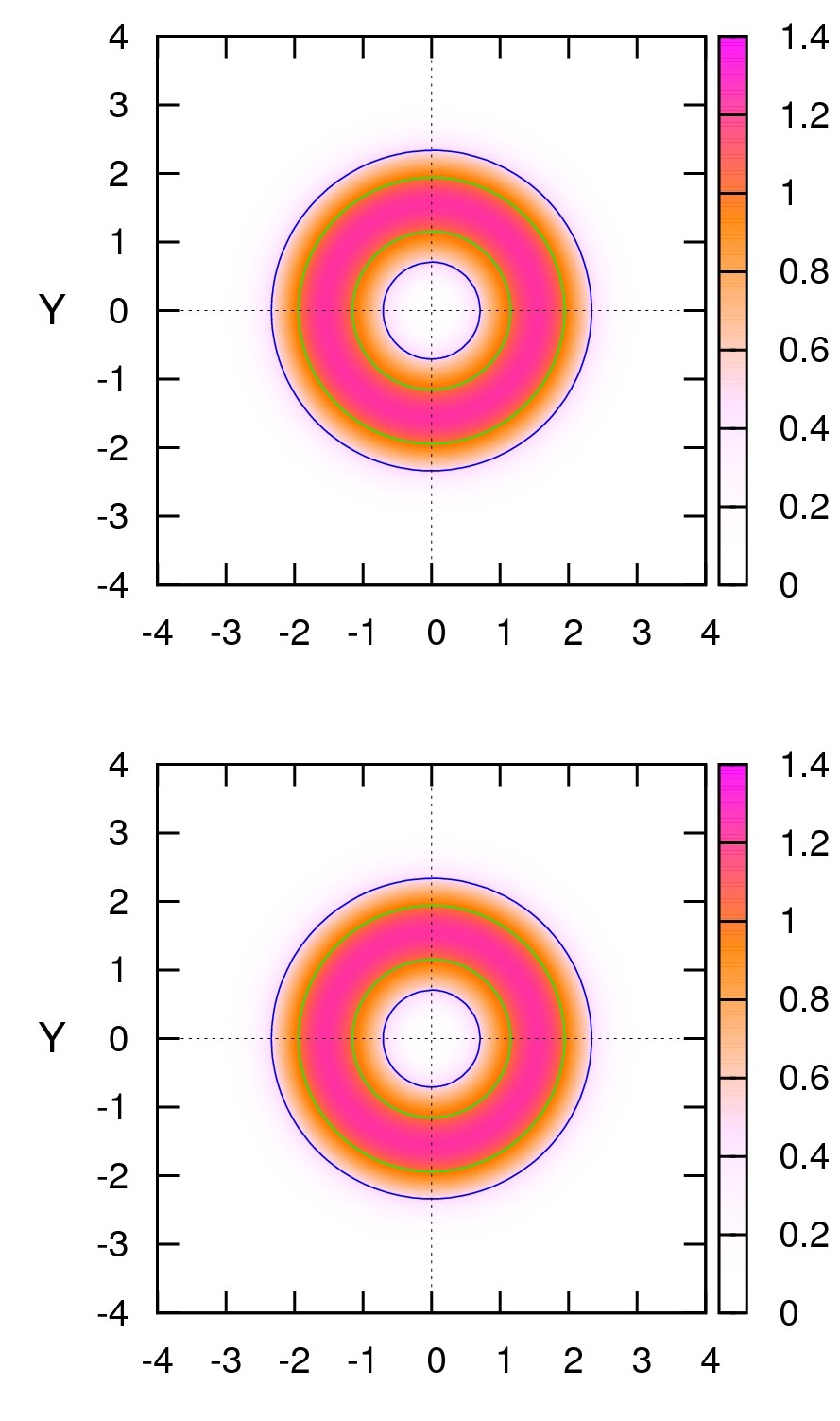}\hspace{0.2 cm} %
\includegraphics[height=5.4cm,angle=0,bb=00 00 325
725]{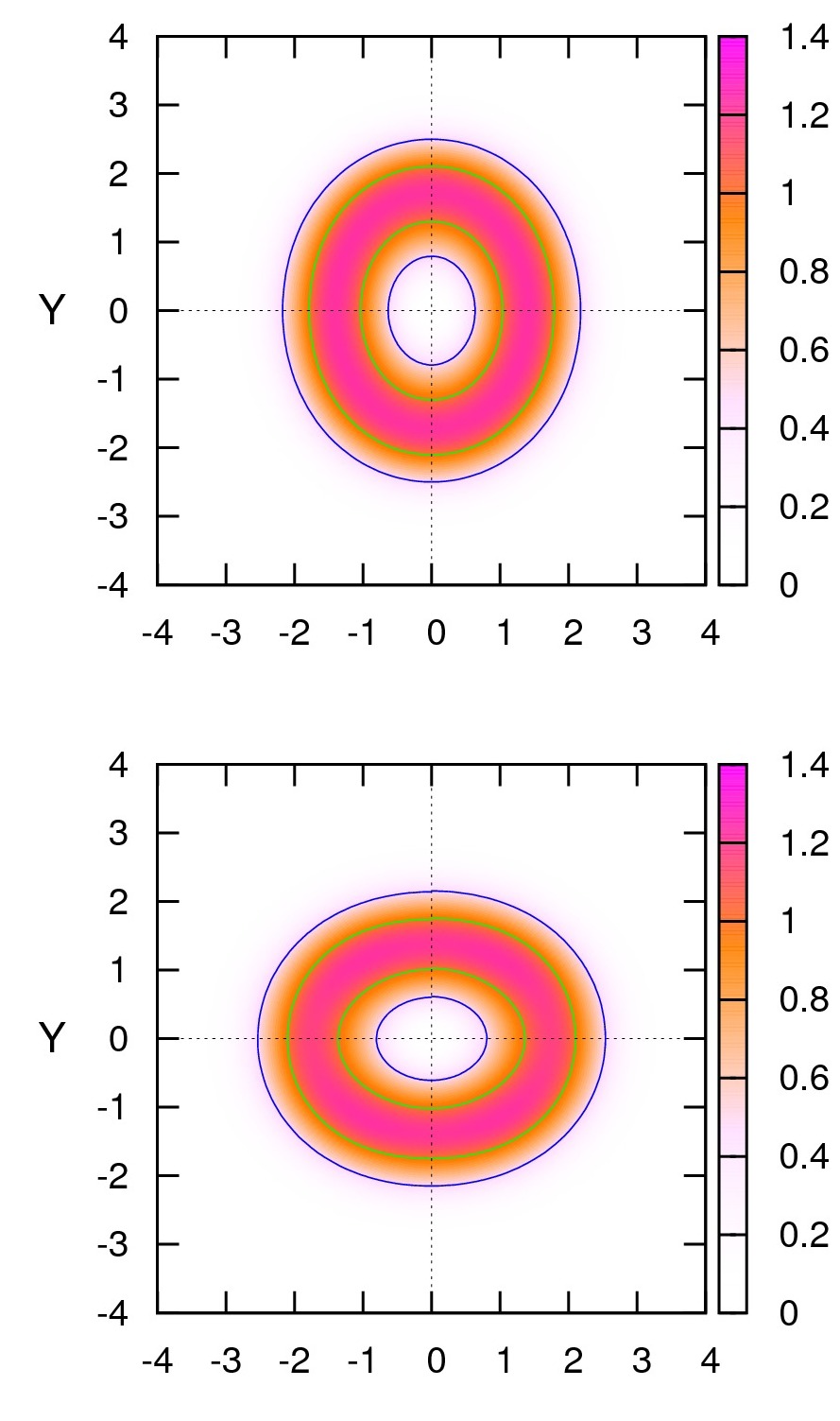}\hspace{0.2 cm} %
\includegraphics[height=5.4cm,angle=0,bb=00 00 325
725]{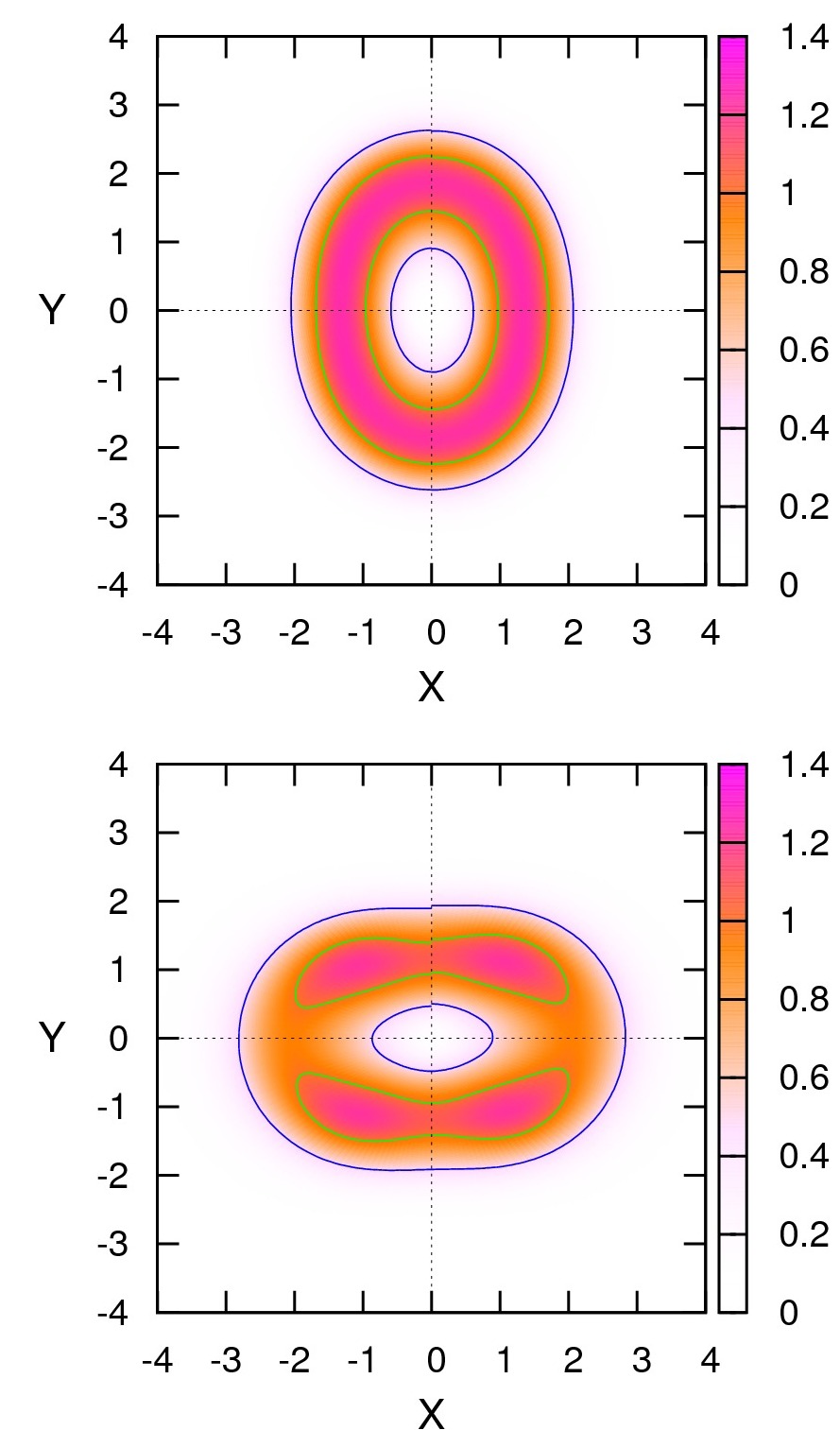}\hspace{0.2 cm} %
\includegraphics[height=5.4cm,angle=0,bb=00 00 325
725]{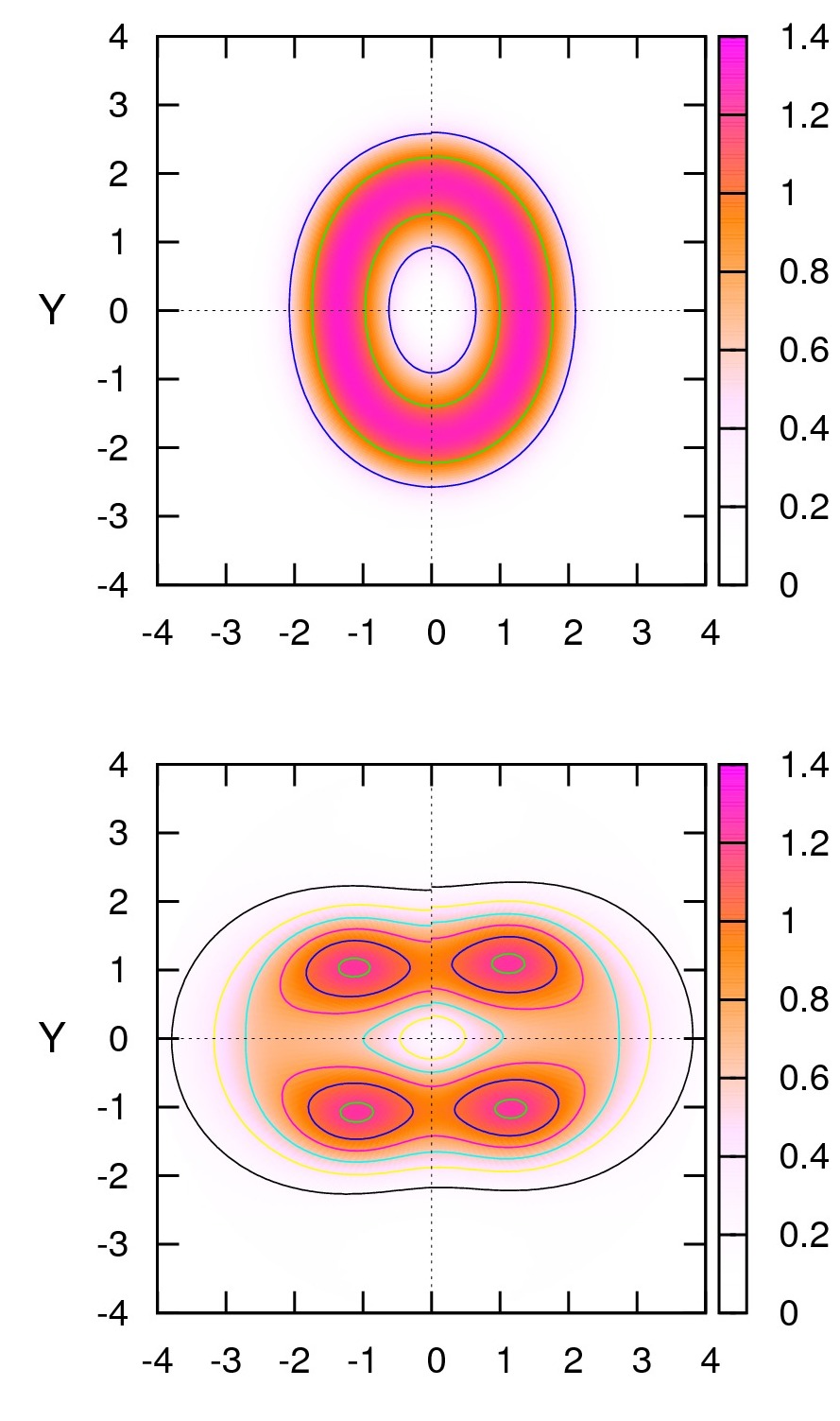}\hspace{0.2 cm} %
\includegraphics[height=5.4cm,angle=0,bb=00 00 325
725]{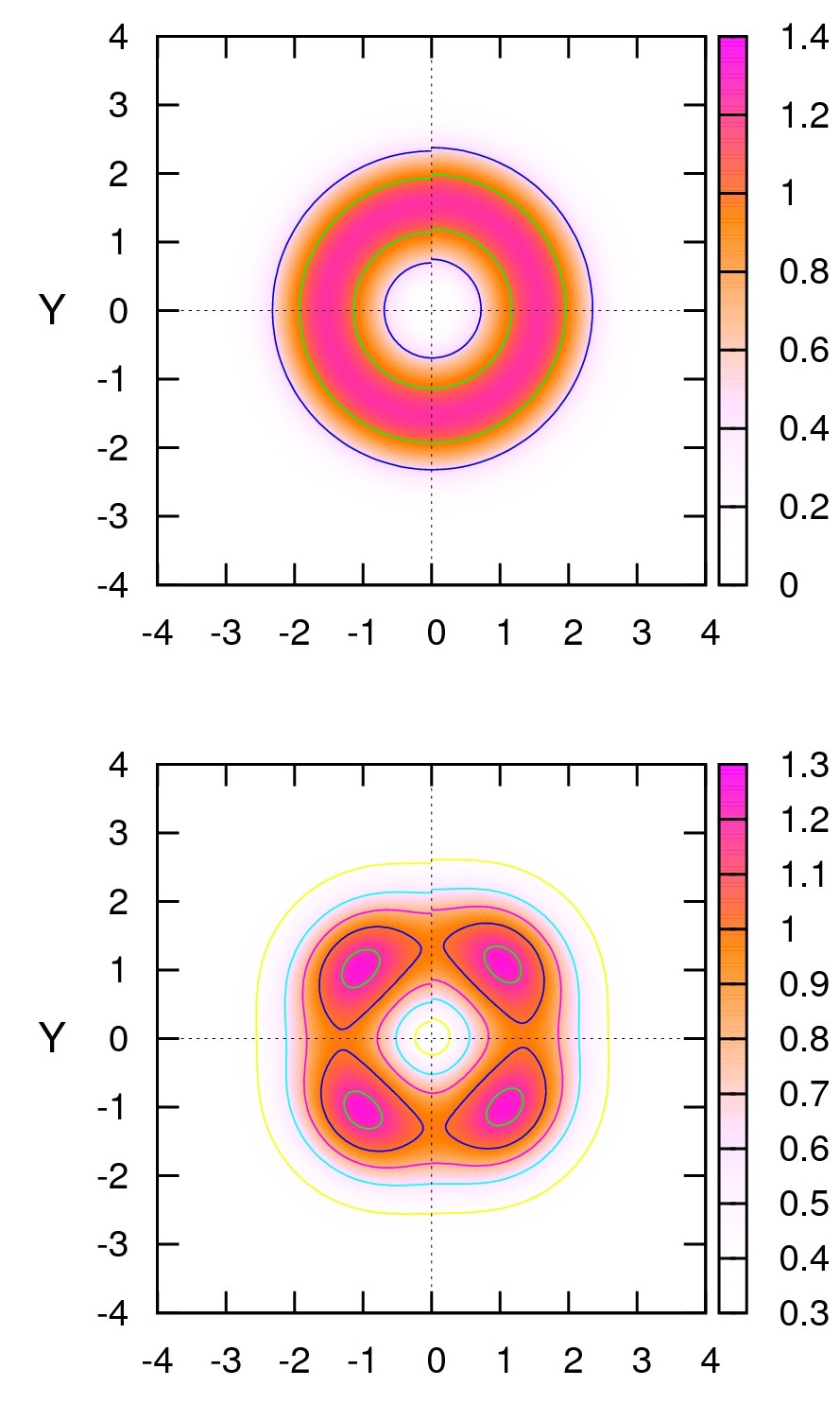}
\caption{(Color online) Contour plots of the energy density of the sectors
in the $(2,2)$ configuration in the model with potential (\protect\ref{pot})
at $\protect\kappa =0,0.2,0.7,1.0,2.0$ (from left to right).}
\label{f-16}
\end{figure}

\begin{figure}[th]
\refstepcounter{fig}\setlength{\unitlength}{1cm} \centering
\vspace{3.3 cm} \hspace{-0.0 cm} \includegraphics[height=5.2cm,angle=0,bb=00
00 325 725]{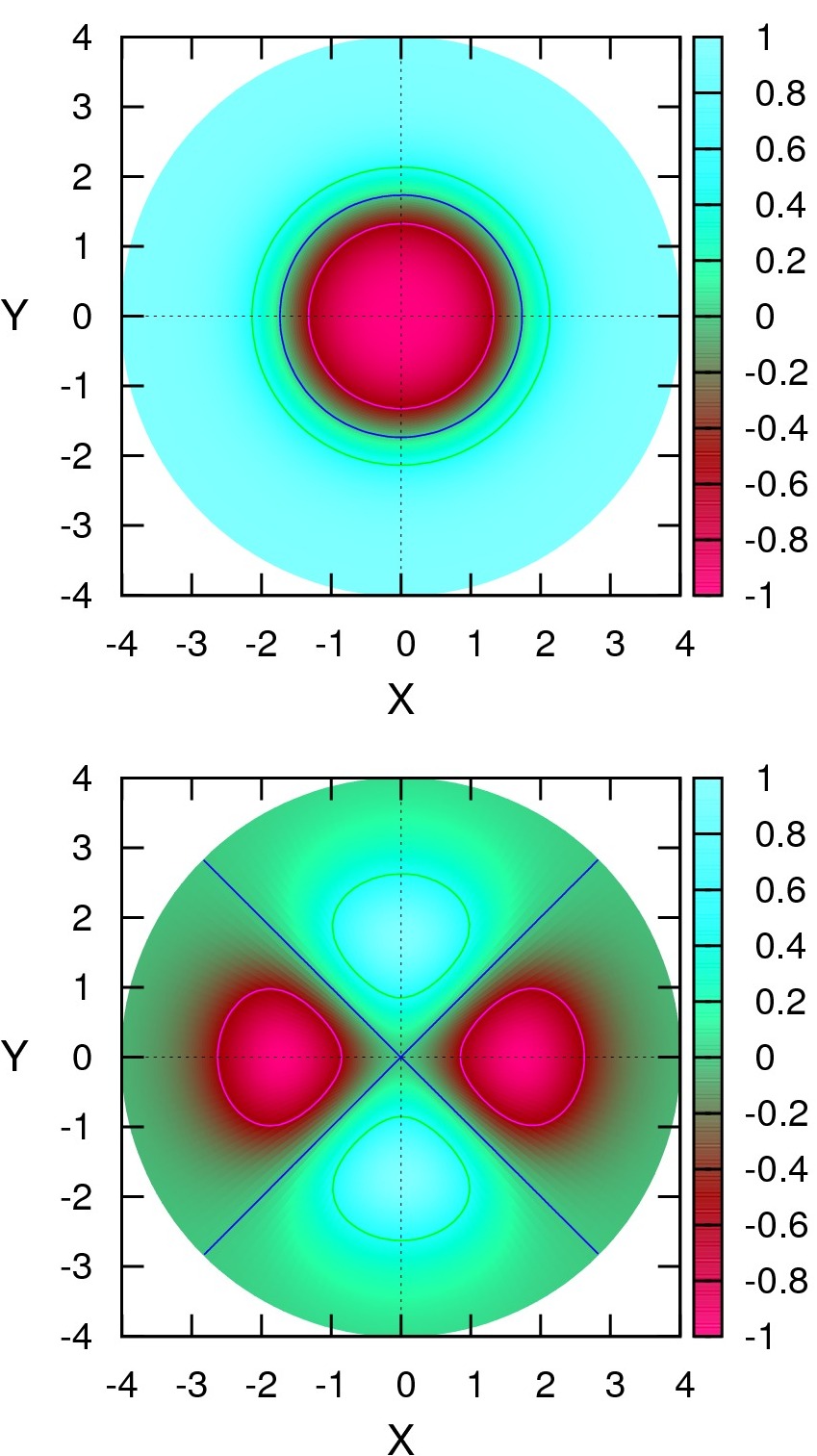}\hspace{0.2 cm} %
\includegraphics[height=5.2cm,angle=0,bb=00 00 325
725]{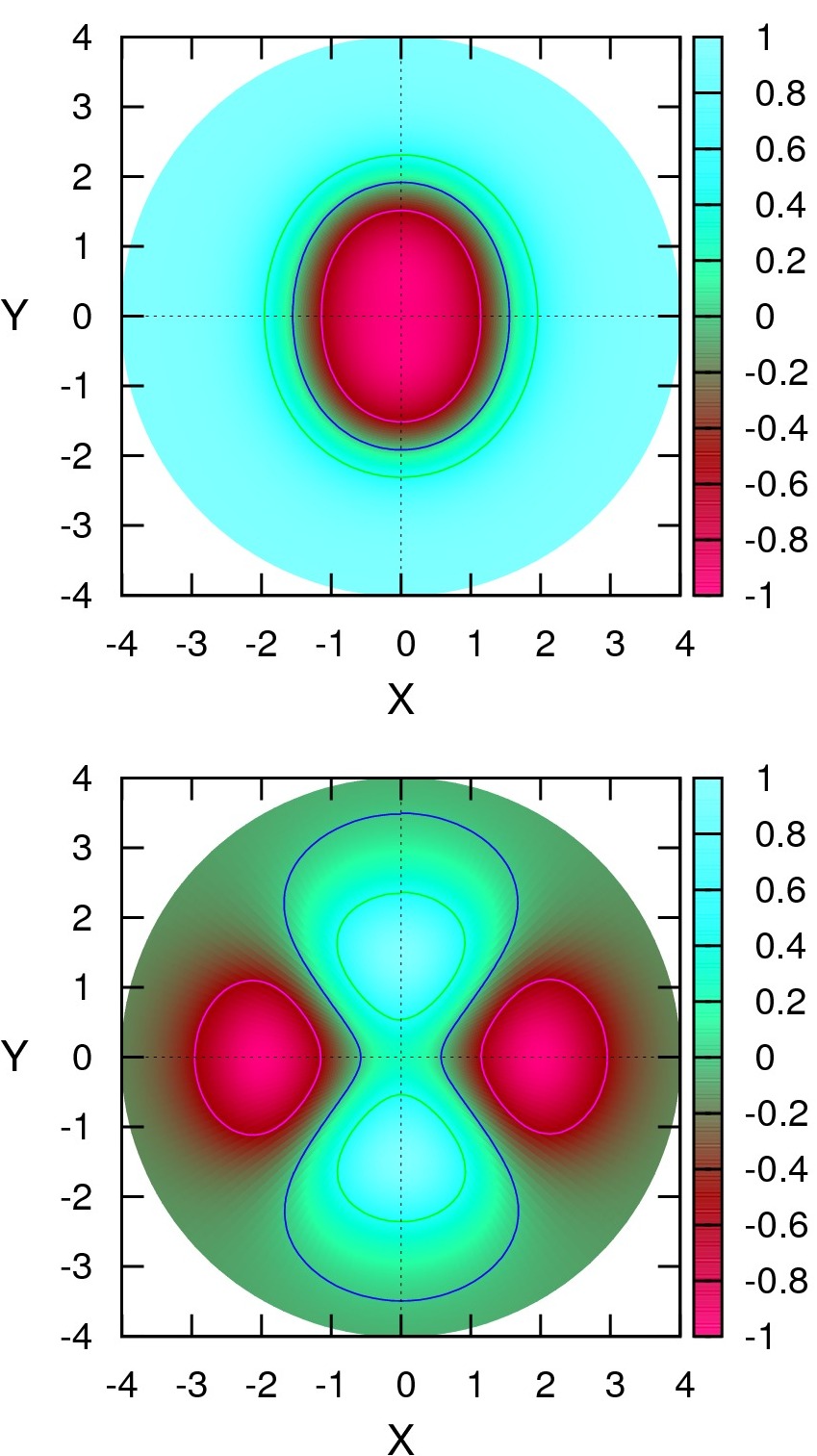}\hspace{0.2 cm} %
\includegraphics[height=5.2cm,angle=0,bb=00 00 325
725]{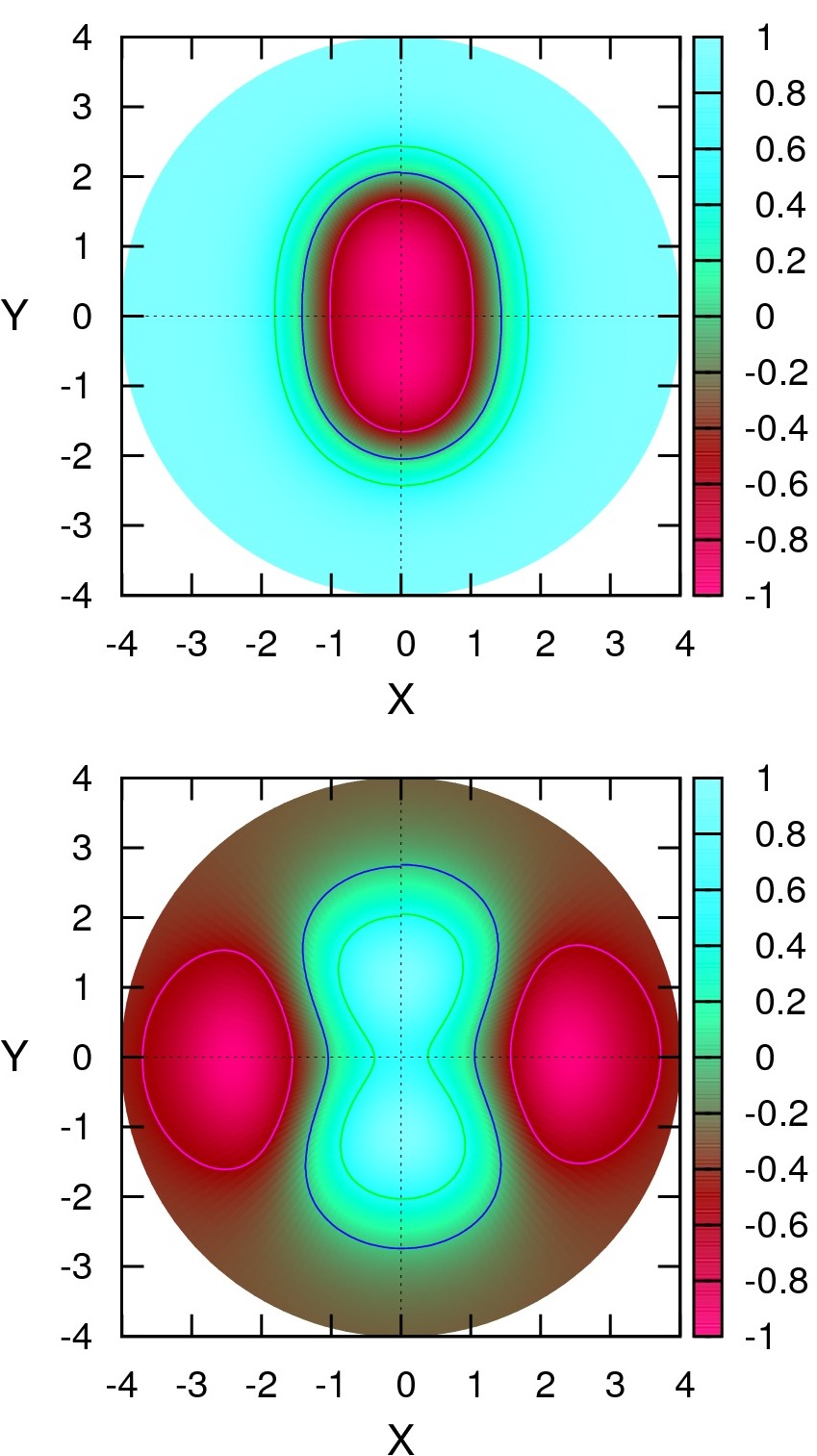}\hspace{0.2 cm} %
\includegraphics[height=5.2cm,angle=0,bb=00 00 325
725]{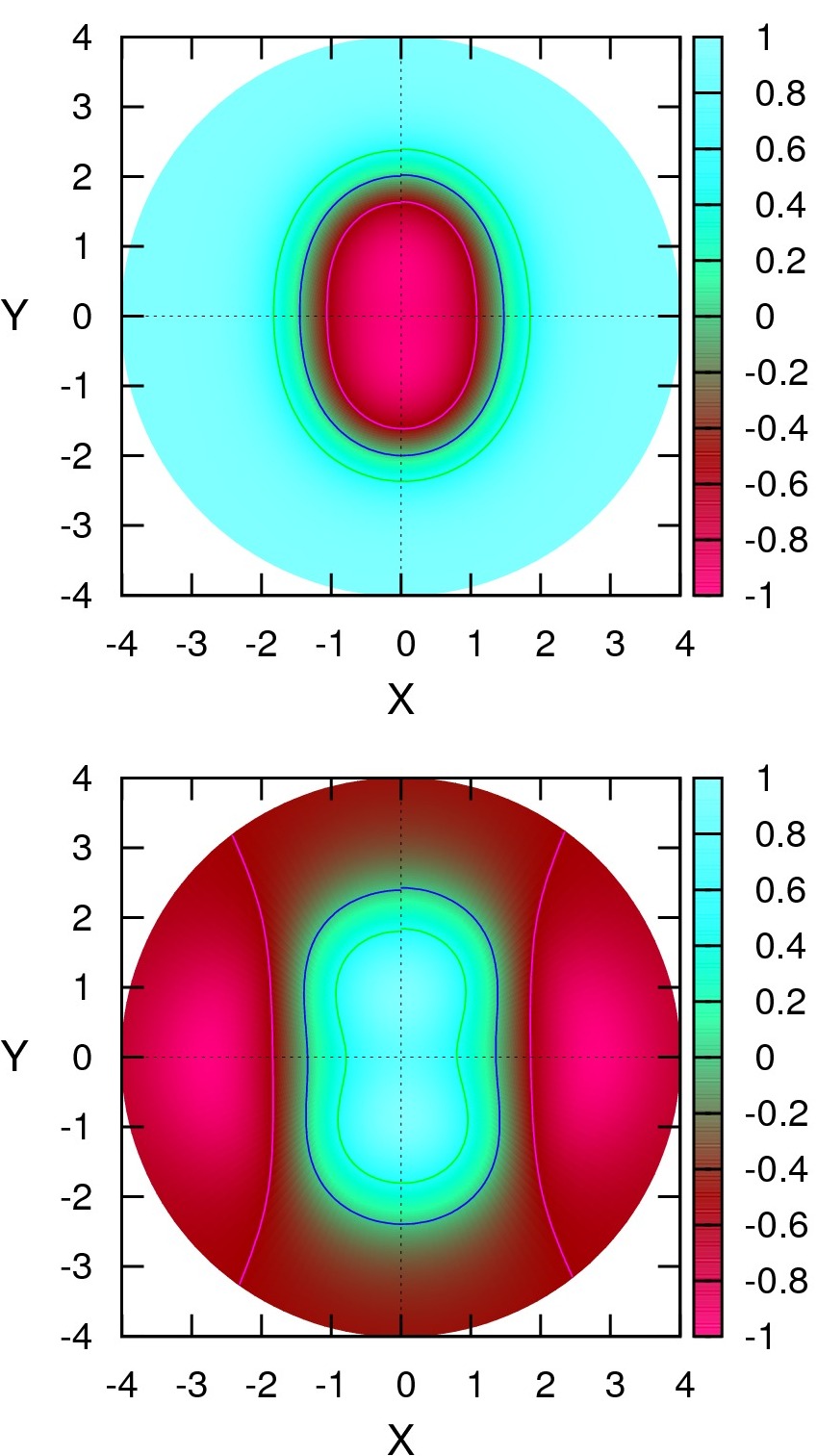}\hspace{0.2 cm} %
\includegraphics[height=5.2cm,angle=0,bb=00 00 325
725]{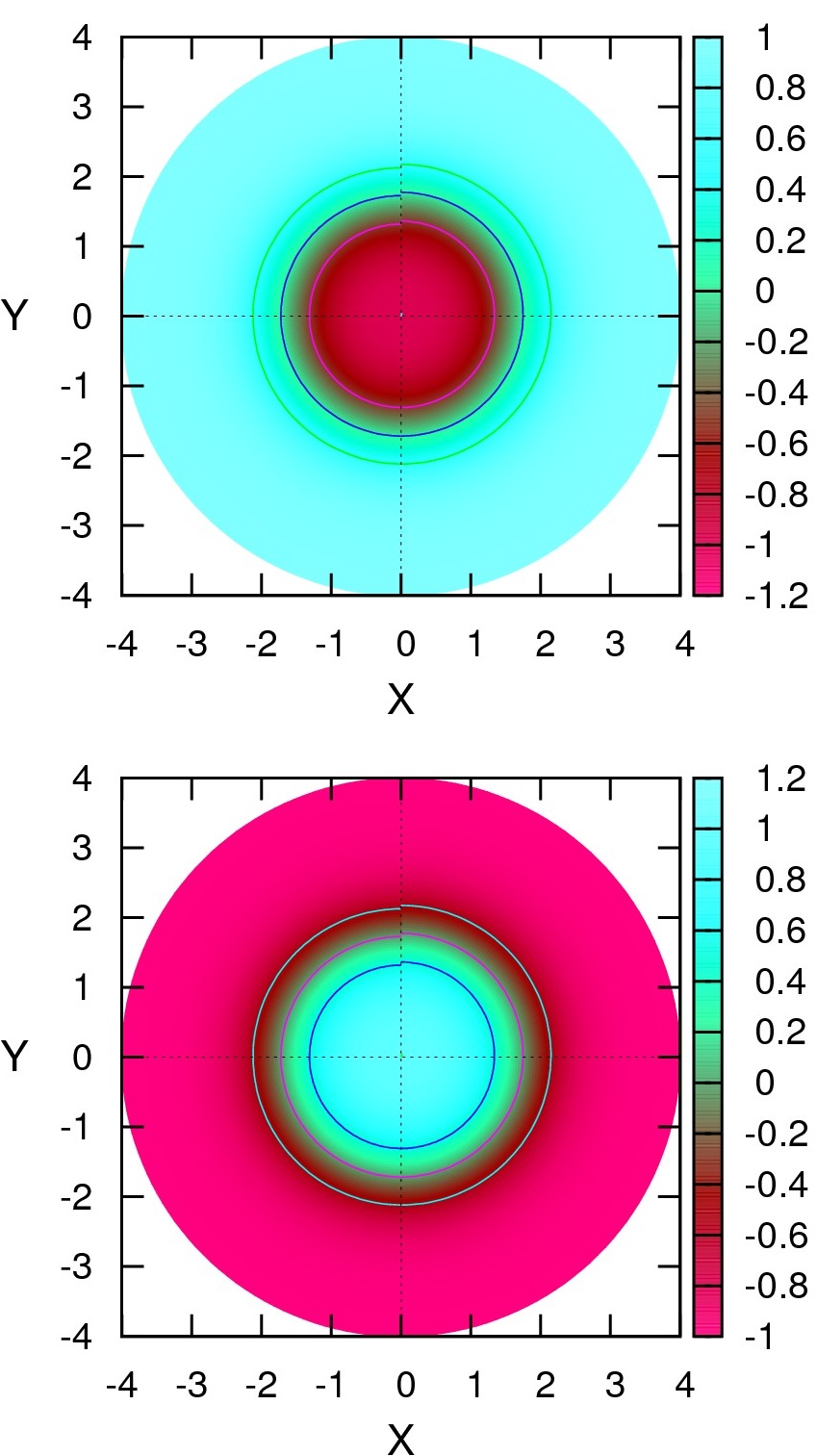}
\caption{(Color online) Contour plots of coupled components $\protect\phi %
_{3}^{(1)}$ and $\protect\phi _{1}^{(2)}$ (the upper and lower rows,
respectively) of the $(2,2)$ configuration in the model with potential (%
\protect\ref{pot}) at $\protect\kappa =0,0.2,0.7,1.0,2.0$ (from left to
right).}
\label{f-4second}
\end{figure}

\begin{figure}[th]
\refstepcounter{fig} \setlength{\unitlength}{1cm} \centering
\vspace{3.3 cm} \hspace{-0.0 cm} \includegraphics[height=5.2cm,angle=0,bb=00
00 325 725]{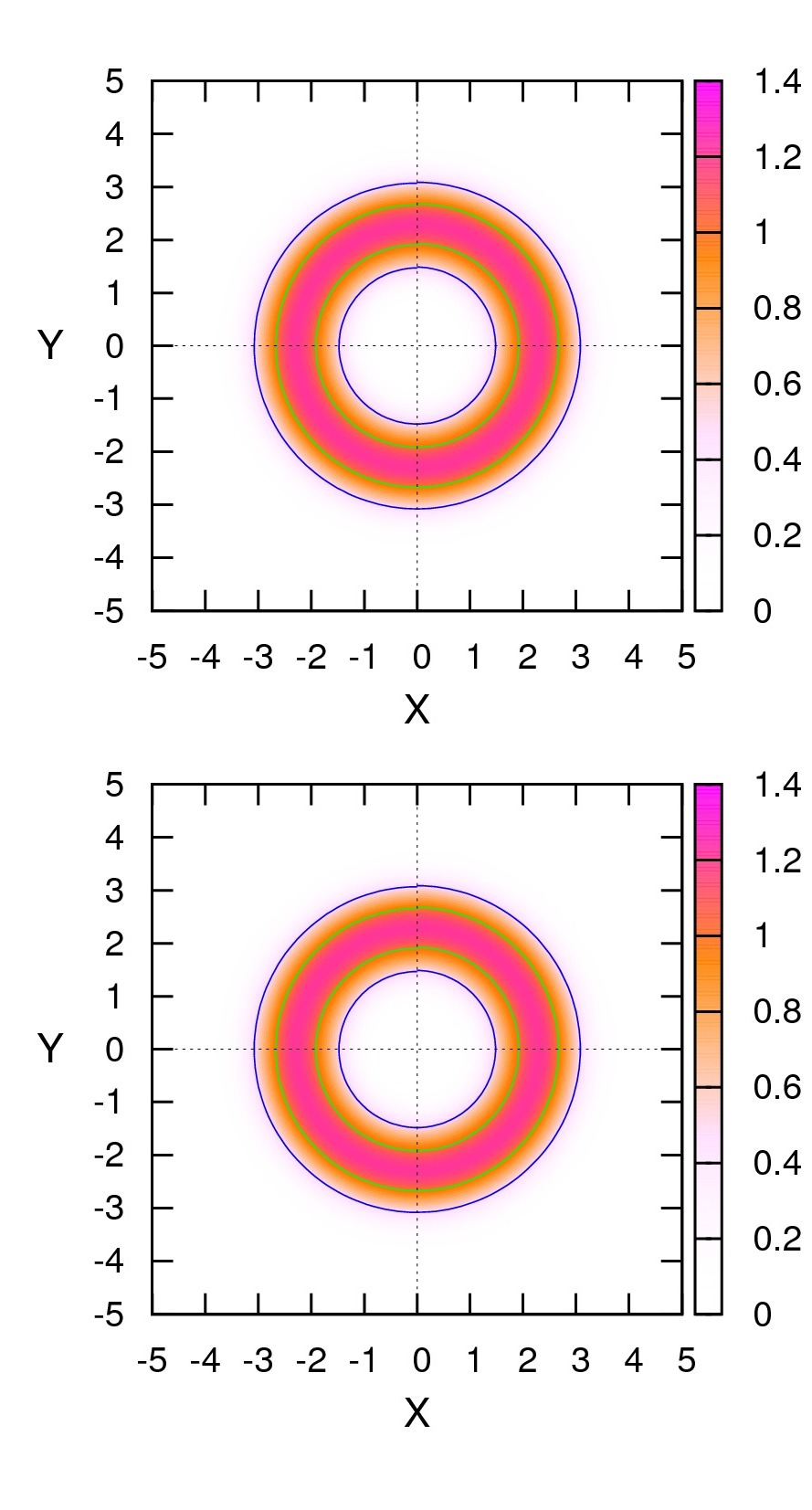}\hspace{0.2 cm} %
\includegraphics[height=5.2cm,angle=0,bb=00 00 325
725]{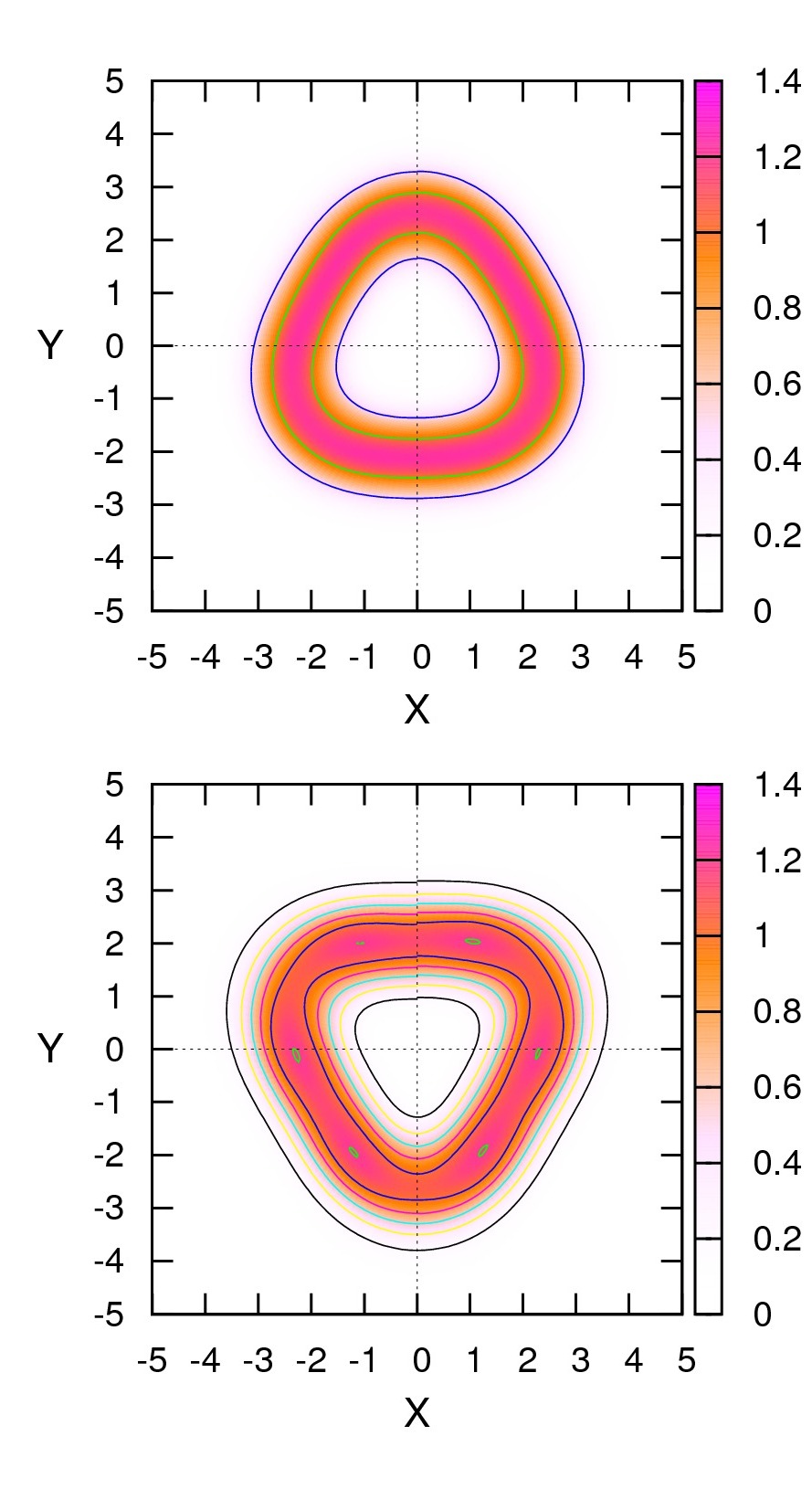}\hspace{0.2 cm} %
\includegraphics[height=5.2cm,angle=0,bb=00 00 325
725]{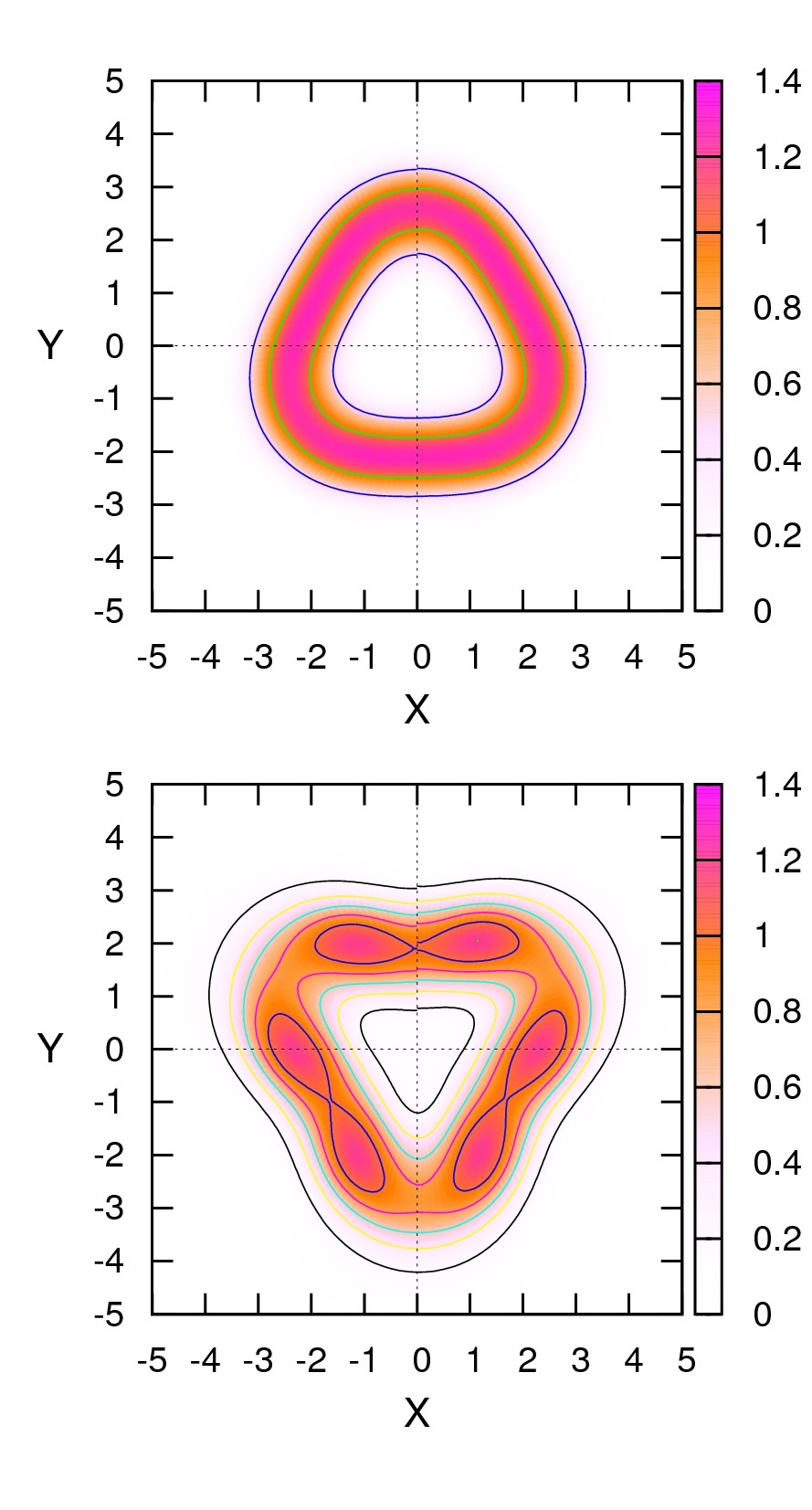}\hspace{0.2 cm} %
\includegraphics[height=5.2cm,angle=0,bb=00 00 325
725]{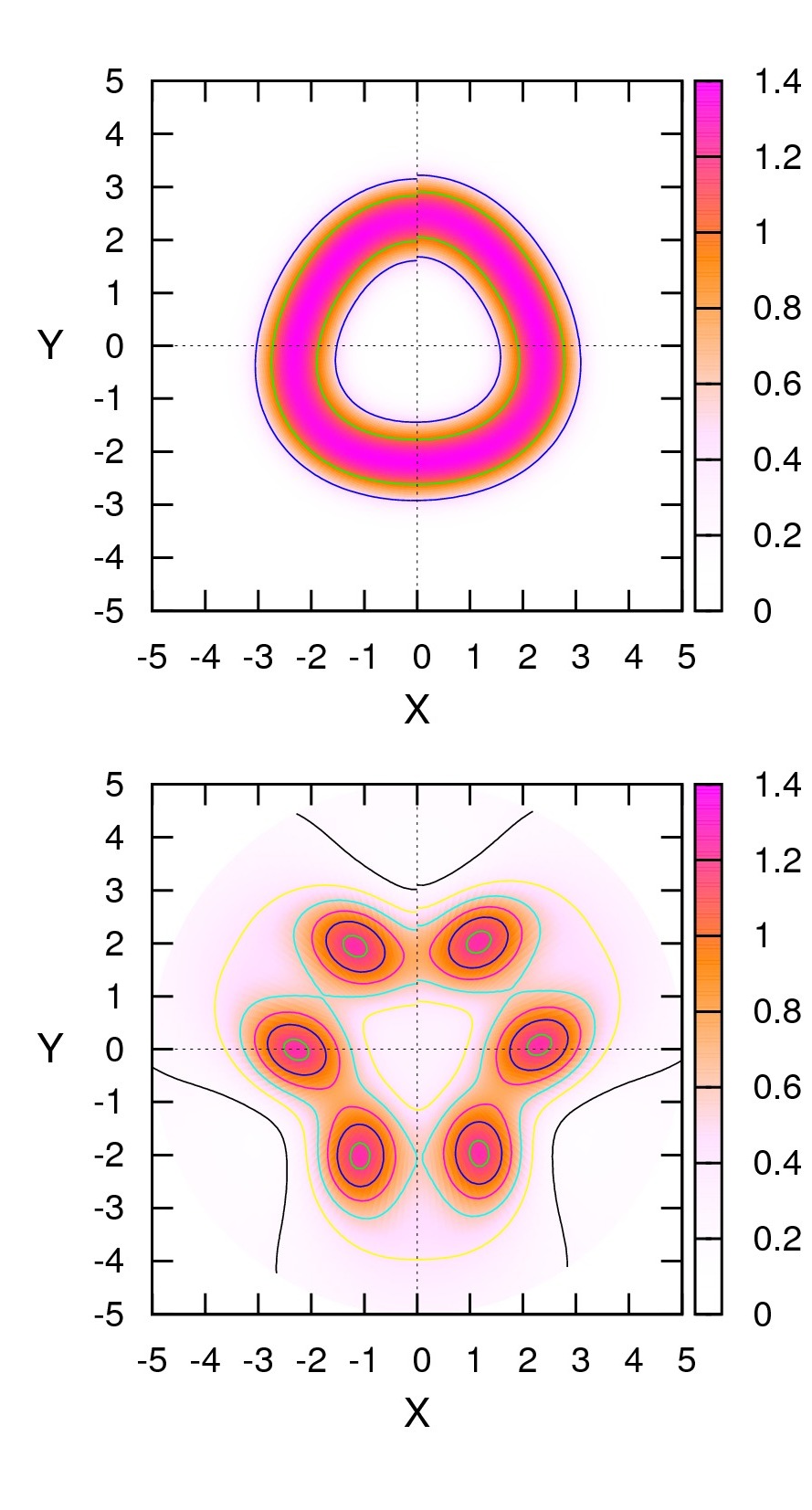}\hspace{0.2 cm} %
\includegraphics[height=5.2cm,angle=0,bb=00 00 325
725]{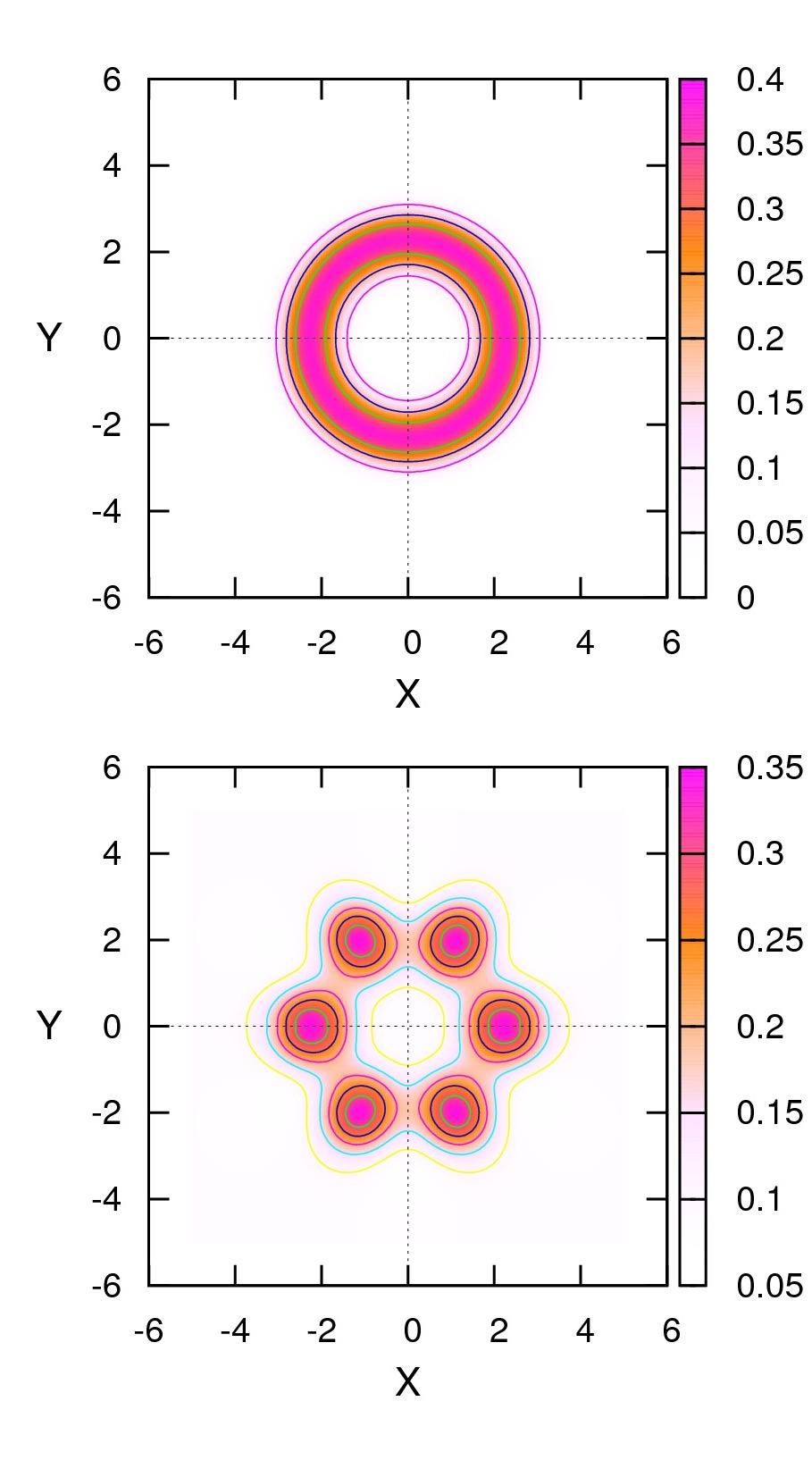}
\caption{(Color online) Contour plots of the energy density of the sectors
in the $(3,3)$ configuration in the model with potential (\protect\ref{pot})
at $\protect\kappa =0,0.4,0.8,1.5,2.0$ (from left to right).}
\label{f-3}
\end{figure}

\begin{figure}[th]
\refstepcounter{fig} \setlength{\unitlength}{1cm} \centering
\vspace{3.3 cm} \hspace{-0.0 cm} \includegraphics[height=5.2cm,angle=0,bb=00
00 325 725]{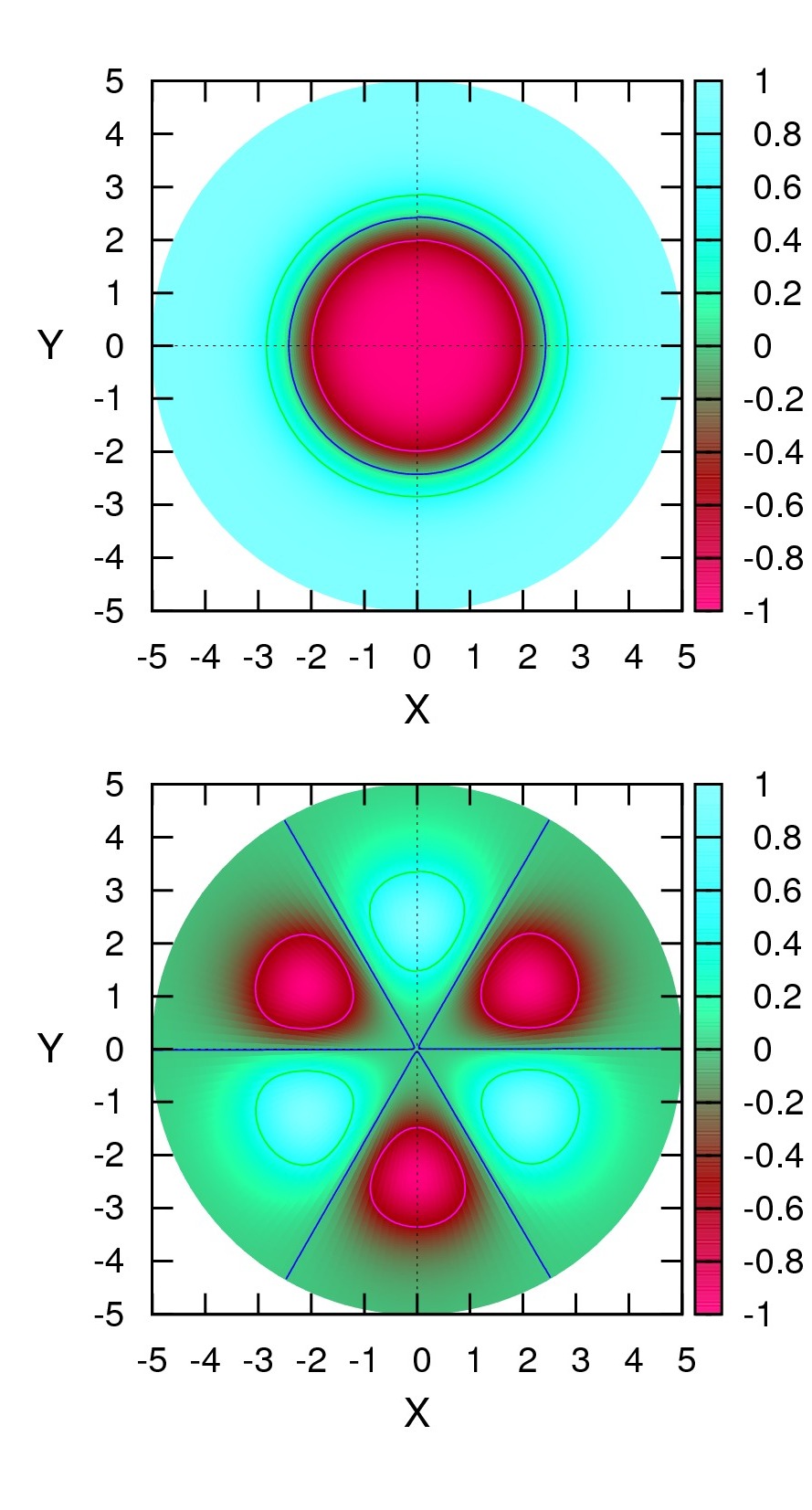}\hspace{0.2 cm} %
\includegraphics[height=5.2cm,angle=0,bb=00 00 325
725]{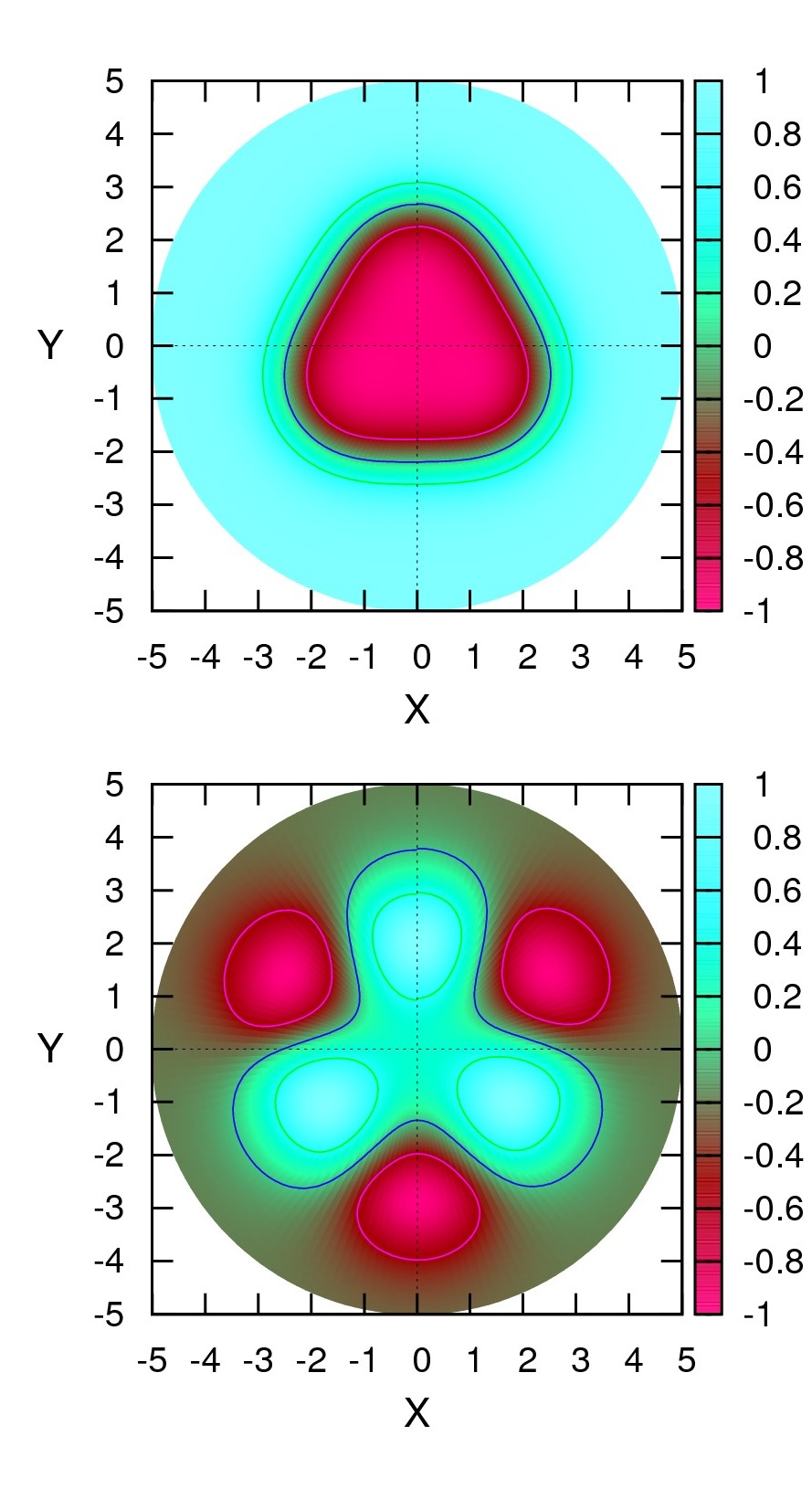}\hspace{0.2 cm} %
\includegraphics[height=5.2cm,angle=0,bb=00 00 325
725]{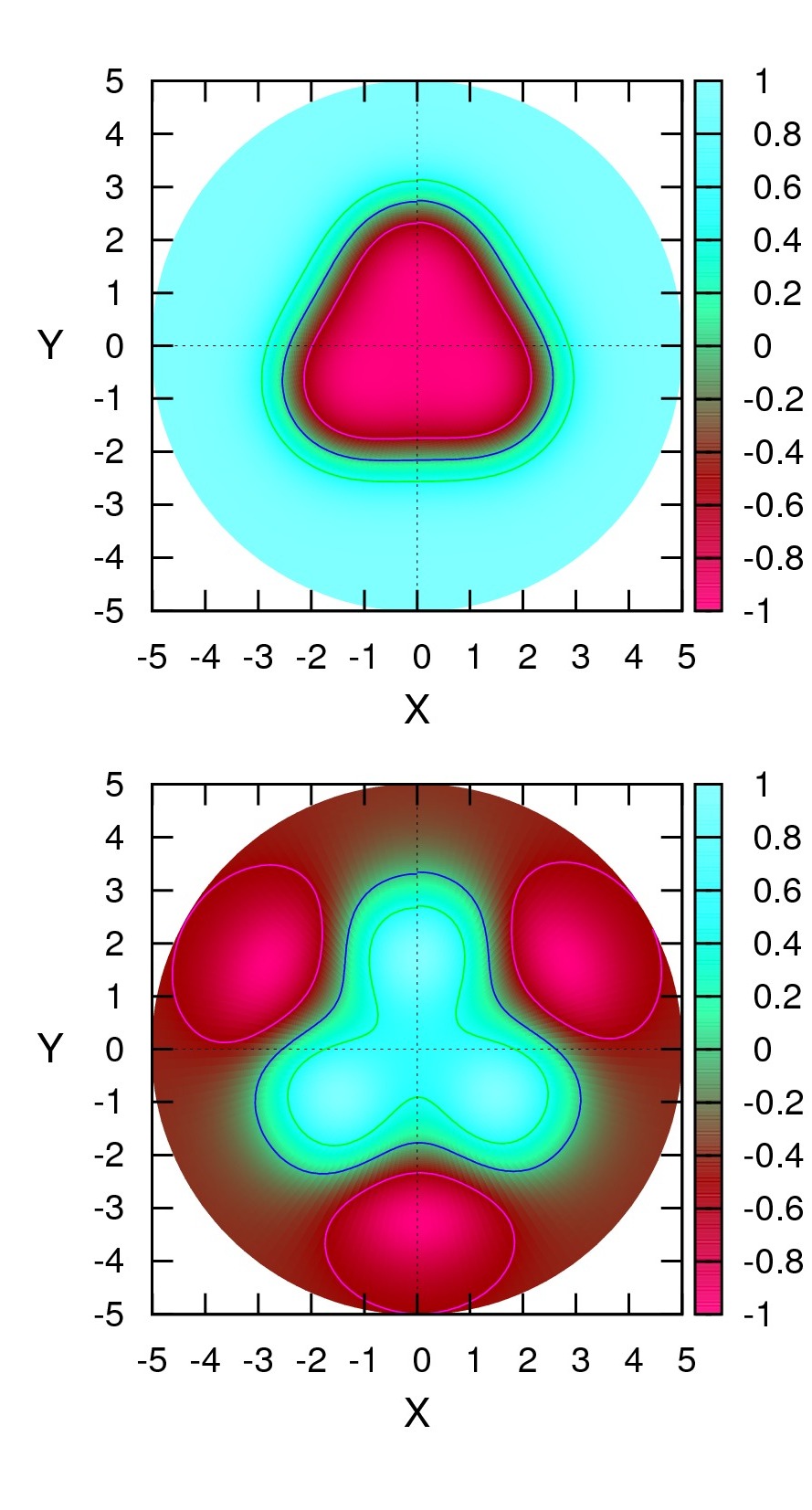}\hspace{0.2 cm} %
\includegraphics[height=5.2cm,angle=0,bb=00 00 325
725]{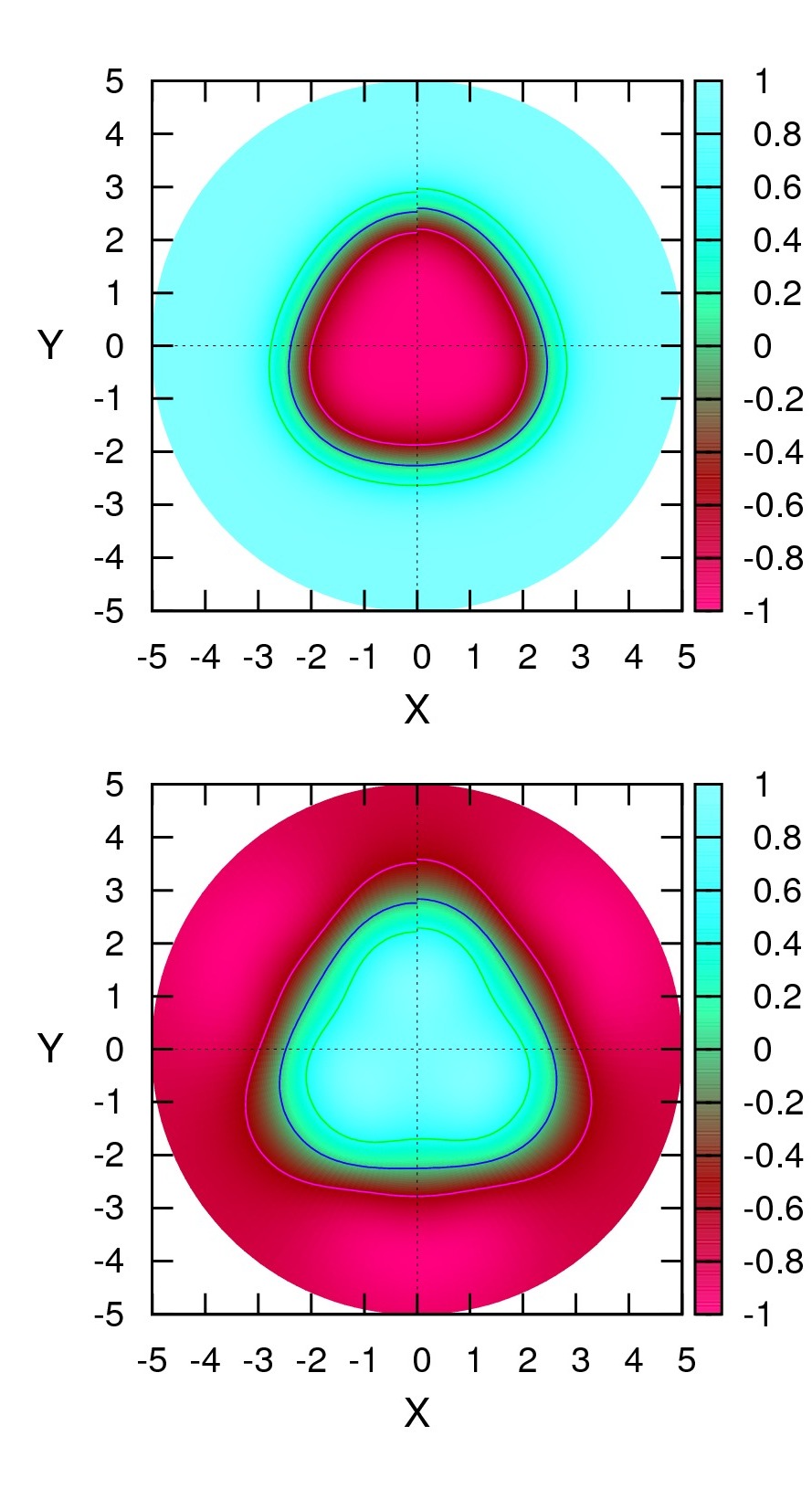}\hspace{0.2 cm} %
\includegraphics[height=5.2cm,angle=0,bb=00 00 325
725]{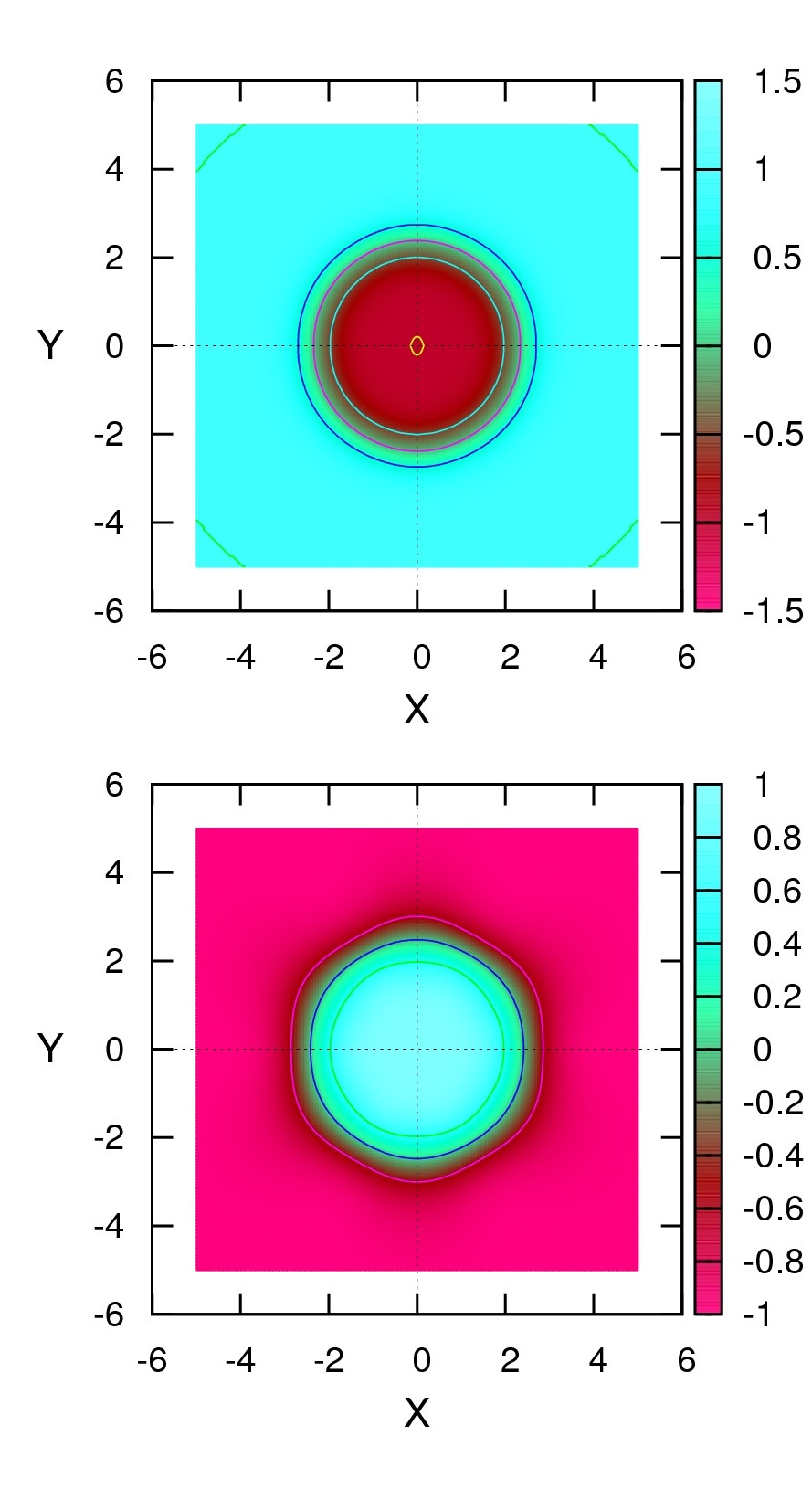}
\caption{(Color online) Contour plots of coupled components $\protect\phi %
_{3}^{(1)}$ and $\protect\phi _{1}^{(2)}$ (the upper and lower rows,
respectively) of the $(3,3)$ configuration in the model with potential (%
\protect\ref{pot}) at $\protect\kappa =0,0.4,0.8,1.5,2.0$ (from left to
right).}
\label{f-6}
\end{figure}

\begin{figure}[th]
\refstepcounter{fig} \setlength{\unitlength}{1cm} \centering
\vspace{3.3 cm} \hspace{-0.0 cm} \includegraphics[height=5.2cm,angle=0,bb=00
00 325 725]{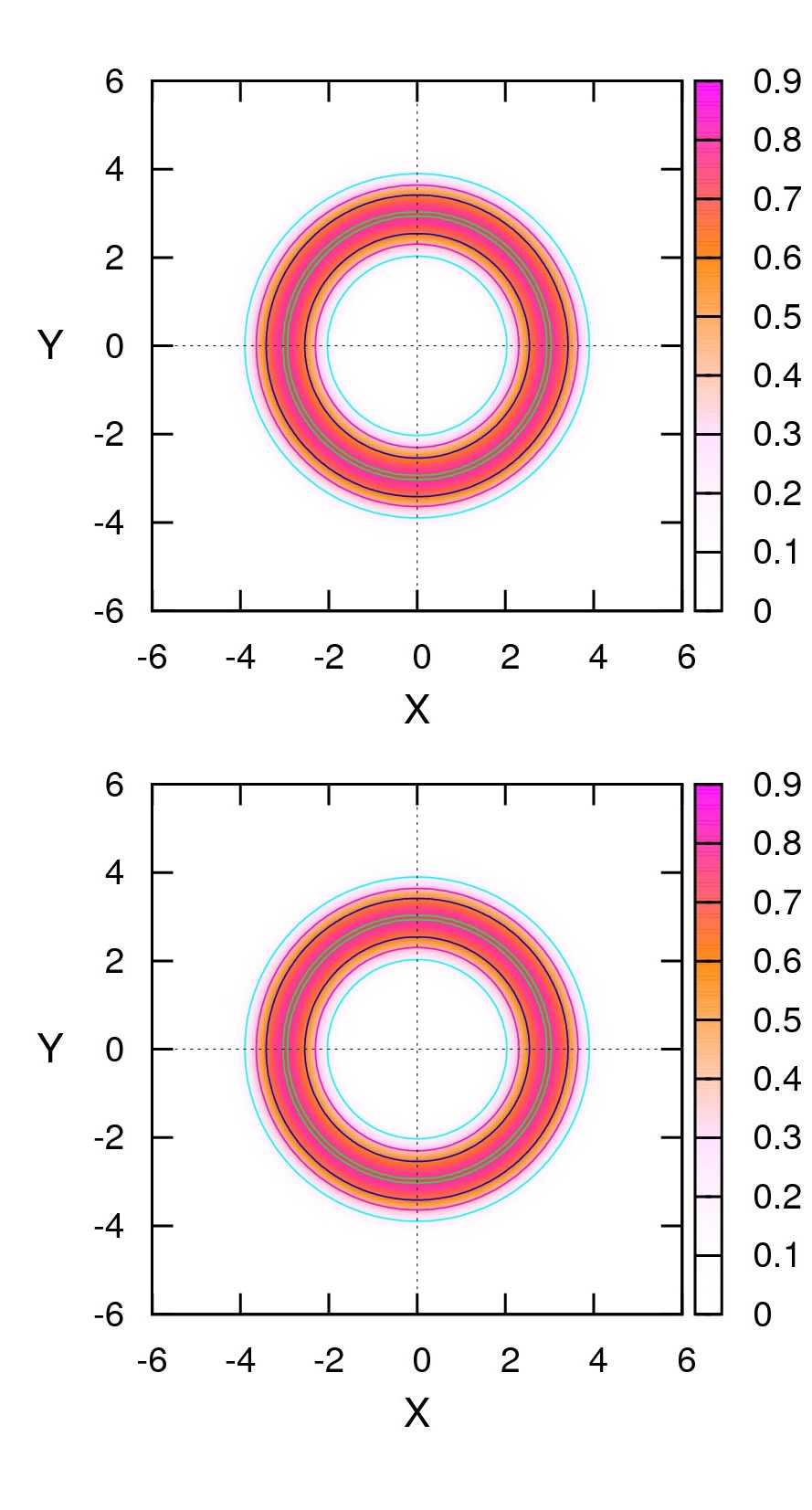}\hspace{0.2 cm} %
\includegraphics[height=5.2cm,angle=0,bb=00 00 325
725]{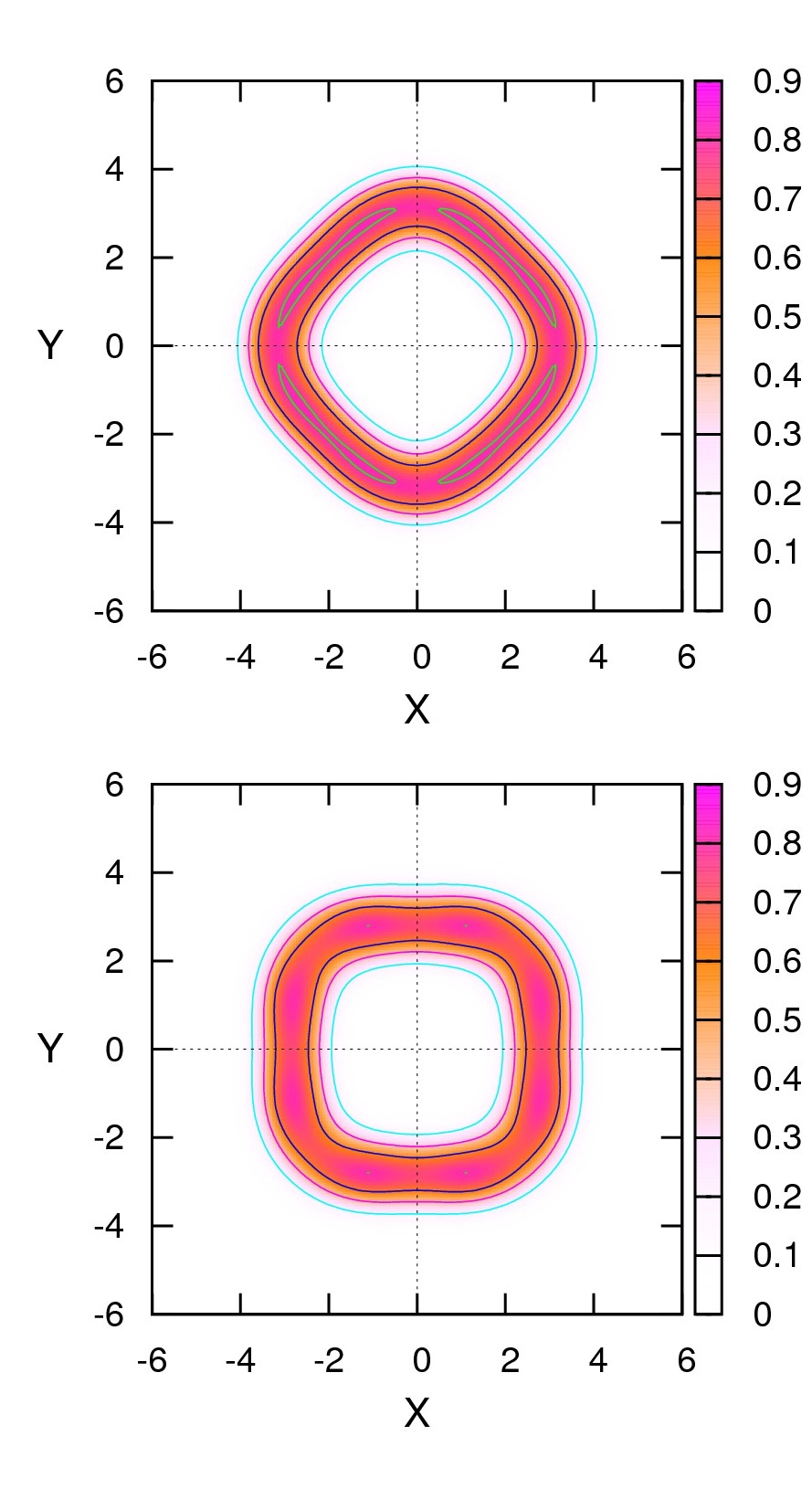}\hspace{0.2 cm} %
\includegraphics[height=5.2cm,angle=0,bb=00 00 325
725]{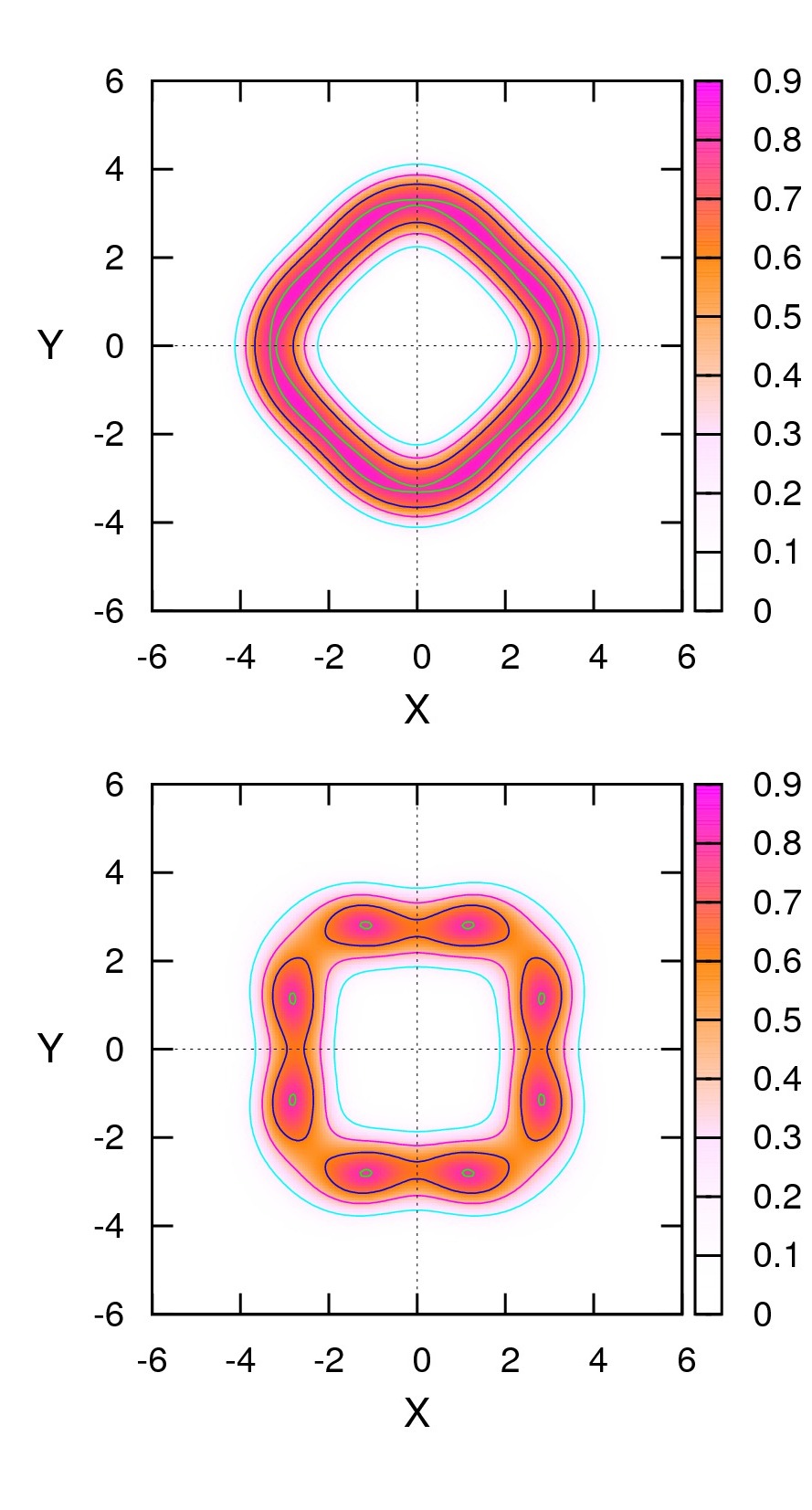}\hspace{0.2 cm} %
\includegraphics[height=5.2cm,angle=0,bb=00 00 325
725]{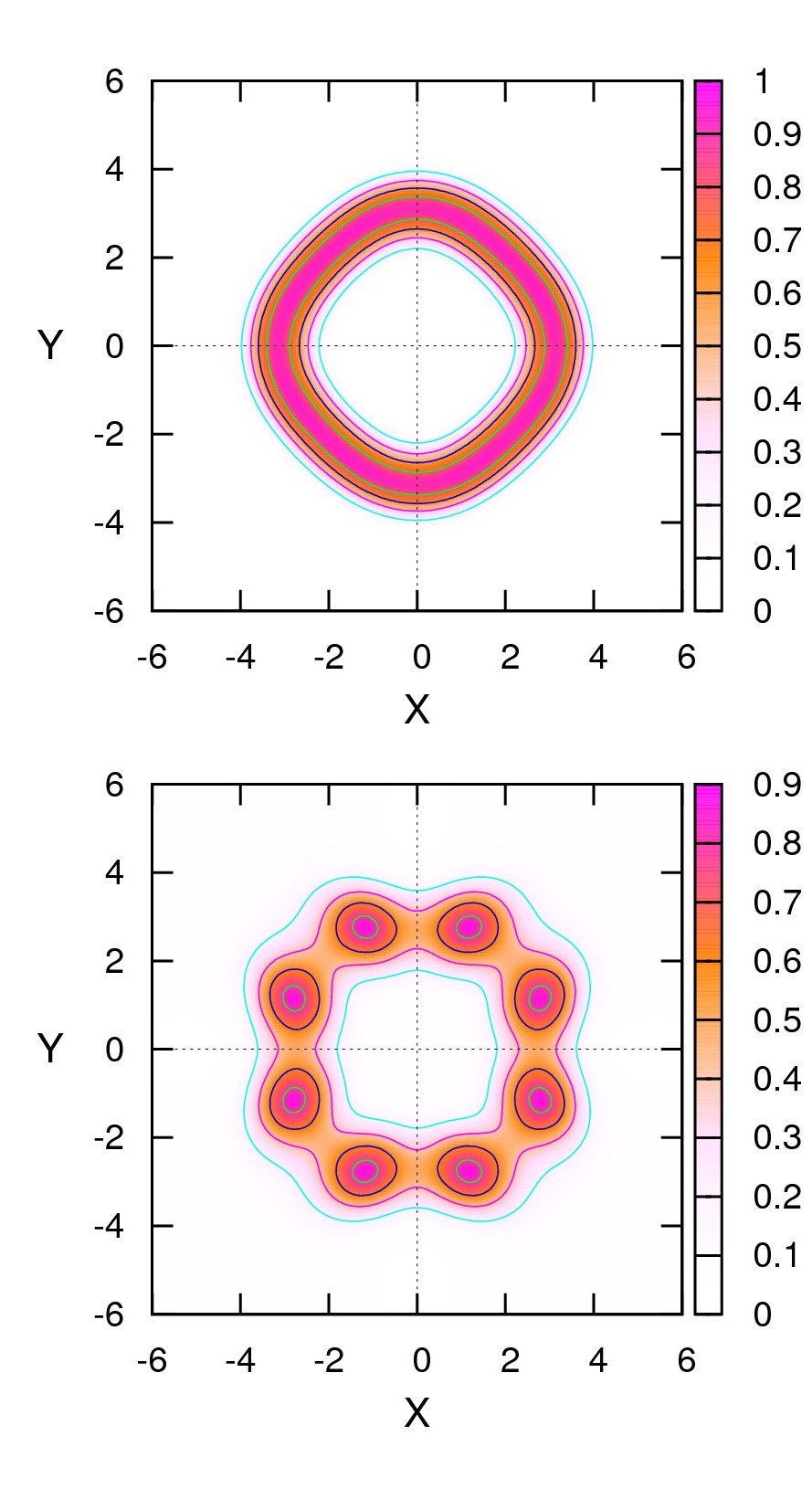}\hspace{0.2 cm} %
\includegraphics[height=5.2cm,angle=0,bb=00 00 325
725]{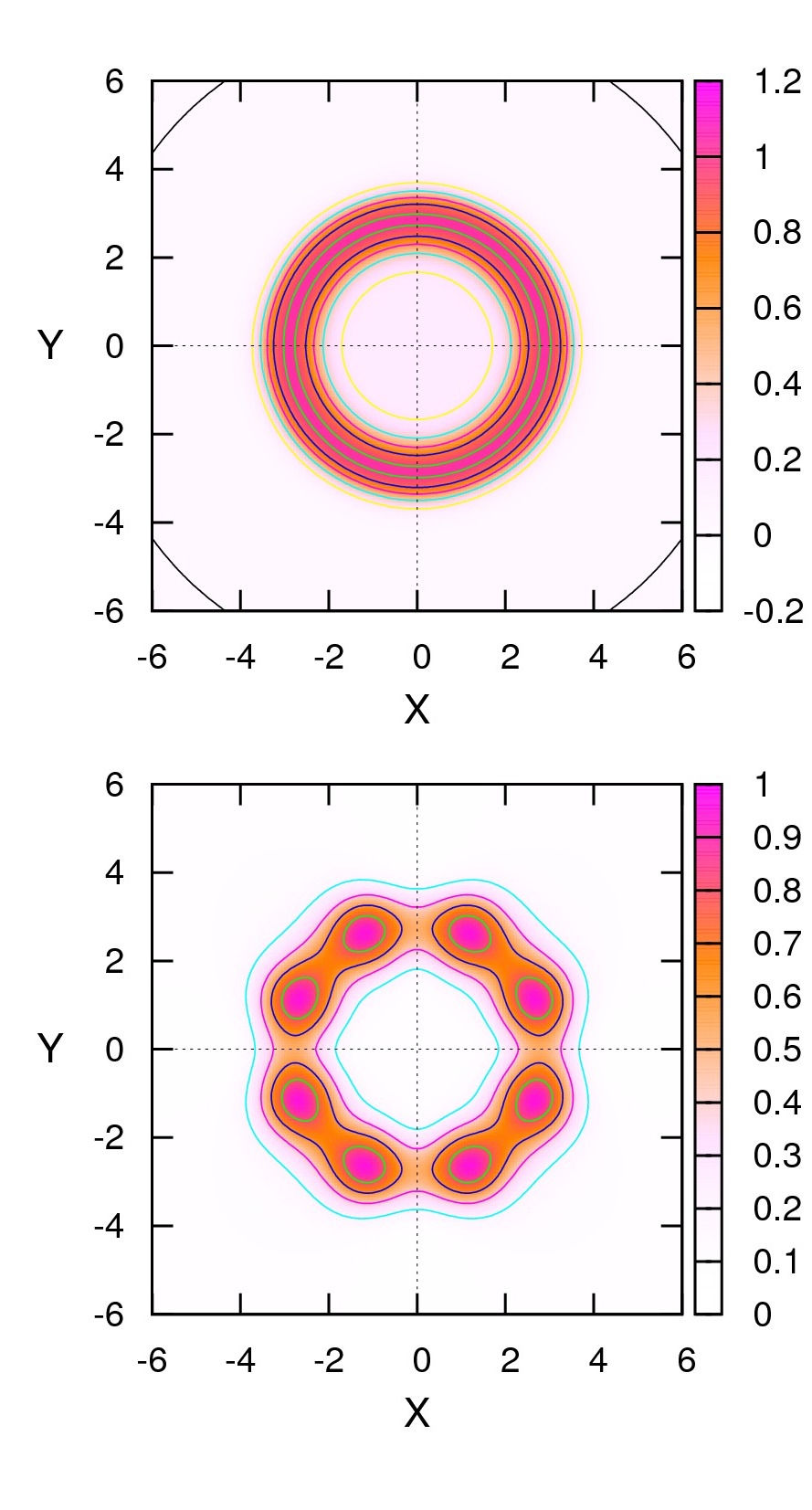}
\caption{(Color online) Contour plots of the energy density of the
components of the $(4,4)$ configuration in the model with potential (\protect
\ref{pot}) at $\protect\kappa =0,0.4,0.8,1.5,2.0$ (from left to right).}
\label{f-10}
\end{figure}

\begin{figure}[th]
\refstepcounter{fig} \setlength{\unitlength}{1cm} \centering
\vspace{3.3 cm} \hspace{-0.0 cm} \includegraphics[height=5.2cm,angle=0,bb=00
00 325 725]{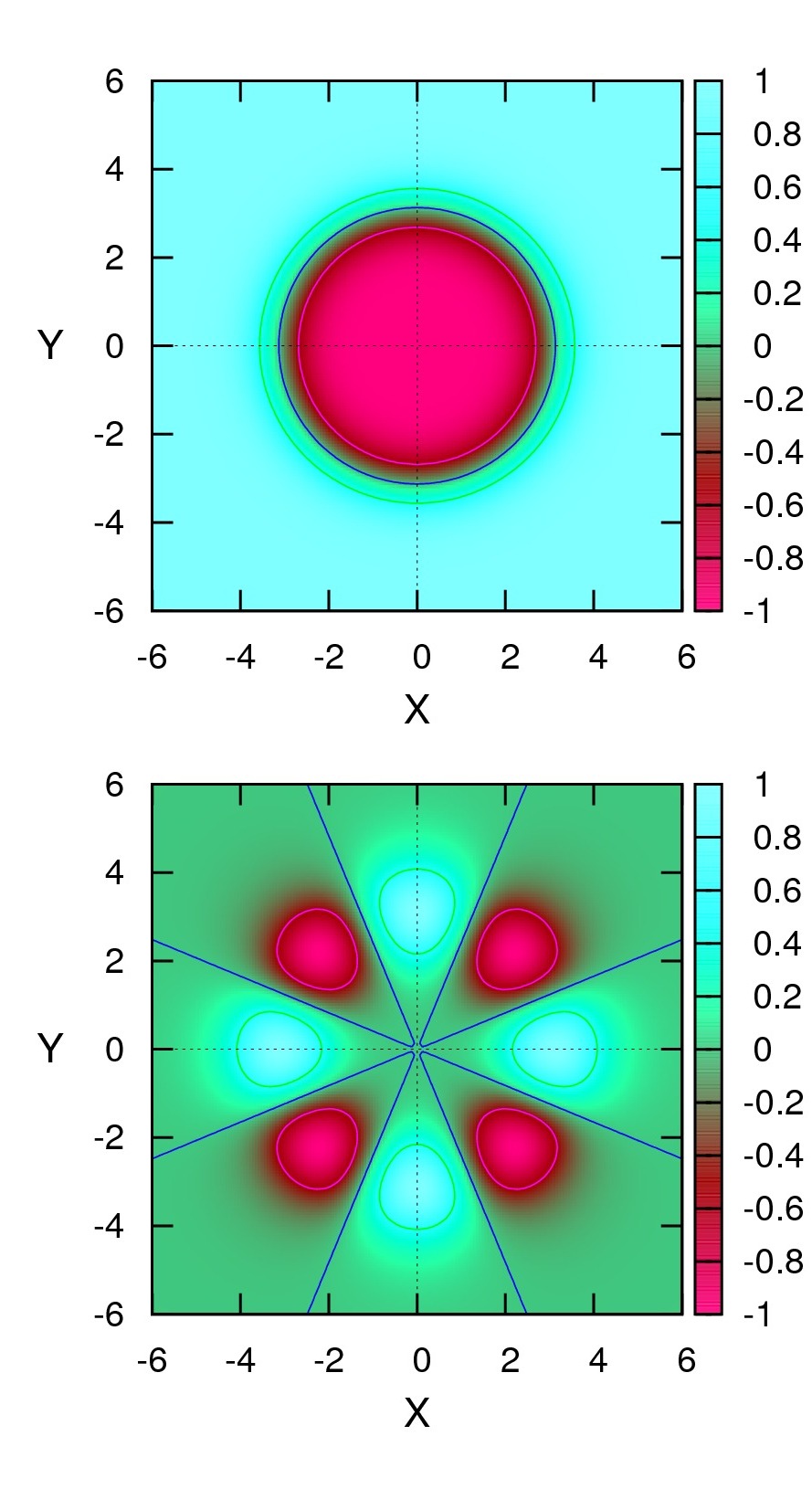}\hspace{0.2 cm} %
\includegraphics[height=5.2cm,angle=0,bb=00 00 325
725]{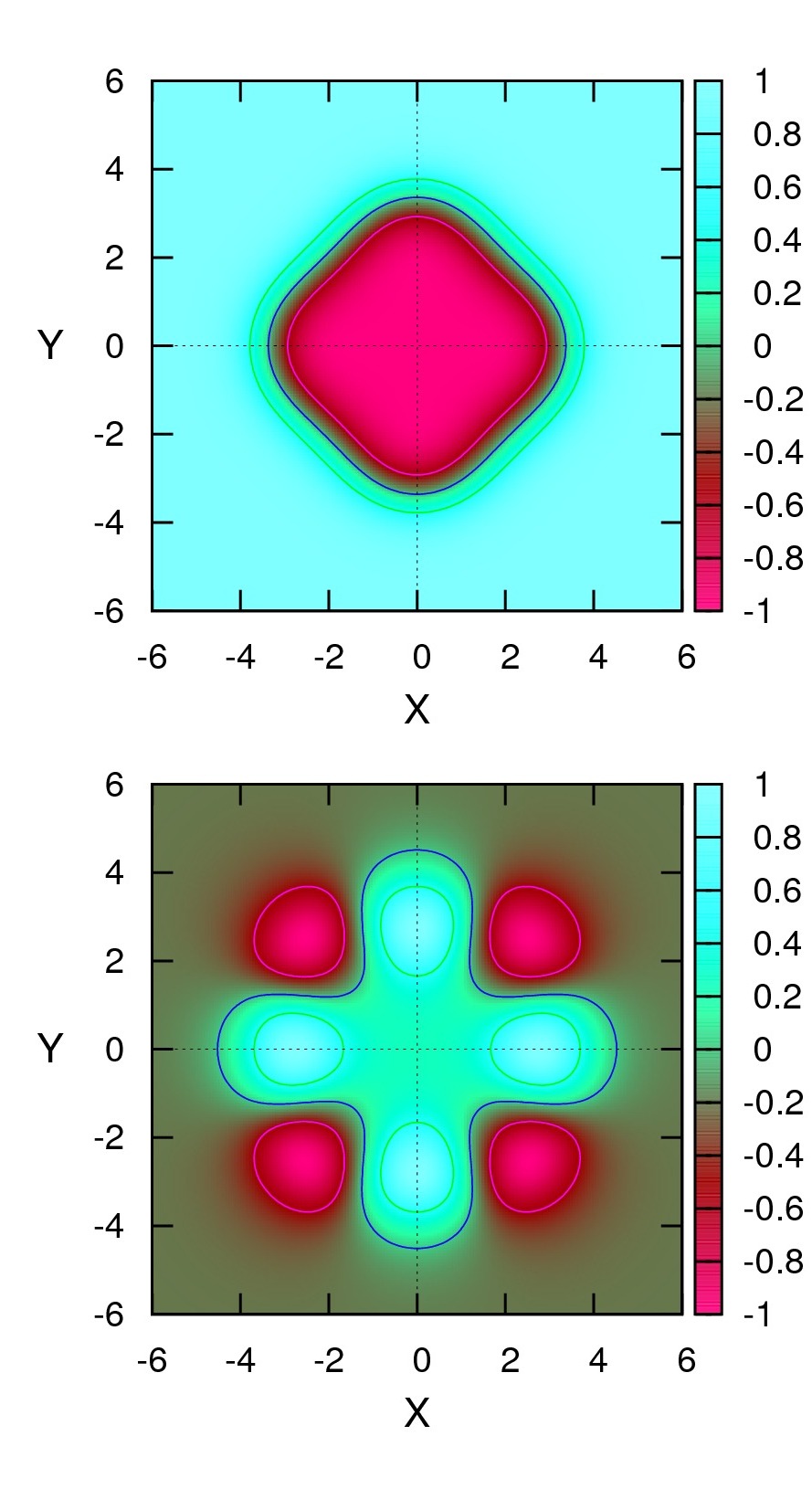}\hspace{0.2 cm} %
\includegraphics[height=5.2cm,angle=0,bb=00 00 325
725]{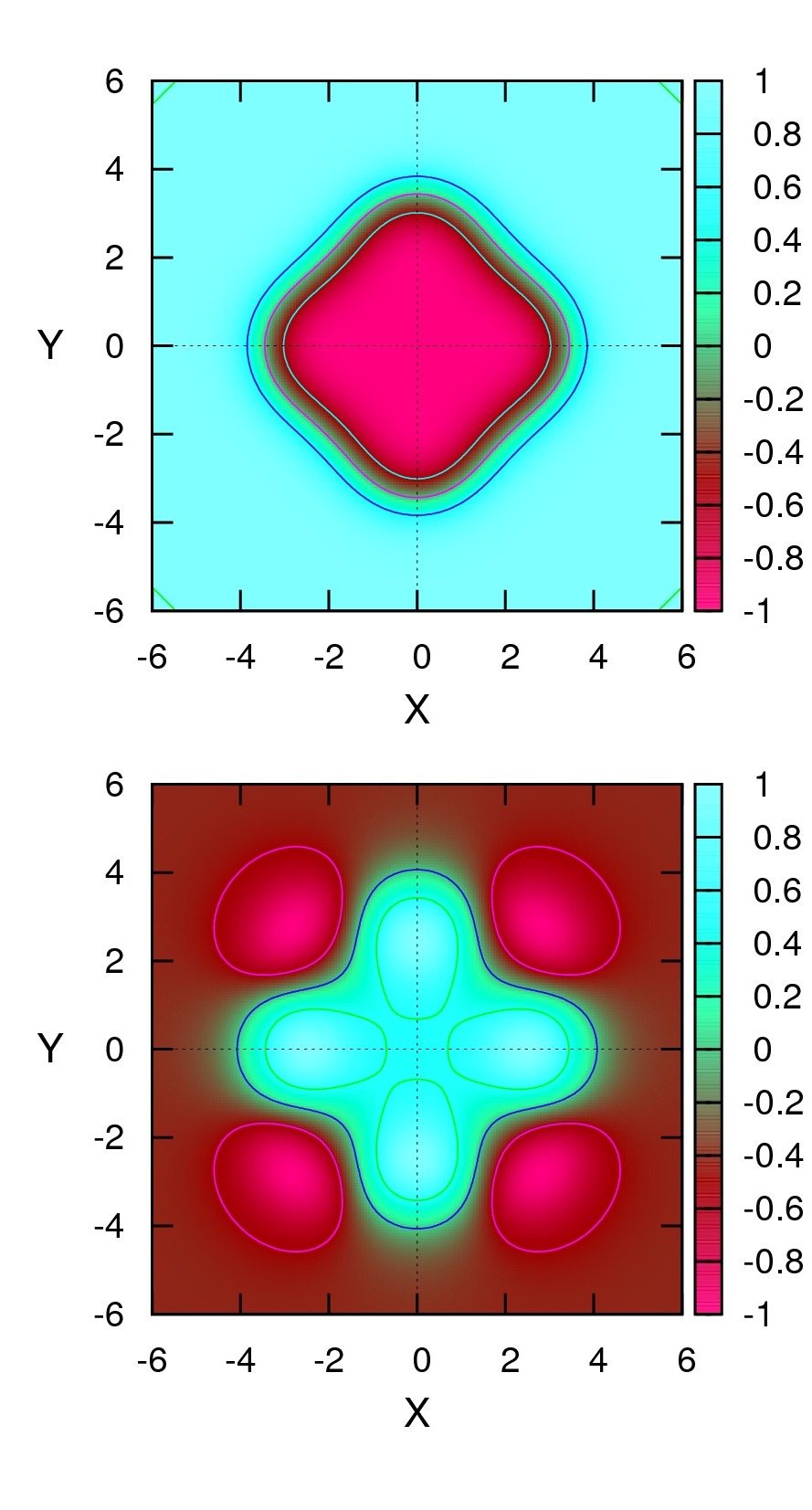}\hspace{0.2 cm} %
\includegraphics[height=5.2cm,angle=0,bb=00 00 325
725]{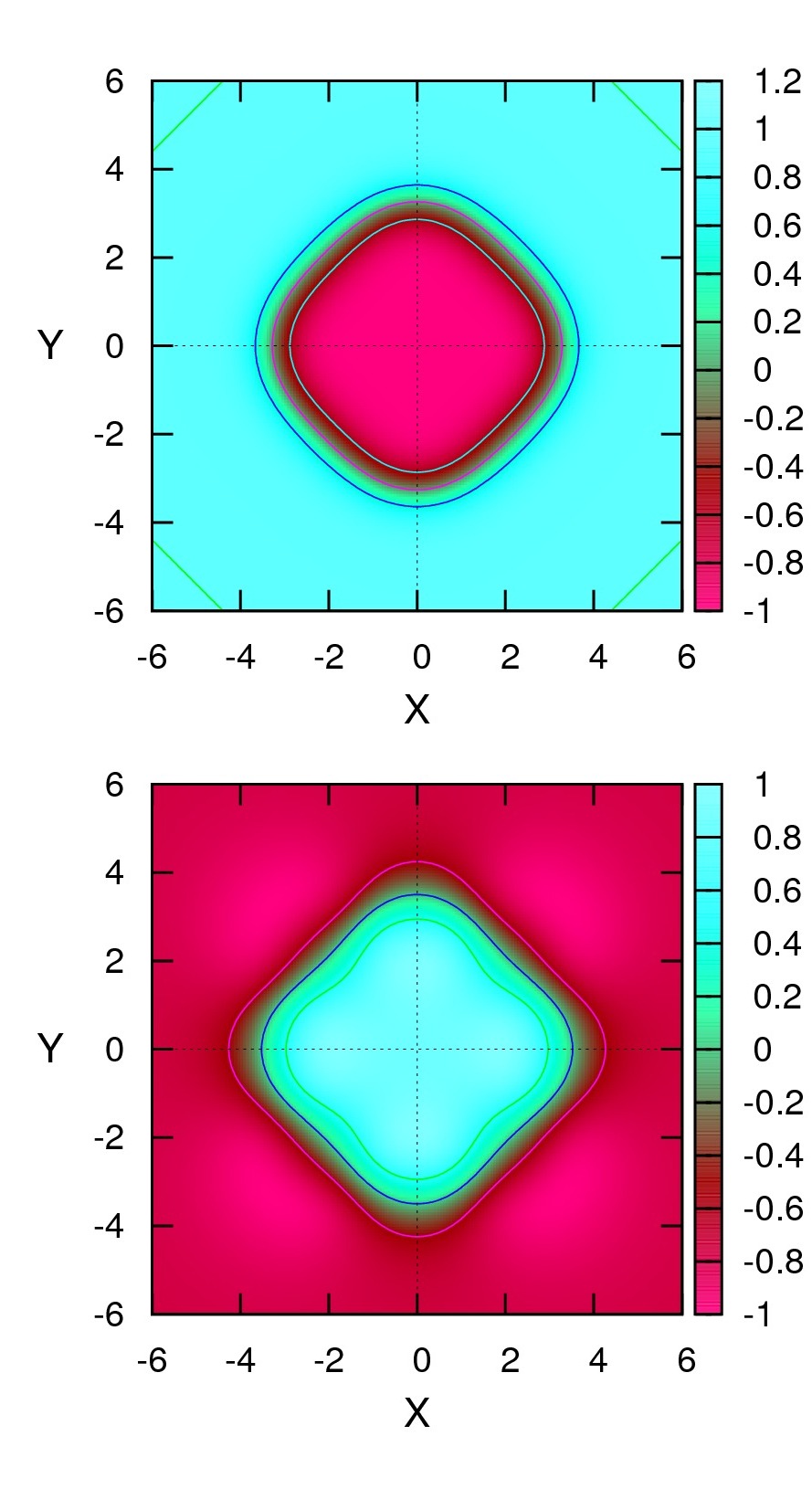}\hspace{0.2 cm} %
\includegraphics[height=5.2cm,angle=0,bb=00 00 325
725]{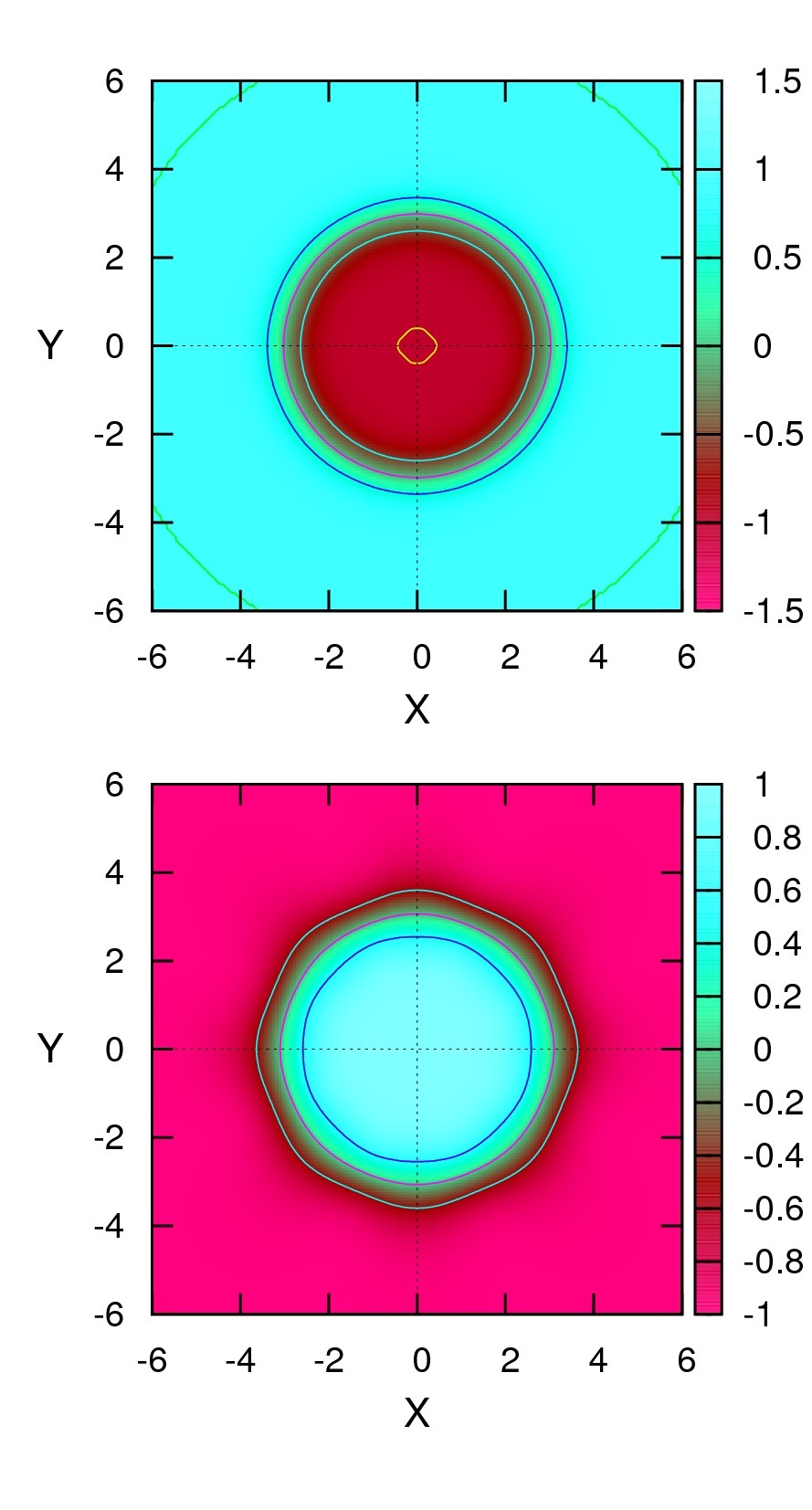}
\caption{(Color online) Contour plots of coupled components $\protect\phi %
_{3}^{(1)}$ and $\protect\phi _{1}^{(2)}$ (the upper and lower rows,
respectively) of the $(4,4)$ configuration in the model with potential (%
\protect\ref{pot}) at $\protect\kappa =0,0.4,0.8,1.5,2.0$ (from left to
right).}
\label{f-9}
\end{figure}

Hereafter, we restrict the consideration for the coupled BSMs with fixed
values of the mass parameters, $\mu _{1}=\mu _{2}=1$, unless stated
otherwise (recall $\mu _{2}=1$ was already fixed above), and coupling
constant $\kappa $ gradually increasing from zero. In Figs. \ref{f-1}-\ref%
{f-9} we present the so found contour plots of the energy densities in the
two sectors of the system, given by functions $L_{1}(x,y)$ and $L_{2}(x,y)$,
and the coupled field components, $\phi _{1}^{(2)},~\phi _{3}^{(1)}$, which
illustrate typical configurations of the $\left( B^{(1)},B^{(2)}\right)
=(1,1)$, $(2,2)$, $(3,3)$ and $(4,4)$ types.

First, we consider in detail the simplest configuration, of the $(1,1)$
type. As the inter-core coupling, $\kappa $, increases, the components in
the two cores rapidly start to separate, the distance between them attaining
a maximum at $\kappa \approx 0.15$. This distance is, actually, slightly
smaller than the size of the charge-$1$ baby skyrmion, as seen in the second
row of Fig.~\ref{f-1}.

\begin{figure}[t]
\refstepcounter{fig} 
\par
\begin{center}
\hspace{-9.7 cm} \includegraphics[height=7.9cm,angle=0,bb=00 -80 00
825]{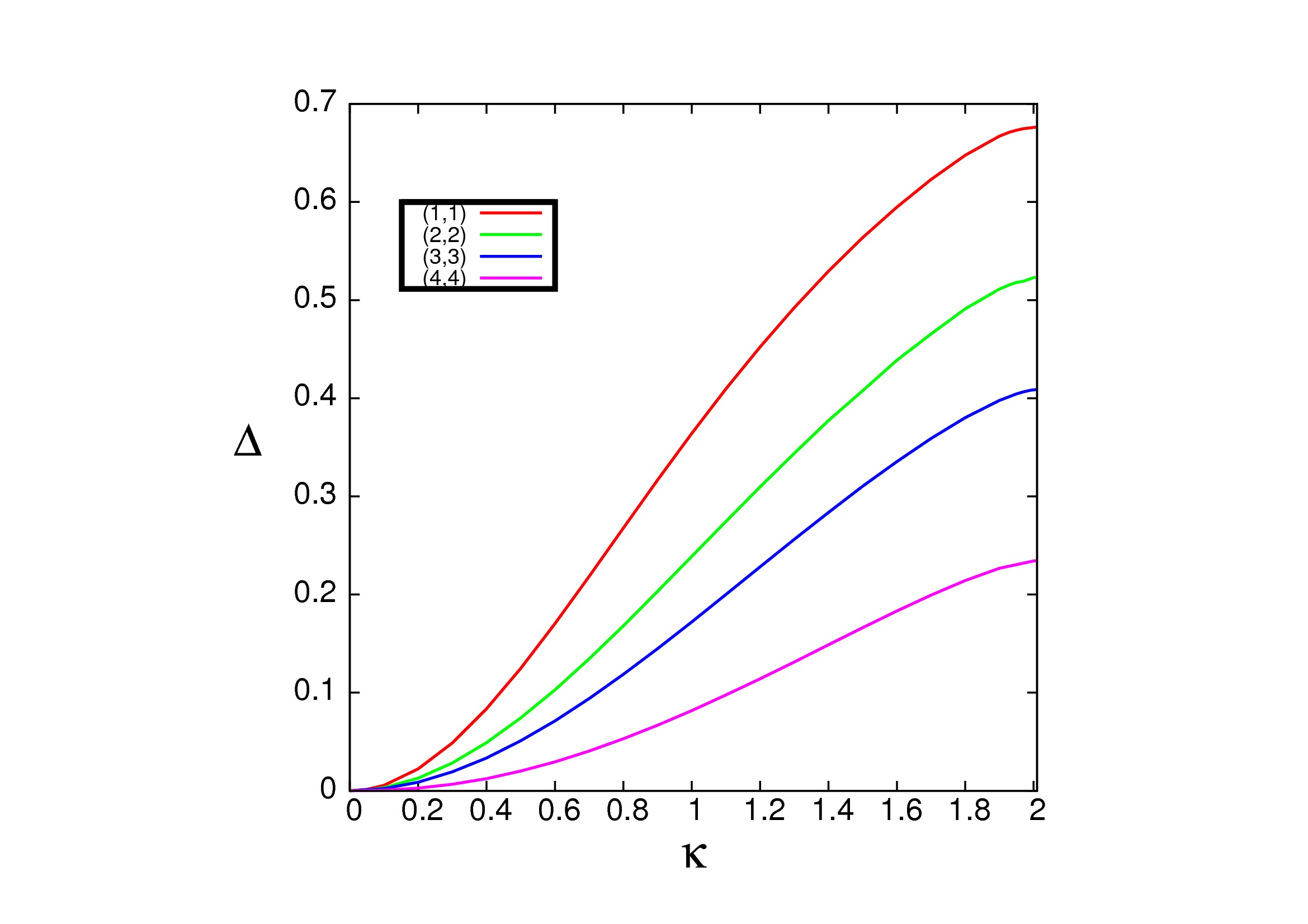}\hspace{7.8 cm} \includegraphics[height=7.9cm,angle=0,bb=00
-80 00 825]{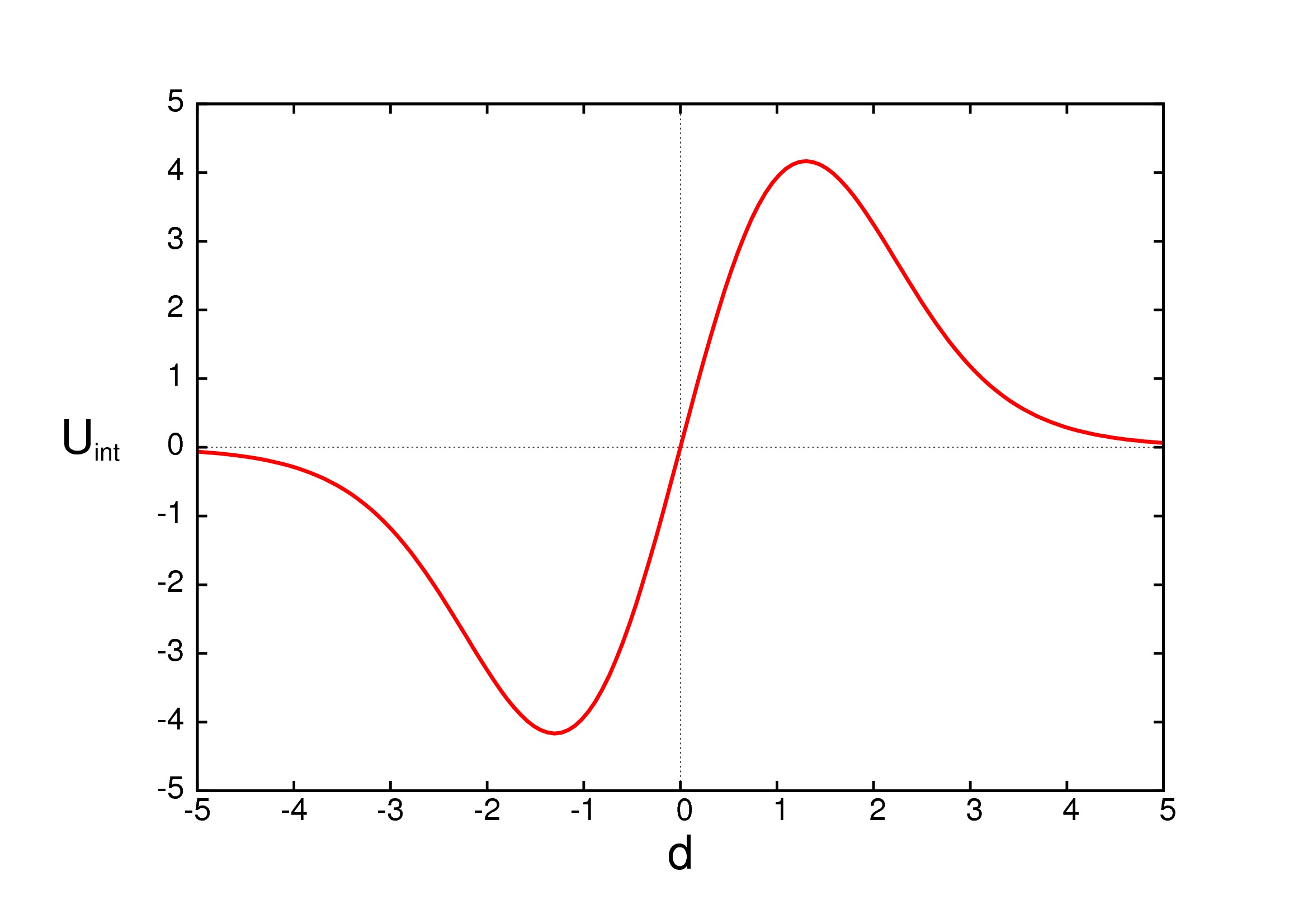}\hspace{0.0 cm} \vspace{1.7 cm}
\end{center}
\caption{(Color online) Left: The symmetry measure, defined as per Eq. (%
\protect\ref{Delta}), for the $(1,1),(2,2),(3,3)$ and $(4,4)$
configurations, versus the coupling constant, $\protect\kappa $. Right: The
effective potential of interaction between the components of the $(1,1)$
configuration, defined as per Eq. (\protect\ref{Uint}), versus the
separation between them, $d$.}
\label{f-12}
\end{figure}

To explain this observation, we note that the coupling remains weak, and the
components in this regime are almost undeformed, which suggests to evaluate
the effective potential of the interaction between the two components as a
function of the separation between them. To this end, following the
well-known approach adopted in the perturbation theory for solitons \cite%
{RMP}, we take unperturbed baby skyrmions of unit charge in the two cores,
separated by distance $d$ in the lateral direction, and calculate the
corresponding interaction potential of the $(1,1)$ configuration:
\begin{equation}
U_{\mathrm{int}}(d)=\kappa \int\!\!\int {\phi _{3}^{(1)}}\left( x,y+\frac{d}{%
2}\right) {\phi _{1}^{(2)}}\left( x,y-\frac{d}{2}\right) ~dxdy  \label{Uint}
\end{equation}%
Numerical evaluation shows that this potential has a minimum at $d=-1.4$,
see the right plot in Fig.~\ref{f-12}, which clearly explains the
spontaneous separation between the components.

When coupling becomes stronger, the asymptotic values of the components $%
\phi _{1}^{(2)}$ and $\phi _{3}^{(2)}$ start to vary and the constituents
begin to deform. The baby skyrmion in the first sector with $B^{(1)}=1$
remains almost rotationally invariant, with a small local minimum of the
energy density at the soliton's center, while the soliton with $B^{(2)}=1$
in the second sector features broken rotational invariance, see the third
row in Figs.~\ref{f-1} and \ref{f-19}. This transition from the separated
but intrinsically symmetric configuration to the one with the broken
symmetry between the components of the state of the $\left( 1,1\right) $
type is a specific realization of the SBB\ in the present system. Below,
realizations of the SBB\ in configurations of the $\left( 2,2\right) $, $%
\left( 3,3\right) $ and $\left( 4,4\right) $ types are presented too.

Further increase of the coupling leads to breaking of the rotational
symmetry of the skyrmion in the first sector too, and it becomes symmetric
with respect to the reflection, $x\rightarrow -x$. Finally, as the coupling
constant approaches the critical value (\ref{cr}), components $\phi
_{1}^{(2)}$ and $\phi _{3}^{(2)}$ swap their roles, as shown in Fig.~\ref%
{f-19}. Indeed, the strong coupling between component $\phi _{3}^{(1)}$ in
the first sector, which is subject to the fixed vacuum-asymptotic behavior, $%
\phi _{3}^{(1)}\rightarrow -1$ at $\rho =0$ and $\phi _{3}^{(1)}\rightarrow 1
$ as $\rho \rightarrow \infty $, and the \textquotedblleft flexible"
component in the second sector, $\phi _{1}^{(2)}$, forces the latter one to
interpolate between the \textquotedblleft upper" vacuum value, $\phi
_{1}^{(2)}\rightarrow 1$, at the origin, and the \textquotedblleft lower"
vacuum value, $\phi _{1}^{(2)}\rightarrow -1$, as $\rho \rightarrow \infty $%
. As the asymptotic behavior of component $\phi _{2}^{(2)}$ is fixed, in
this limit configuration components $\phi _{1}^{(2)}$ and $\phi _{3}^{(2)}$
are actually swapped.

In the left plot of Fig.~\ref{f-12}, we represent the results of the
analysis of the SBB in the systematic form, displaying the inter-core
symmetry-breaking energy measure,
\begin{equation}
\Delta =\left( E^{(2)}-E^{(1)}\right) /\left( E^{(2)}+E^{(1)}\right) ,
\label{Delta}
\end{equation}%
where $E^{(a)}=\int\!\!\int L_{a}dxdy$ is defined as per Eq.~(\ref{Lag}), as
a function of the coupling constant, $\kappa $, for the static
configurations with identical topological charges in both sectors.

Evidently, the asymmetry is growing from zero to a maximal value which
corresponds to the critical coupling (\ref{cr}). Note that the asymmetry
between the sectors decreases with the increase of the common topological
charge of both sectors.

In the uncoupled double-vacuum system, the baby skyrmions are always
rotationally invariant. As the coupling strength, $\kappa $, increases, the
symmetry gets broken and the norm of the soliton in the second sector grows
faster, as its symmetry is lower. When the coupling approaches the critical
value (\ref{cr}), the charge-$1$ soliton in the first sector regains the
rotational invariance, featuring an annular shape of its energy-density
distribution. On the other hand, the second sector is composed of two
segments, featuring the discrete dihedral $D_{2}$ symmetry, similar to the
solutions presented in Ref. \cite{Ward} in the single-component model with
an $O(2)$-symmetry-breaking potential, $U(\phi )=\mu ^{2}(1-\phi
_{3}^{2})(1-\phi _{1}^{2})$, cf. Eq.~(\ref{pot}). Further increase of $%
\kappa $ almost does not affect the asymptotic values of the fields, the
asymmetry between the components remaining nearly constant; however as
mentioned above, in that limit the baby skyrmions are, in fact, compactons,
with the fields reaching the vacuum values at a finite distance from the
center of the configuration.

\subsection{States with different topological charges in the sectors and
other forms of the coupling}

\begin{figure}[tbh]
\refstepcounter{fig} \setlength{\unitlength}{1cm} \centering
\vspace{3.3 cm} \hspace{-0.0 cm} \includegraphics[height=5.2cm,angle=0,bb=00
00 325 725]{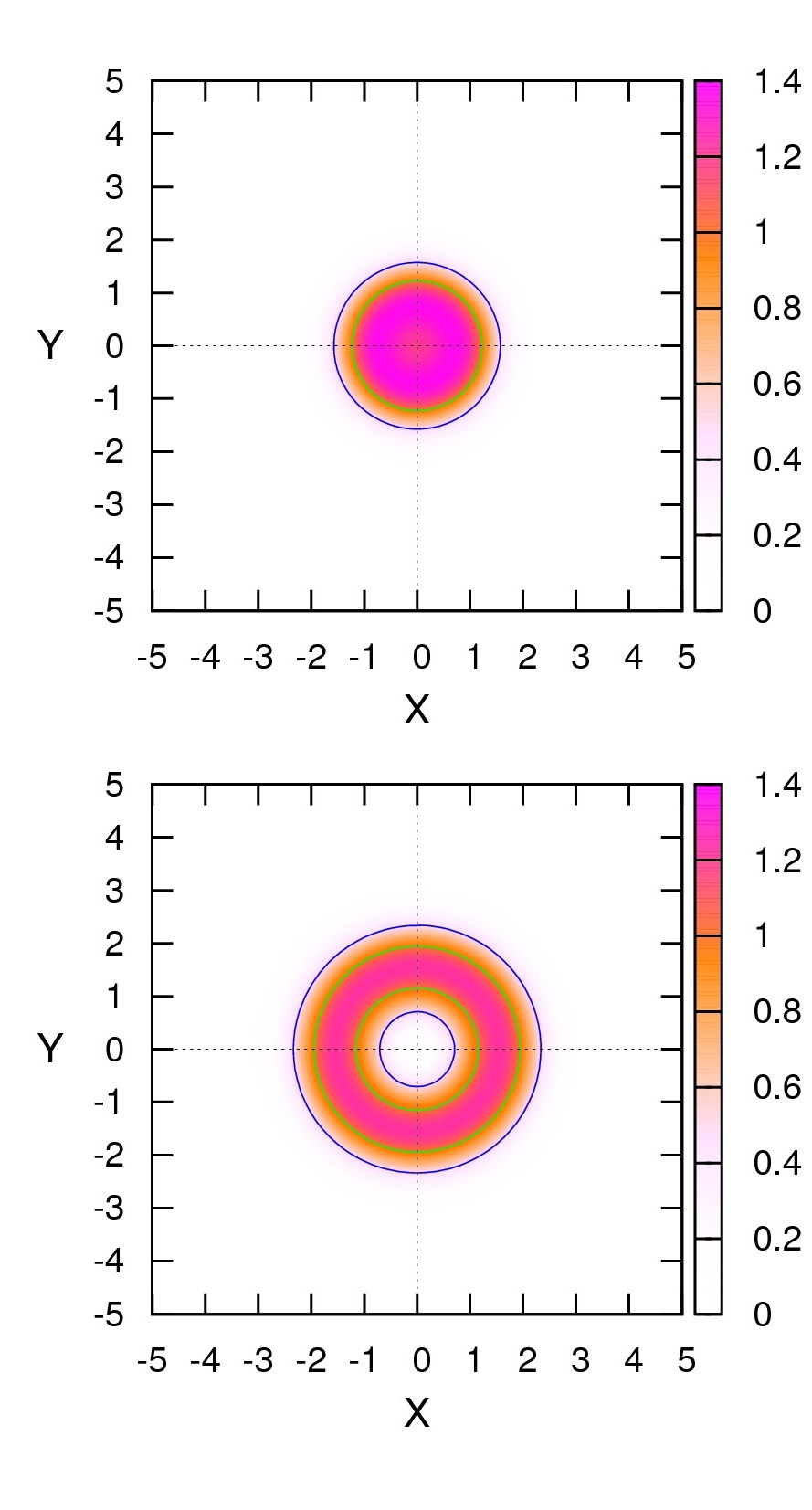}\hspace{0.2 cm} %
\includegraphics[height=5.2cm,angle=0,bb=00 00 325
725]{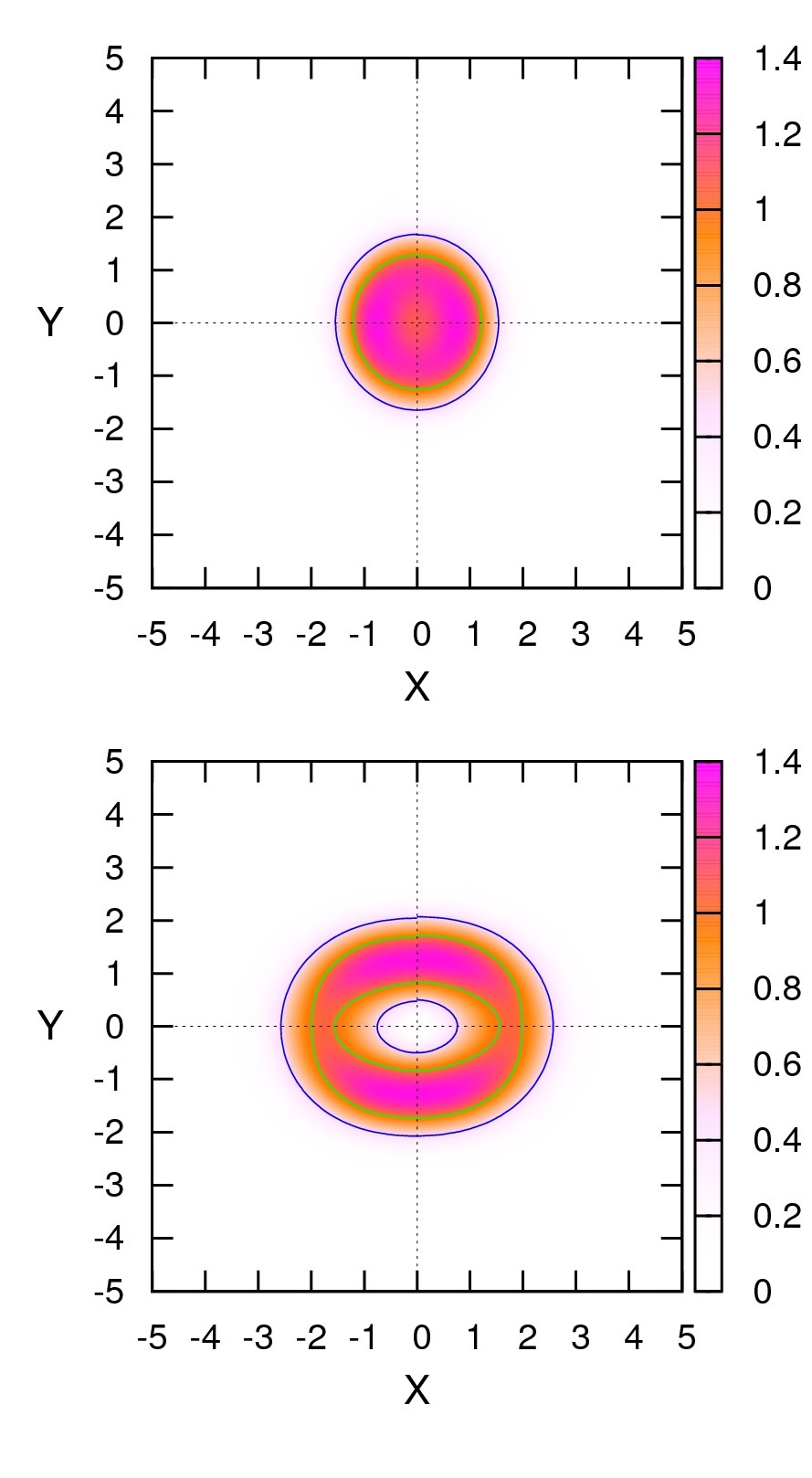}\hspace{0.2 cm} %
\includegraphics[height=5.2cm,angle=0,bb=00 00 325
725]{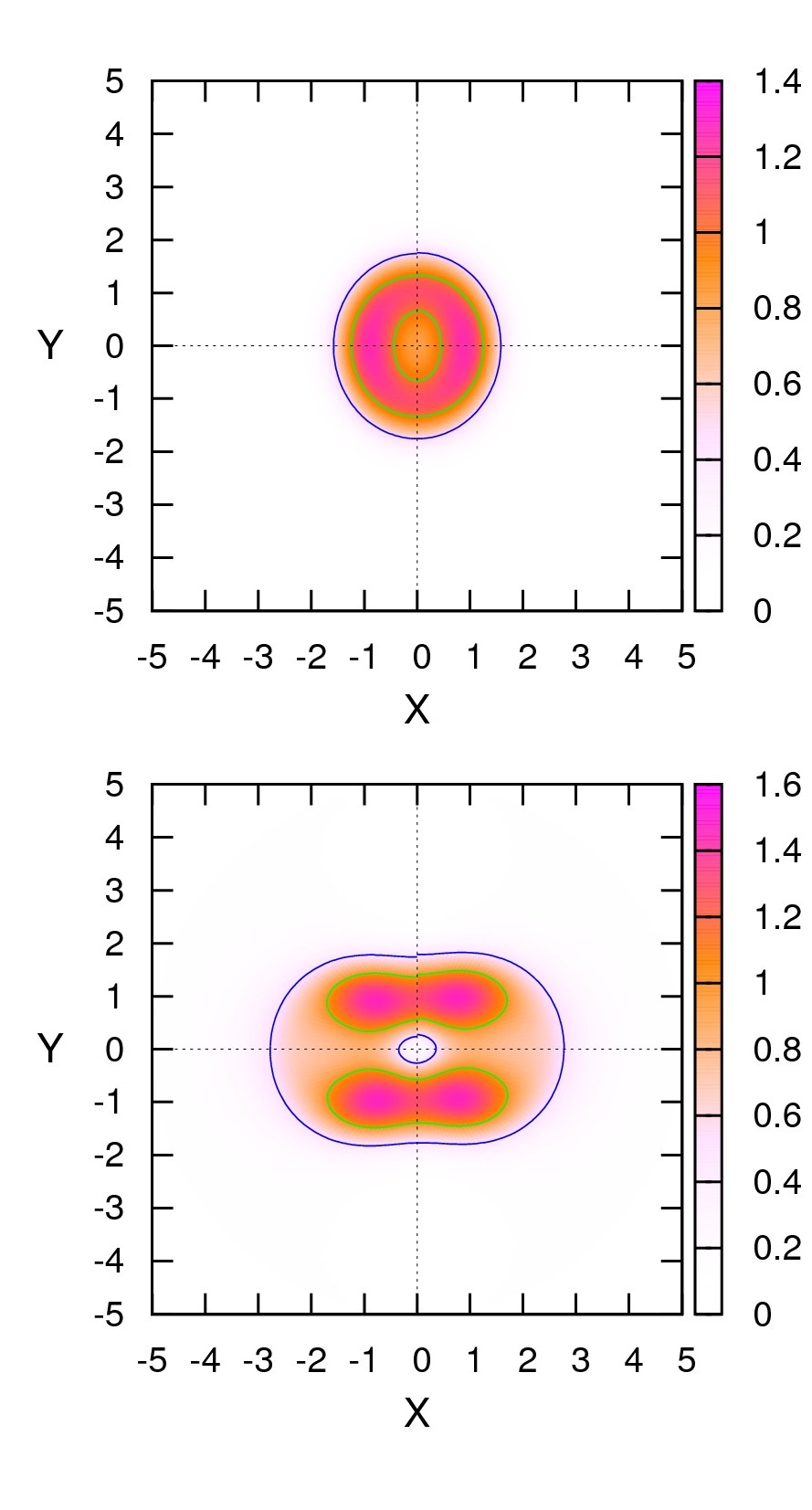}\hspace{0.2 cm} %
\includegraphics[height=5.2cm,angle=0,bb=00 00 325
725]{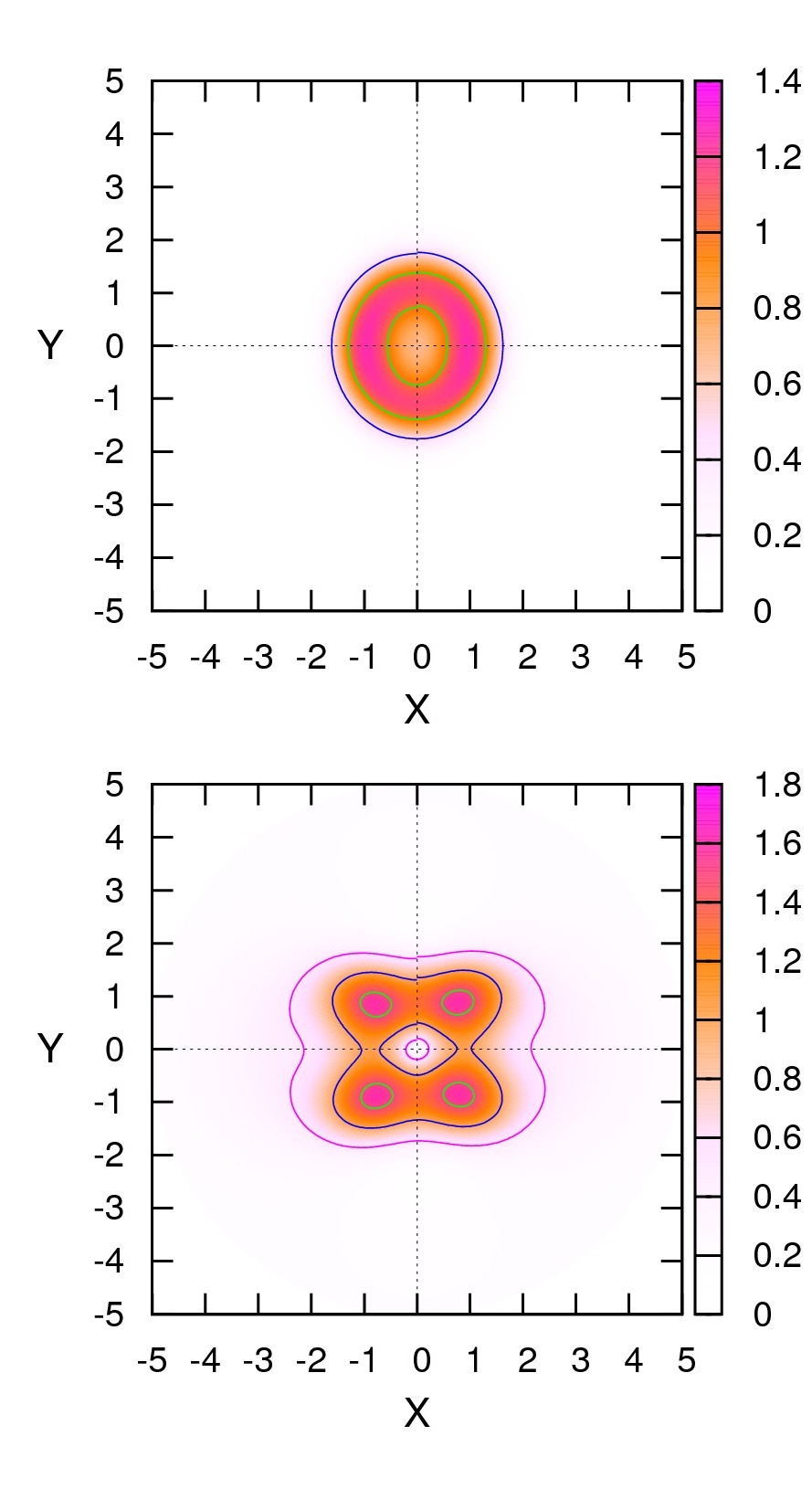}\hspace{0.2 cm} %
\includegraphics[height=5.2cm,angle=0,bb=00 00 325
725]{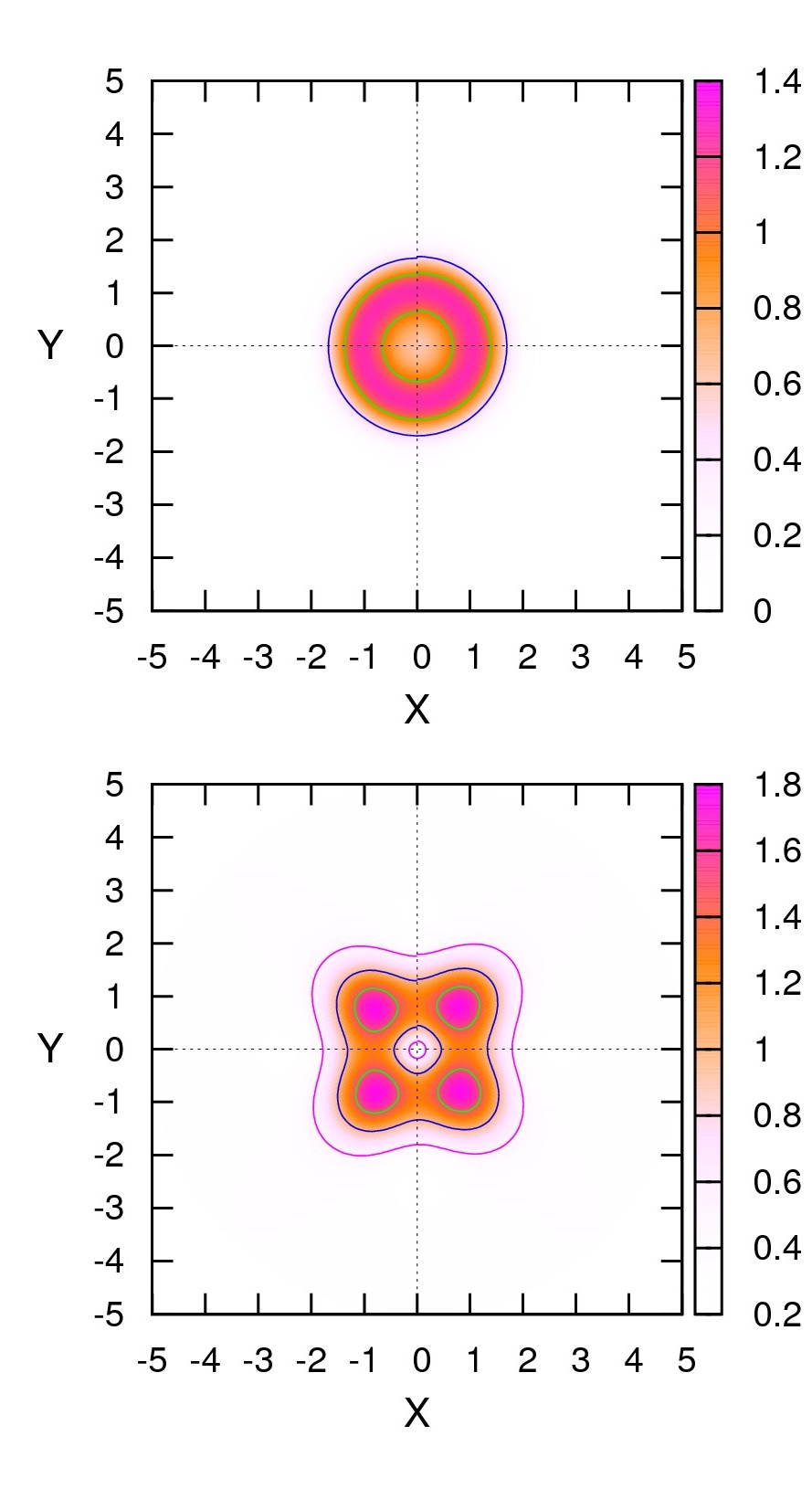}
\caption{(Color online) Contour plots of the energy density of the sectors
in the $(1,2)$ configuration in the model with potential (\protect\ref{pot})
at $\protect\kappa =0,0.4,1.0,1.5,2.0$ (from left to right).}
\label{f-2}
\end{figure}

\begin{figure}[tbh]
\refstepcounter{fig} \setlength{\unitlength}{1cm} \centering
\vspace{3.3 cm} \hspace{-0.0 cm} \includegraphics[height=5.2cm,angle=0,bb=00
00 325 725]{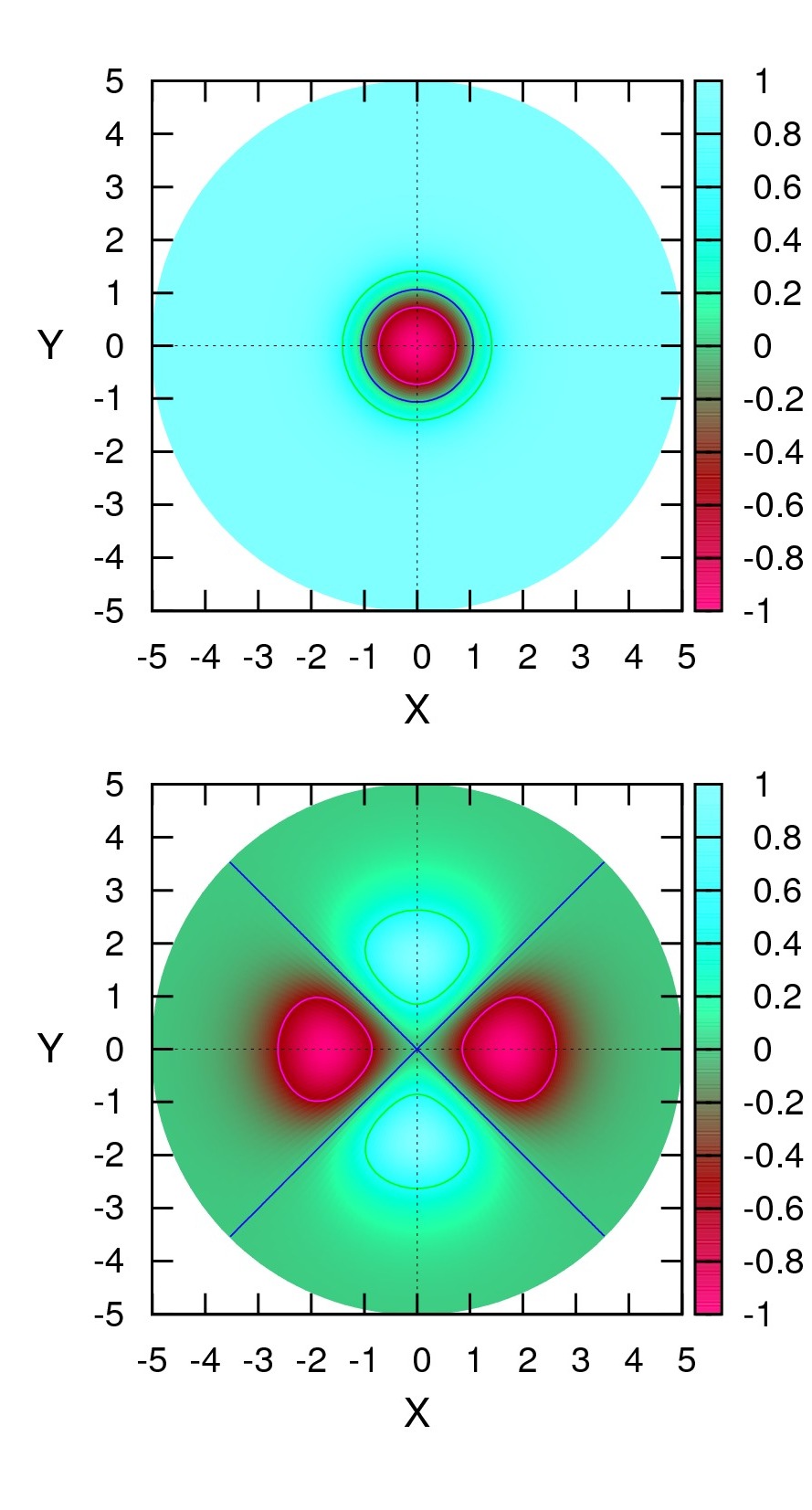}\hspace{0.2 cm} %
\includegraphics[height=5.2cm,angle=0,bb=00 00 325
725]{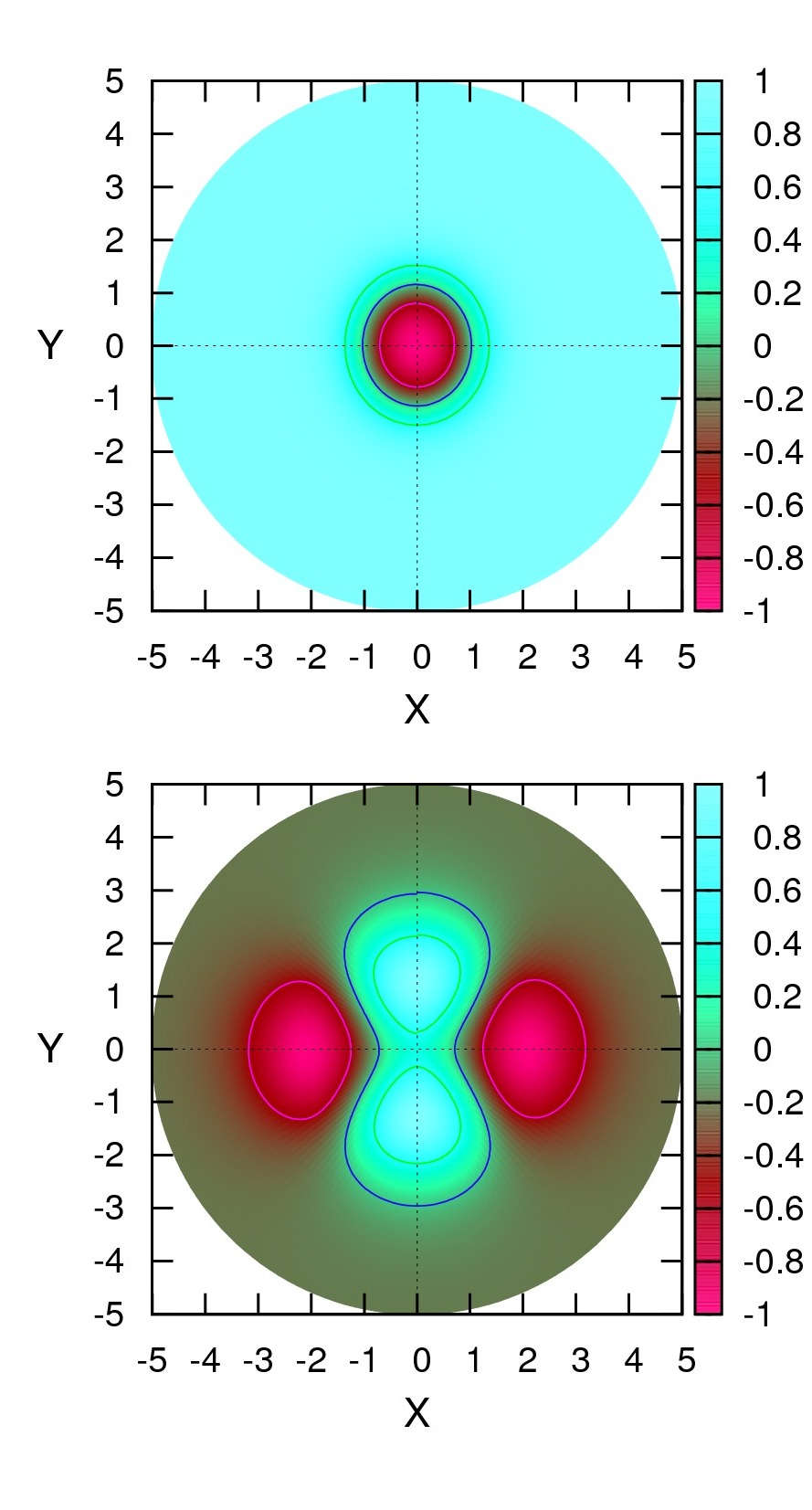}\hspace{0.2 cm} %
\includegraphics[height=5.2cm,angle=0,bb=00 00 325
725]{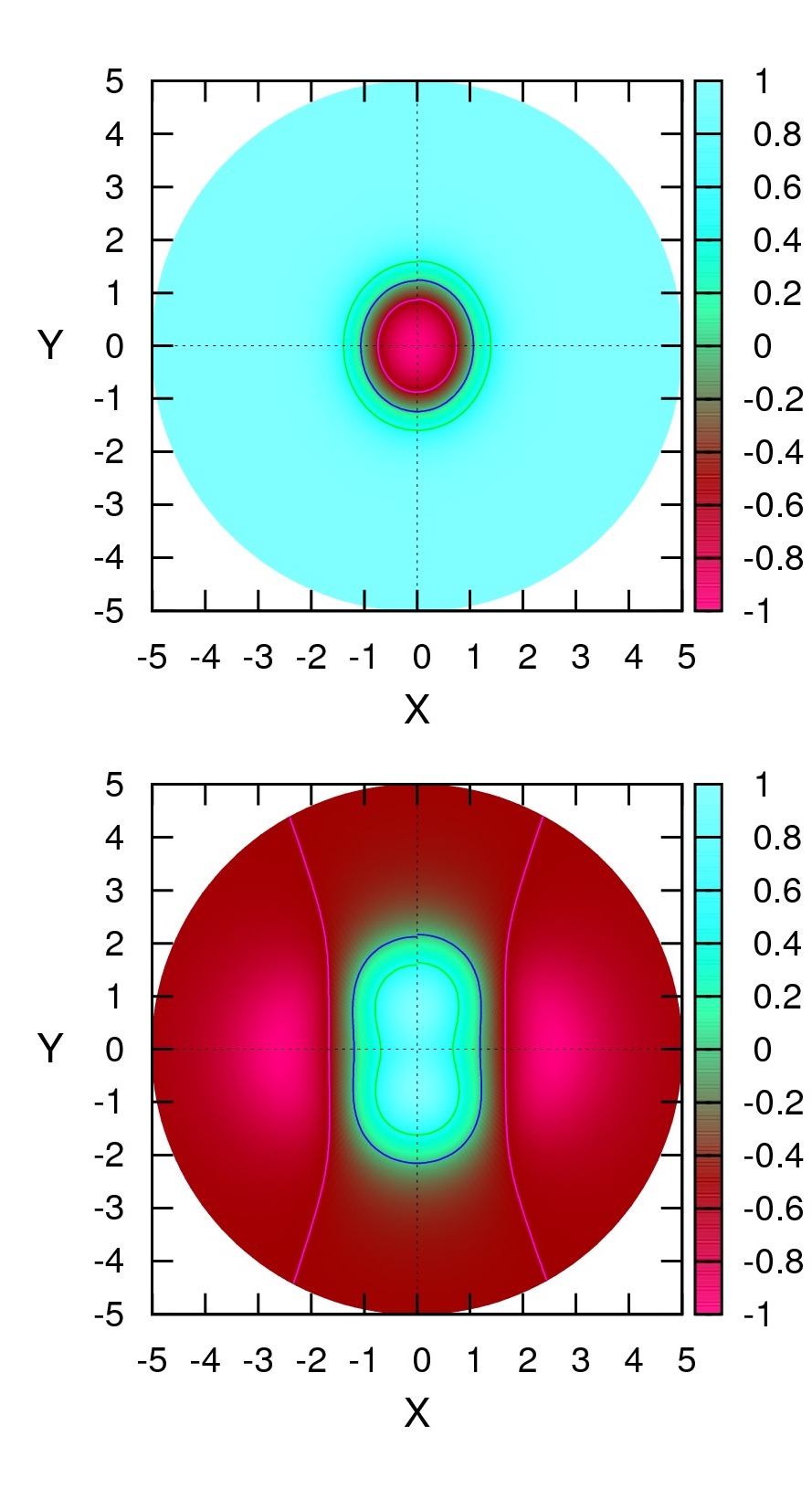}\hspace{0.2 cm} %
\includegraphics[height=5.2cm,angle=0,bb=00 00 325
725]{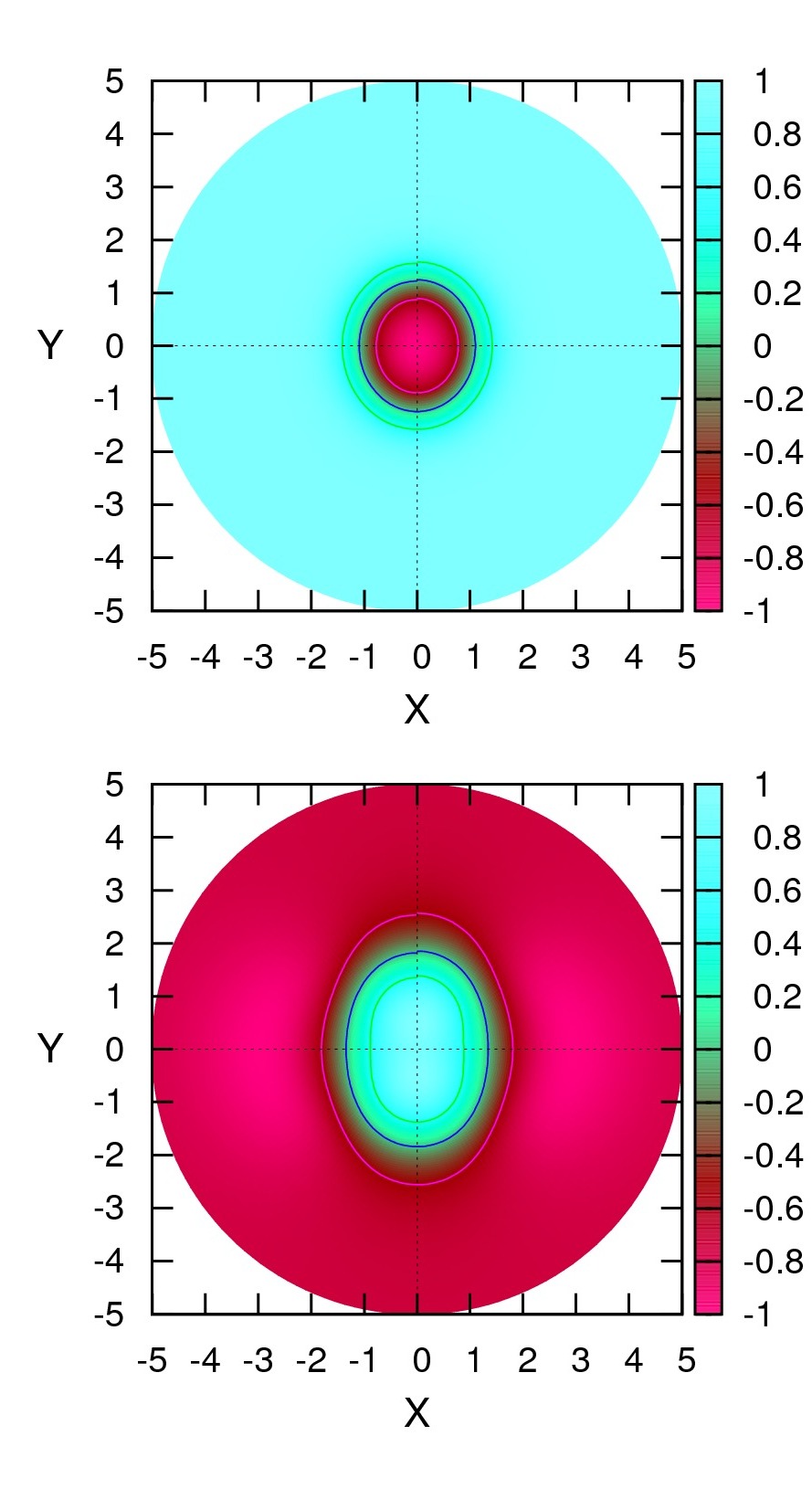}\hspace{0.2 cm} %
\includegraphics[height=5.2cm,angle=0,bb=00 00 325
725]{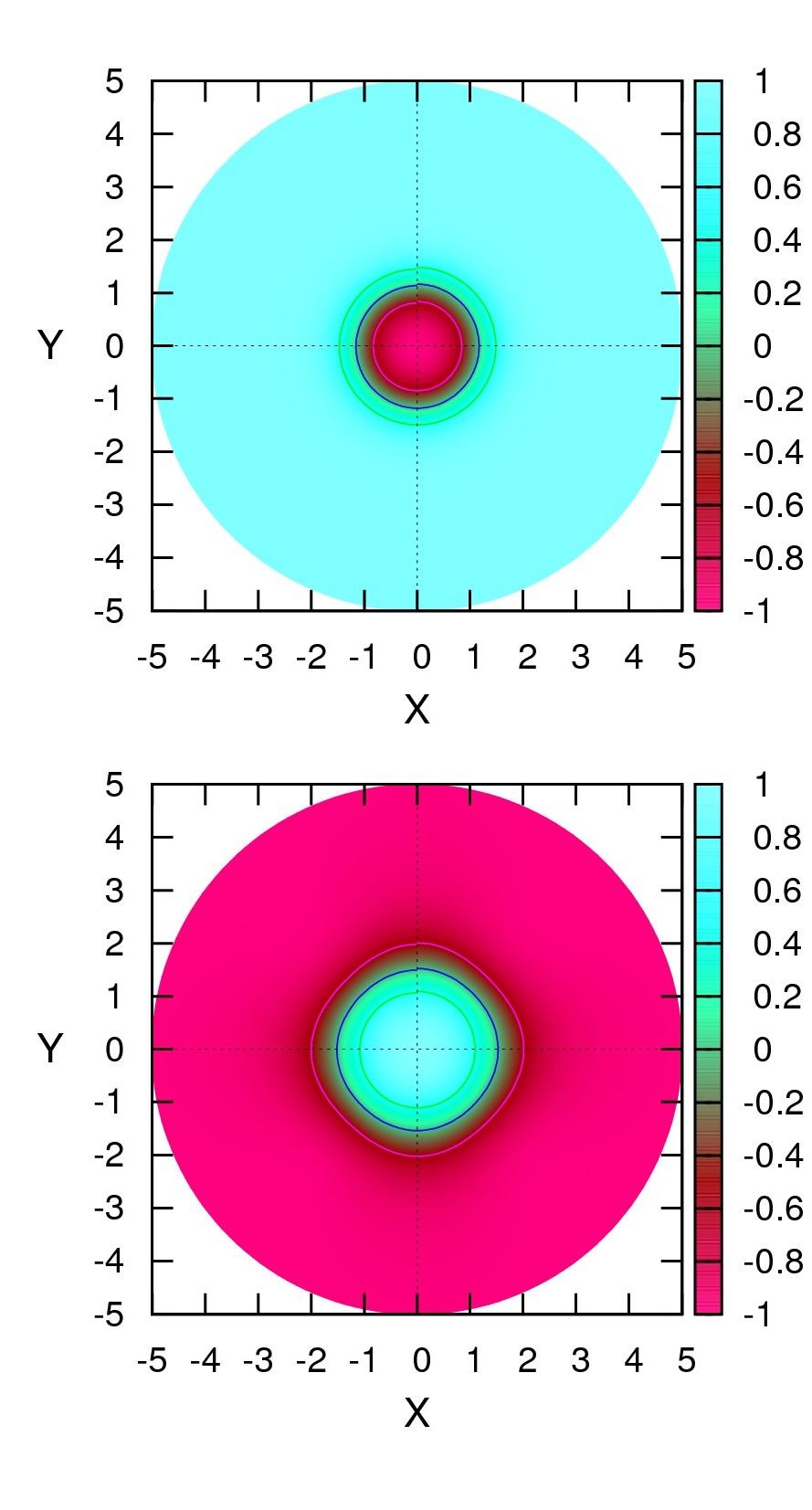}
\caption{(Color online) Contour plots of coupled components $\protect\phi %
_{3}^{(1)}$ and $\protect\phi _{1}^{(2)}$ (the upper and lower rows,
respectively) of the $(1,2)$ configuration in the model with potential (%
\protect\ref{pot}) at $\protect\kappa =0,0.4,1.0,1.5,2.0$ (from left to
right).}
\label{f-5}
\end{figure}

\begin{figure}[tbh]
\refstepcounter{fig}\setlength{\unitlength}{1cm} \centering
\vspace{3.3 cm} \hspace{-0.0 cm} \includegraphics[height=5.2cm,angle=0,bb=00
00 325 725]{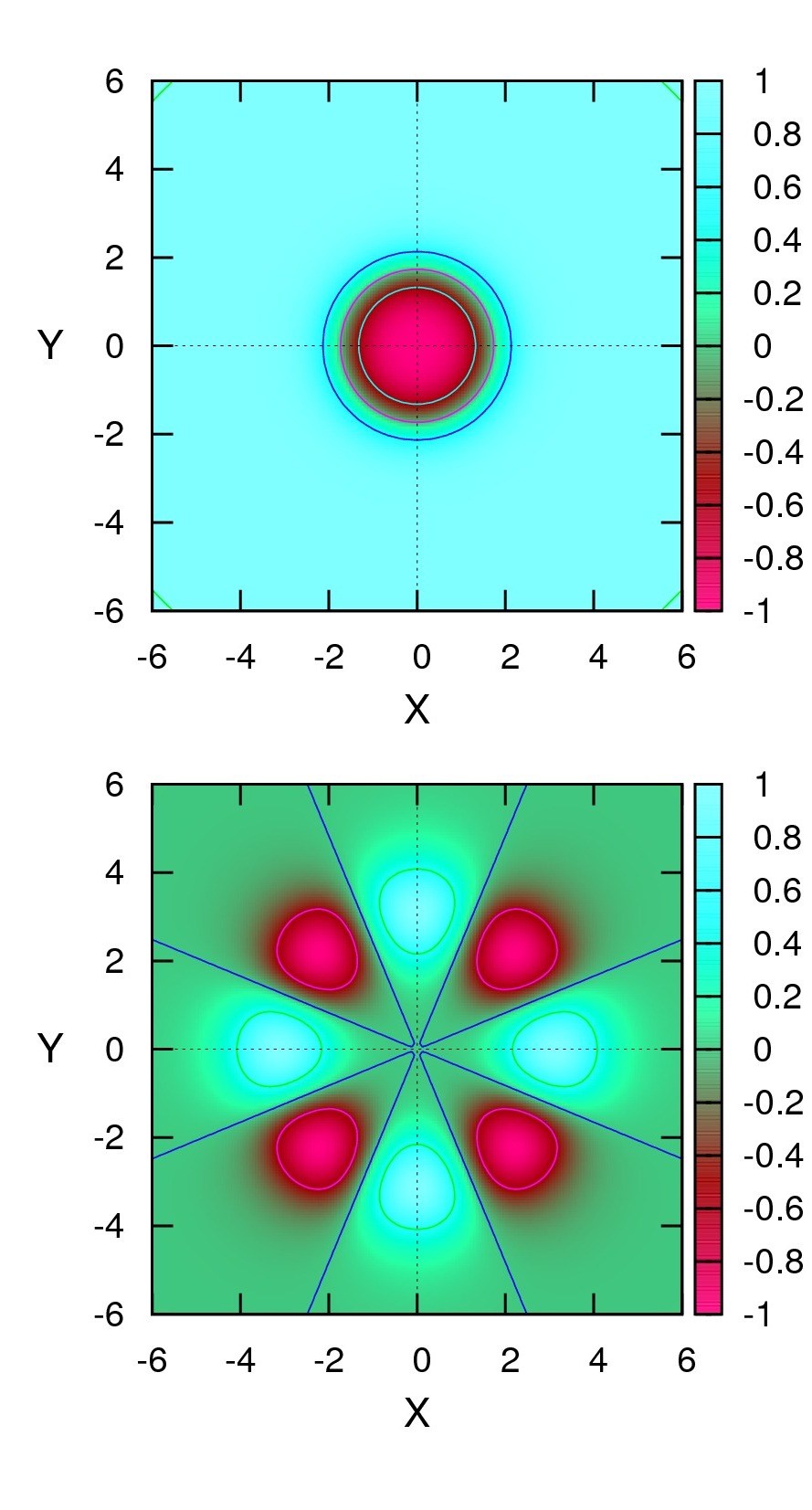}\hspace{0.2 cm} %
\includegraphics[height=5.2cm,angle=0,bb=00 00 325
725]{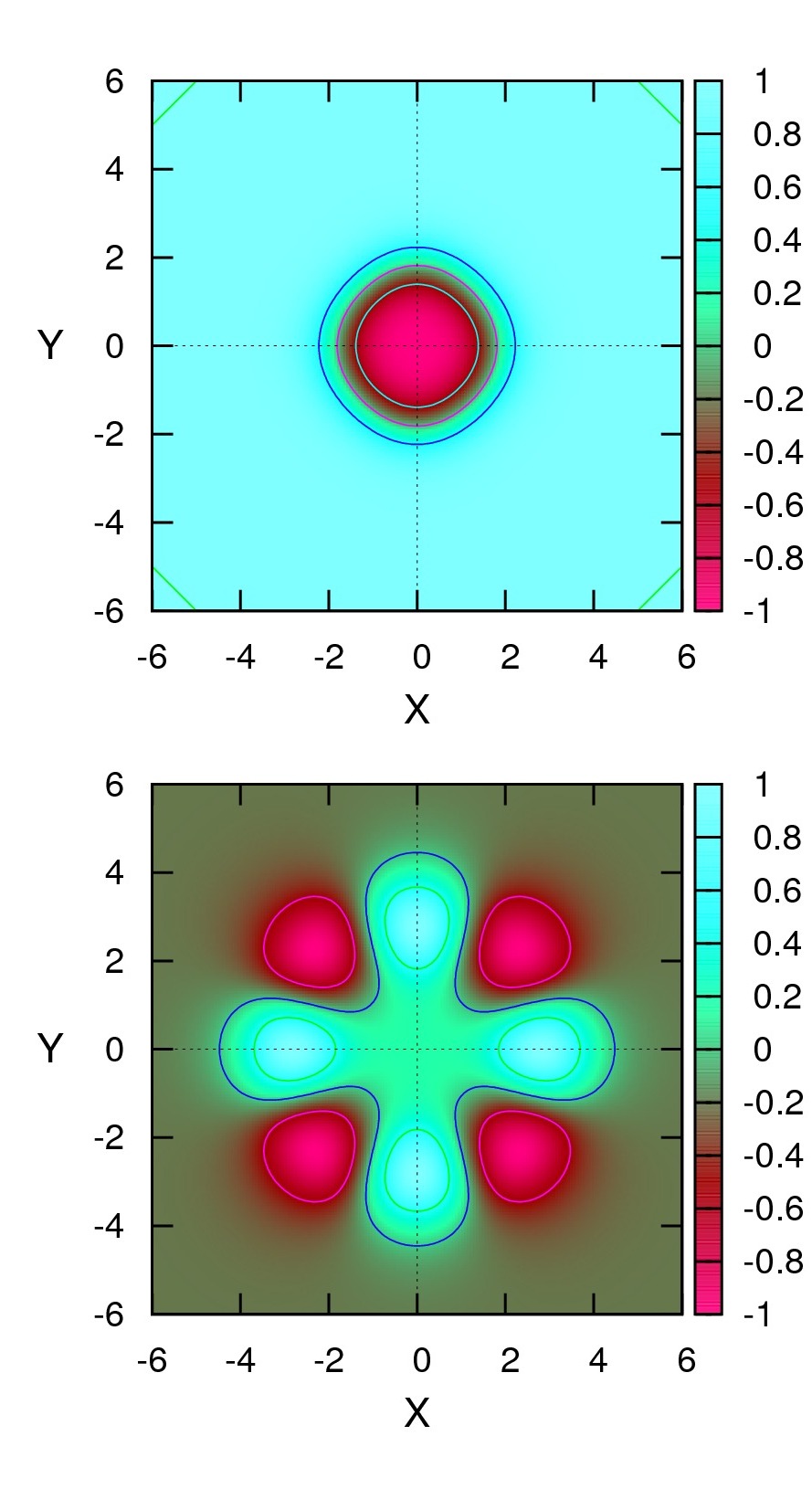}\hspace{0.2 cm} %
\includegraphics[height=5.2cm,angle=0,bb=00 00 325
725]{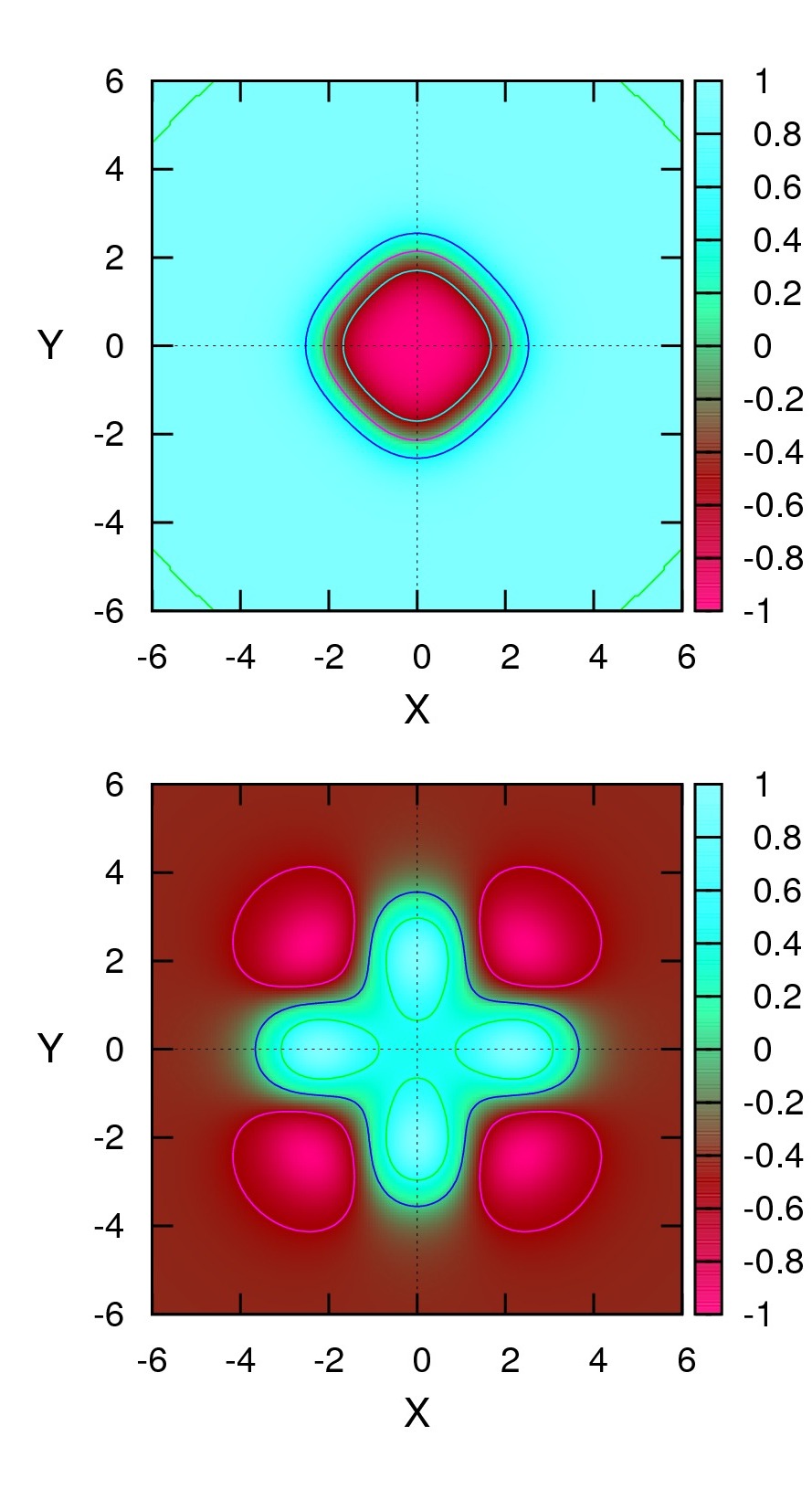}\hspace{0.2 cm} %
\includegraphics[height=5.2cm,angle=0,bb=00 00 325
725]{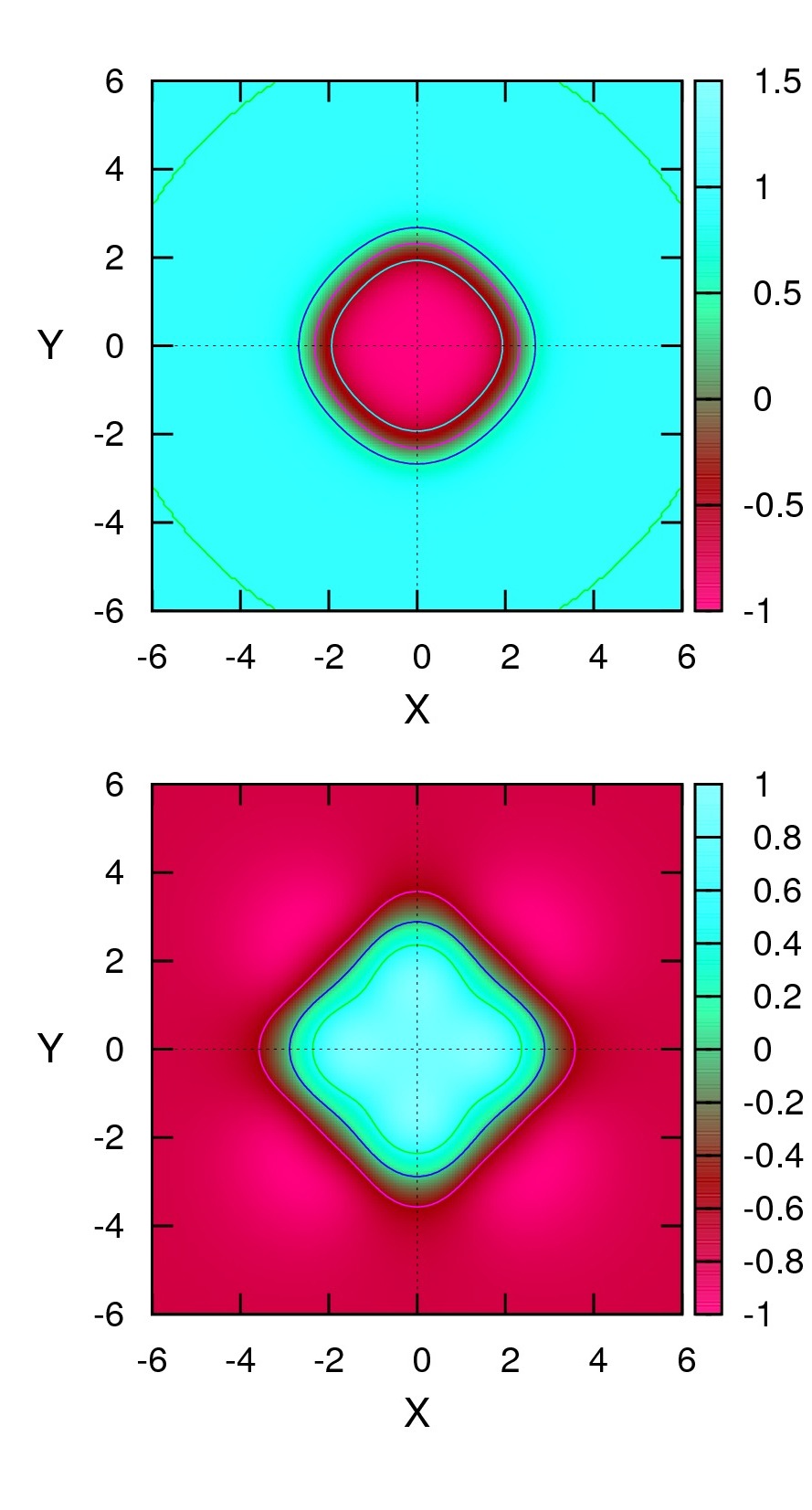}\hspace{0.2 cm} %
\includegraphics[height=5.2cm,angle=0,bb=00 00 325
725]{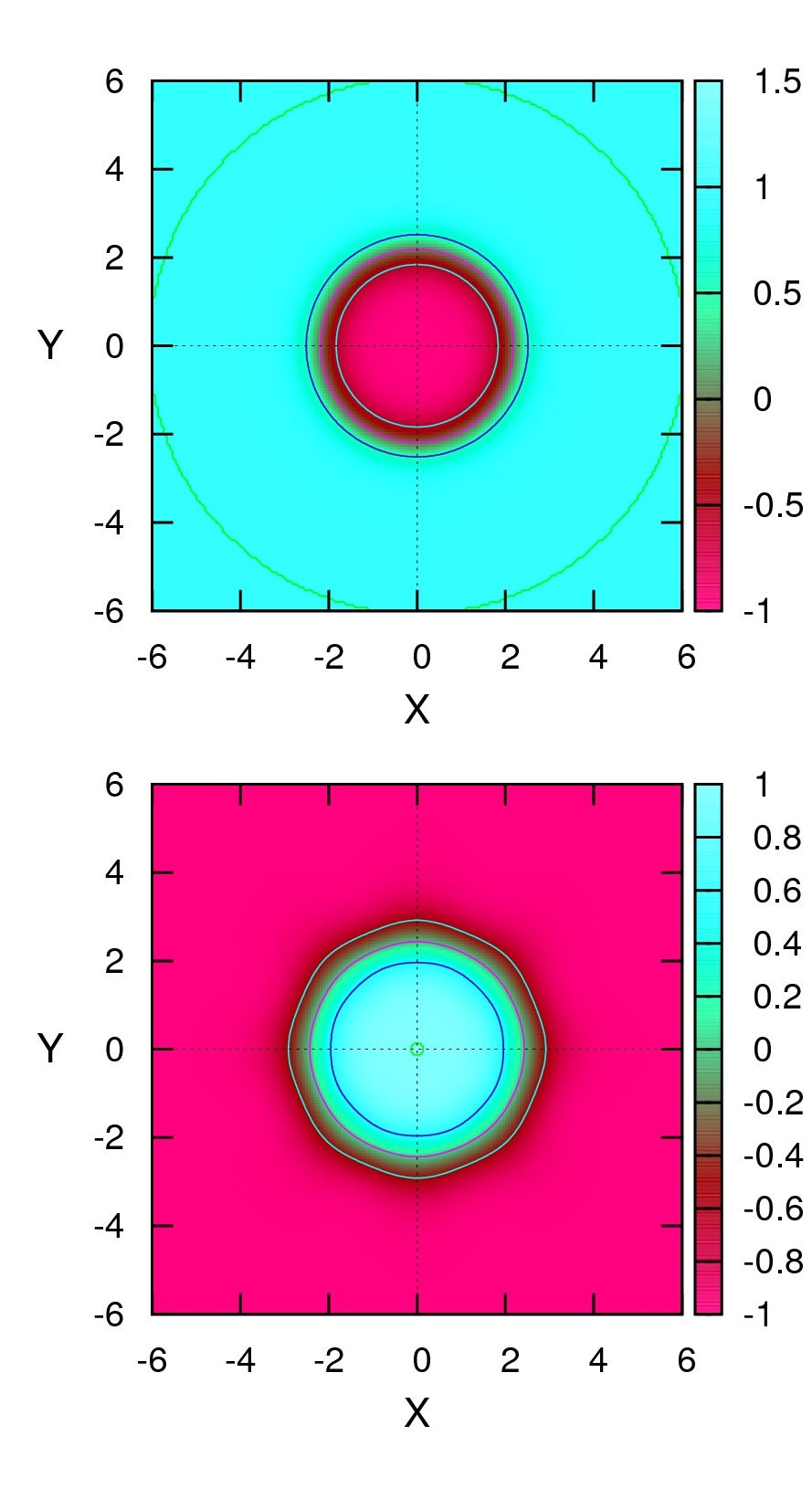}
\caption{(Color online) Contour plots of coupled components $\protect\phi %
_{3}^{(1)}$ and $\protect\phi _{1}^{(2)}$ (the upper and lower rows,
respectively) of the $(2,4)$ configuration in the model with potential (%
\protect\ref{pot}), at $\protect\kappa =0,0.4,0.8,1.5,2.0$ (from left to
right).}
\label{f-8}
\end{figure}

The pattern of the evolution of the coupled configuration with different
topological charges, following the increase of $\kappa $, is somewhat
different from what is outlined above. In Figs.~\ref{f-2},\ref{f-5} and \ref%
{f-7},\ref{f-8} we display contour plots for the energy-density
distributions in the coupled components of the $(1,2)$ and $(2,4)$
configurations, respectively, for a set of values of $\kappa $.

Unlike the configuration $(1,1)$ that we considered above, the increase of $%
\kappa $ from the initial zero value does not cause displacement of the
components from their initial positions. Instead, they start to break the
rotationally symmetric shapes in each sector, evolving towards two different
$D_{2}$-symmetric solitons stretched along the $y$- and $x$-axes in the
first and second sectors, respectively, see the second row in Fig.~\ref{f-2}%
. As the coupling becomes stronger and $\kappa $ increases above a certain
value close to $1$, the dihedral $D_{2}$ symmetry of the soliton in the
second sector breaks down, and two pairs of segments emerge in this charge-$%
2 $ component. Similar to the pattern reported above for the $(1,1)$
configuration, at the critical value (\ref{cr}) of the coupling constant,
components $\phi _{1}^{(2)}$ and $\phi _{3}^{(2)} $ actually swap, $\kappa $
playing the role of the angle of the iso-rotation of the configuration in
the second sector about component $\phi _{2}^{(2)}$. The final configuration
then consist of a charge-$1$ rotationally invariant skyrmion in the first
sector, with an annular shape of the energy-density distribution, and a
charge-$2$ skyrmion with discrete dihedral $D_{4}$ symmetry in the second
sector.

\begin{figure}[tbh]
\refstepcounter{fig} \setlength{\unitlength}{1cm} \centering
\vspace{3.3 cm} \hspace{-0.0 cm} \includegraphics[height=5.2cm,angle=0,bb=00
00 325 725]{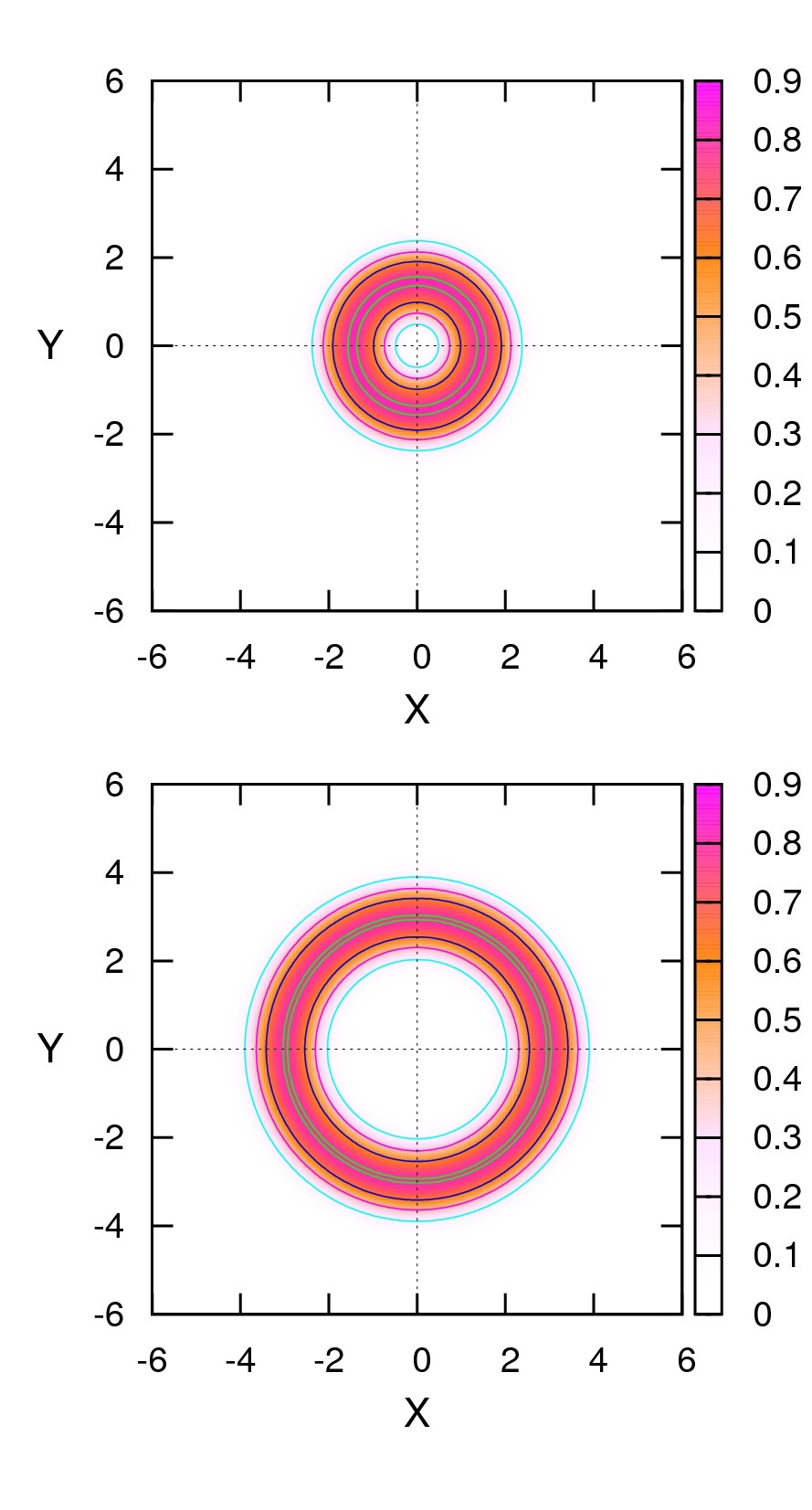}\hspace{0.2 cm} %
\includegraphics[height=5.2cm,angle=0,bb=00 00 325
725]{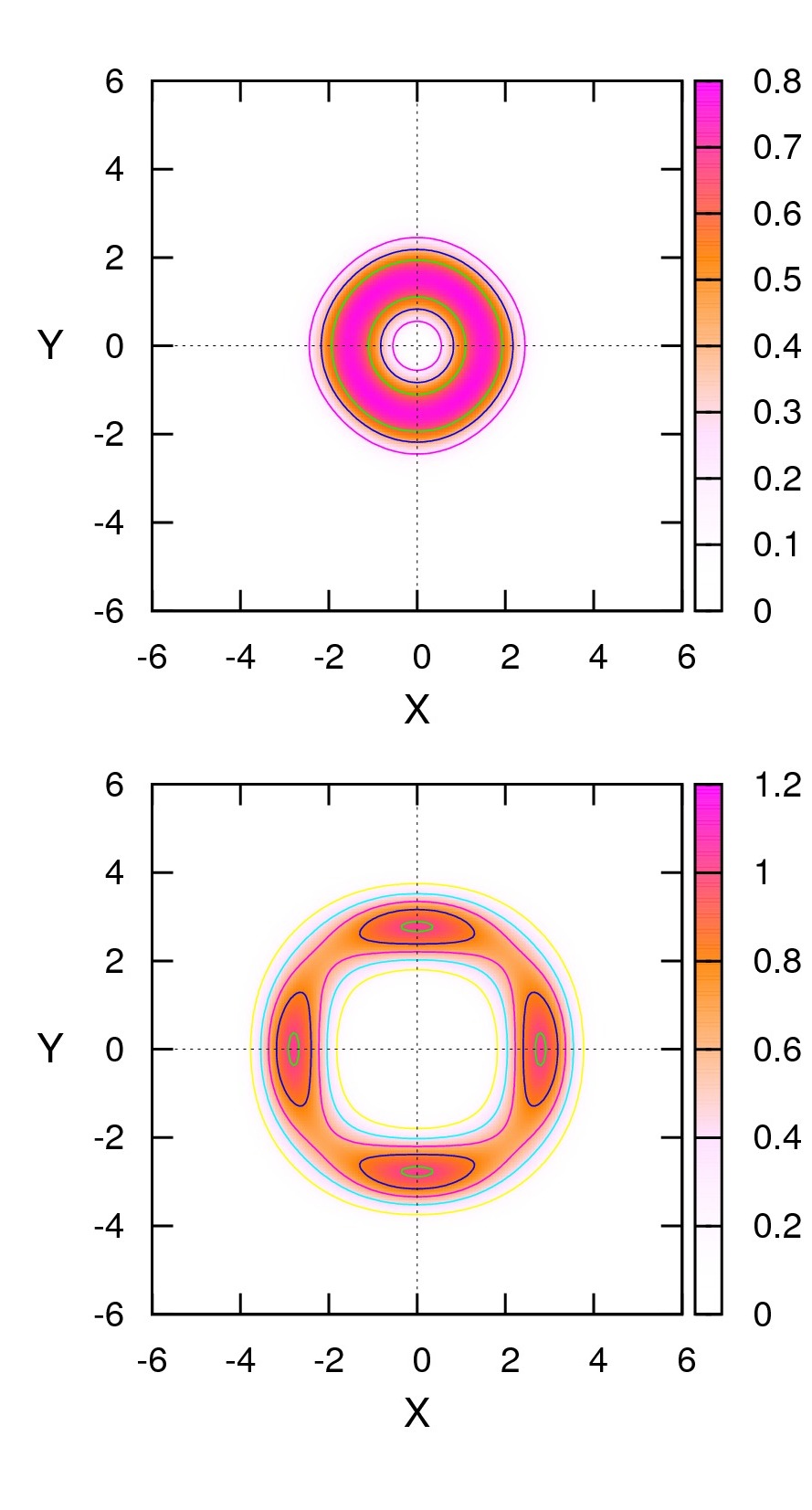}\hspace{0.2 cm} %
\includegraphics[height=5.2cm,angle=0,bb=00 00 325
725]{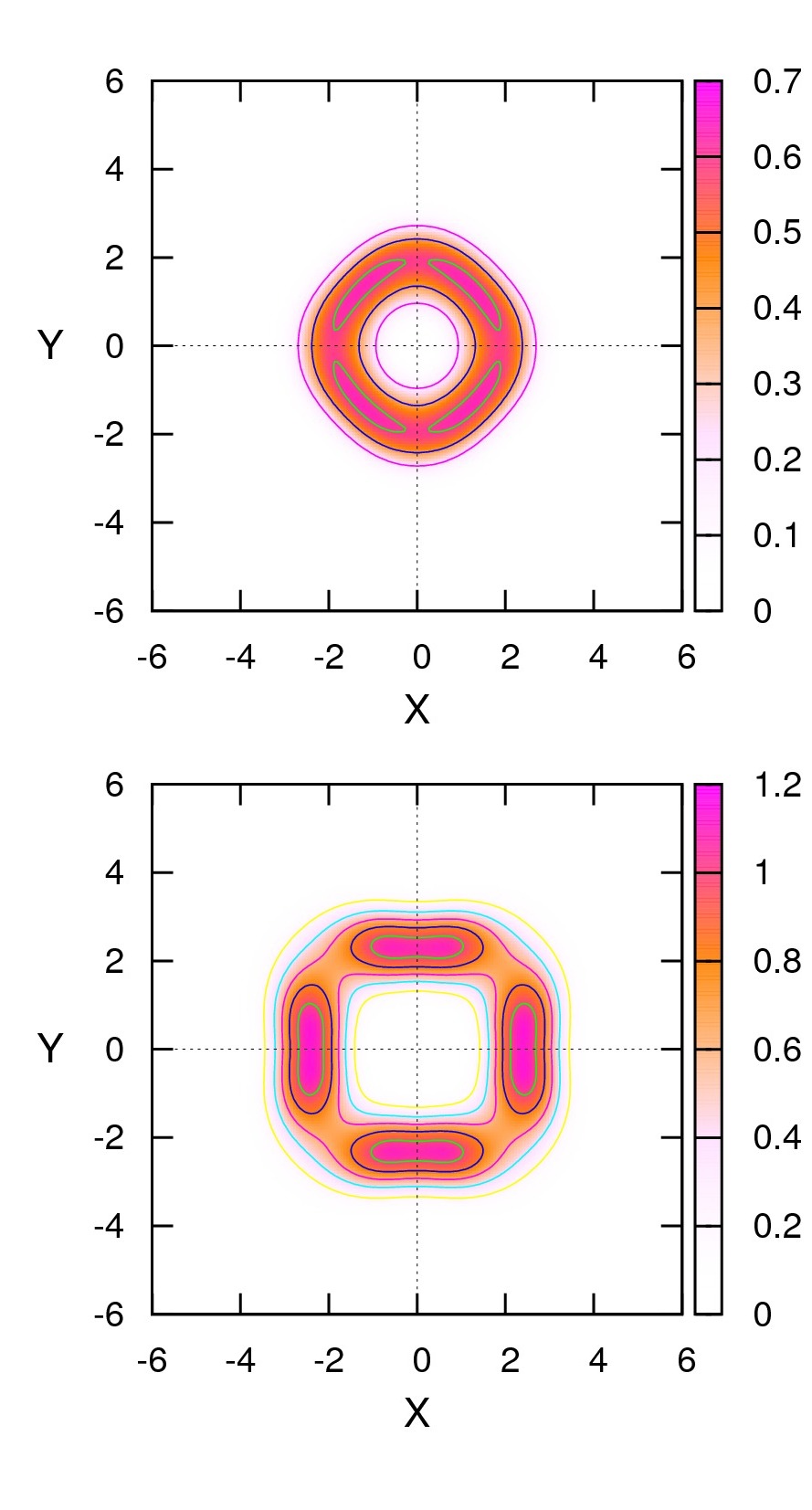}\hspace{0.2 cm} %
\includegraphics[height=5.2cm,angle=0,bb=00 00 325
725]{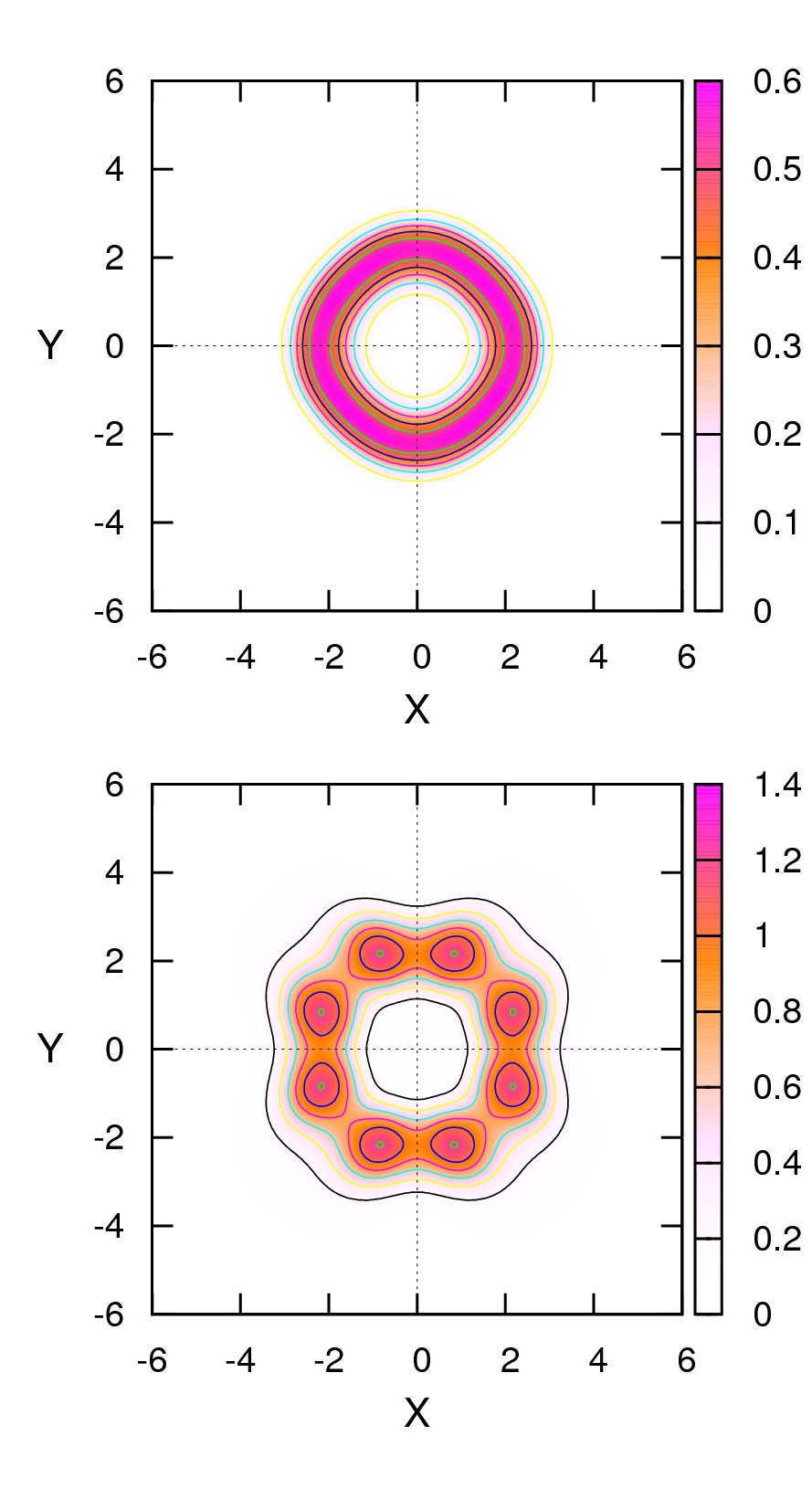}\hspace{0.2 cm} %
\includegraphics[height=5.2cm,angle=0,bb=00 00 325
725]{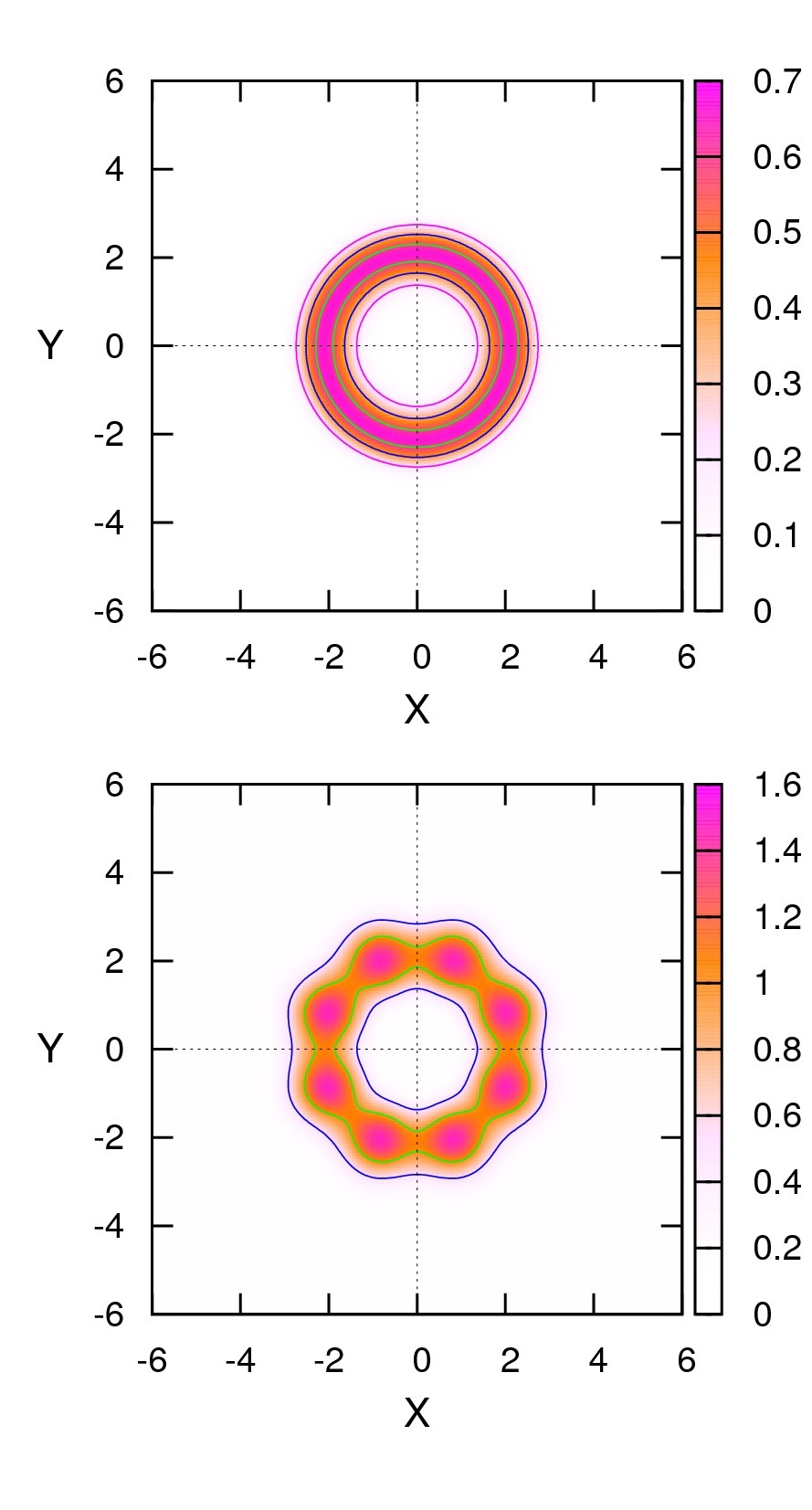}
\caption{(Color online) Contour plots of the energy density of the sectors
of the $(2,4)$ configuration in the model with potential (\protect\ref{pot})
at $\protect\kappa =0,0.4,0.8,1.5,2.0$ (from left to right).}
\label{f-7}
\end{figure}

Similar evolution scenarios were observed for configurations with higher
charges. To summarize those results, we mention that the evolution of the $%
(n,m)$ configuration in the model with potential (\ref{pot}) starts from the
rotationally invariant configurations in both sectors. As the coupling, $%
\kappa $, increases, the energy-density distributions in both components
become polygonal, i.e., symmetric with respect to dihedral group $D_{m}$
with symmetry axes of different orders. The further increase of $\kappa $
induces the permutation of the asymptotic forms of components $\phi
_{1}^{(2)}$ and $\phi _{3}^{(2)}$, and the configuration approaches the
critical limit when the first component with topological charge $B^{(1)}=n$
regains the rotational invariance, while the second component with charge $%
B^{(2)}=m$ is shaped as a necklace built of $2m$ half-skyrmions, which is
symmetric with respect to dihedral group $D_{2m}$. In particular, exactly
this scenario is observed for configurations of $(2,2),(3,3),(2,4)$ and $%
(4,4)$ types in Figs.~\ref{f-16}, \ref{f-3}, \ref{f-10} and~\ref{f-8}.

\begin{figure}[tbh]
\refstepcounter{fig} \setlength{\unitlength}{1cm}
\par
\begin{center}
\vspace{-5.0cm} \hspace{-7.0 cm} %
\includegraphics[height=10.0cm,angle=0,bb=00 00 325
725]{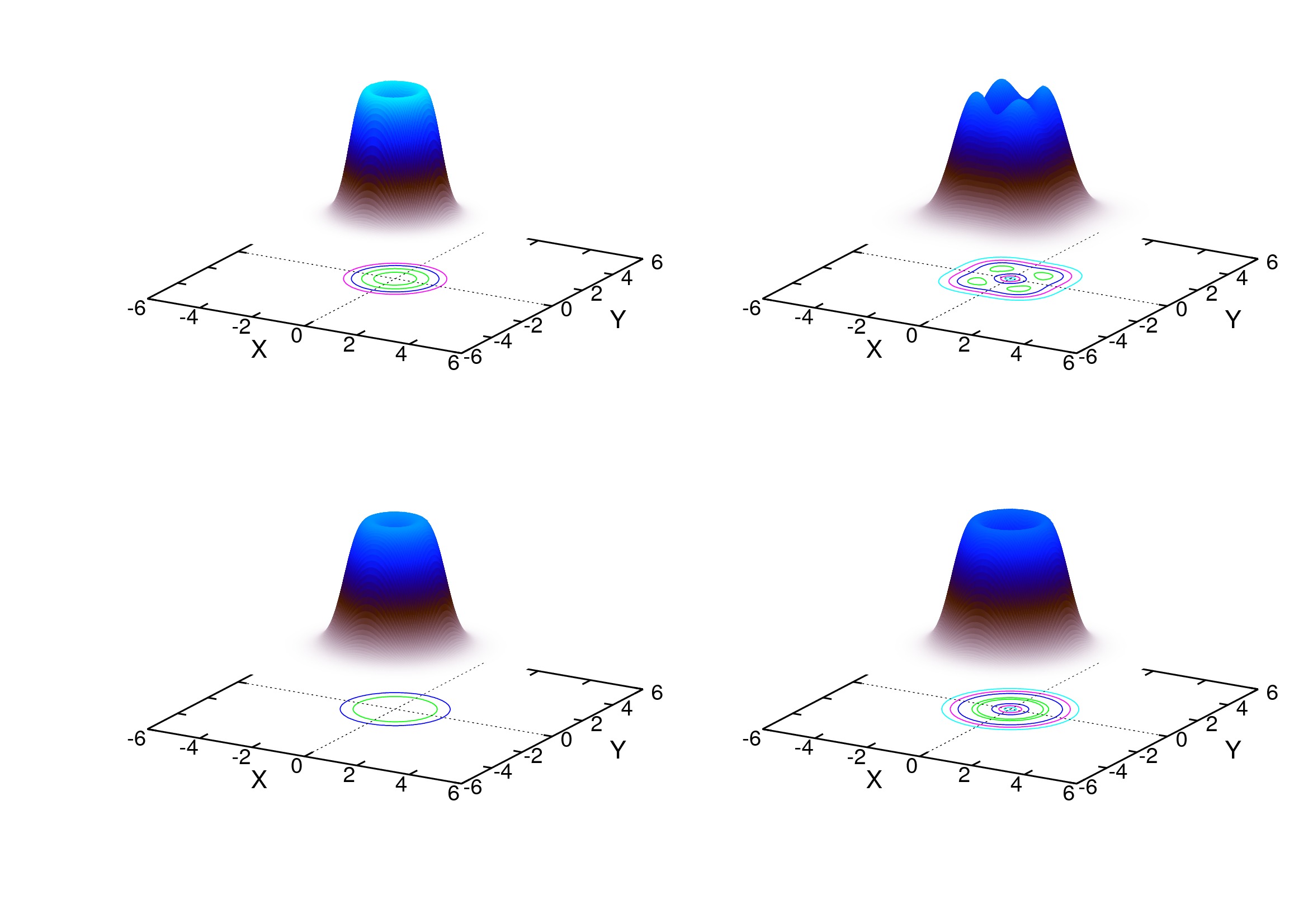}\\[-2.0cm]
\hspace{-1.5 cm}\includegraphics[height=10.0cm,angle=0,bb=00 00 325
725]{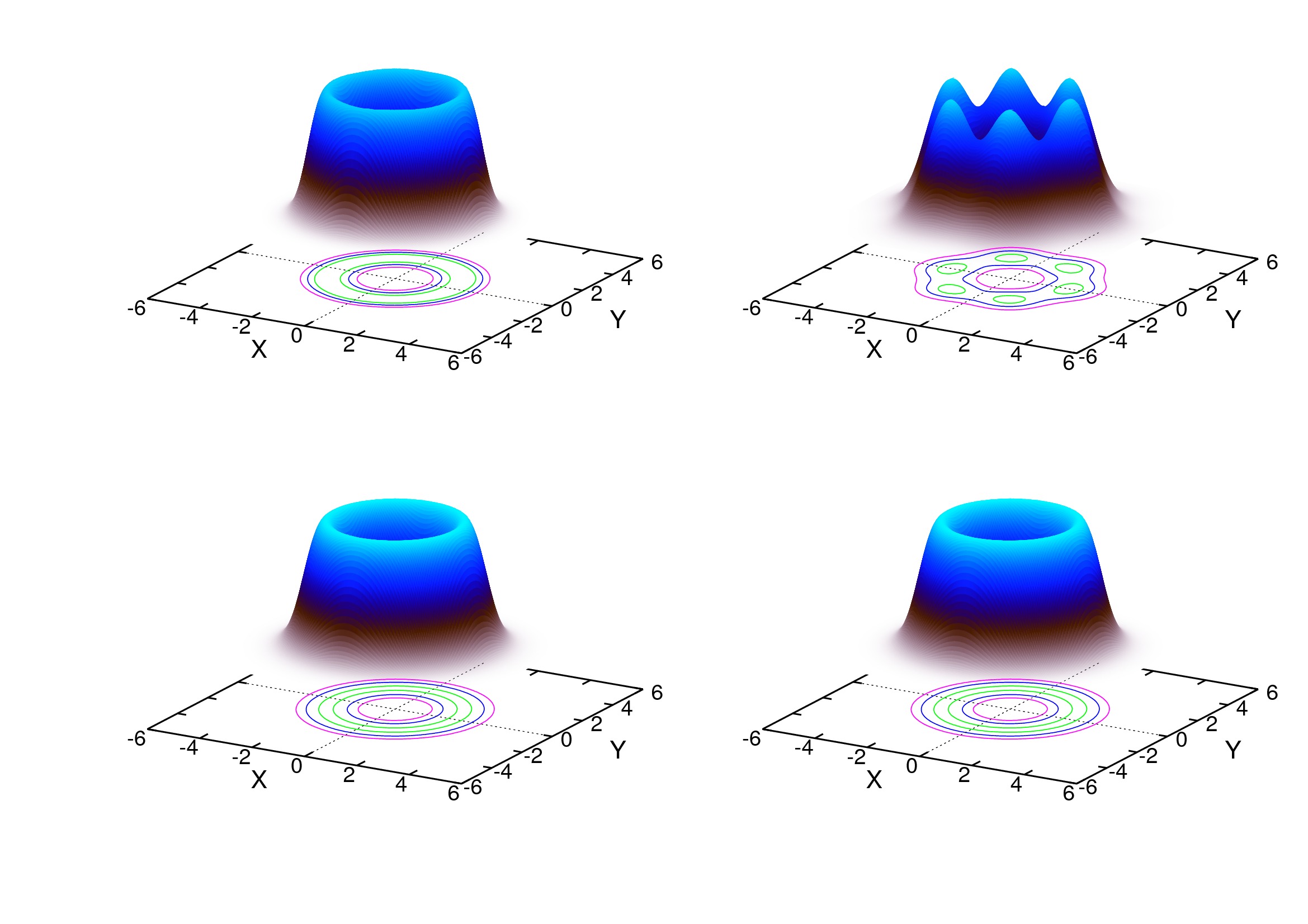} \hspace{-0.0 cm}{%
\mbox{${\vecphi}^{(1)}$\hspace{4cm}${\vecphi}^{(2)}$}}
\end{center}
\caption{(Color online) Energy density plots of the two components in the $%
(1,2)$ and $(3,3)$ configurations (the top and bottom panels, respectively)
for $\protect\kappa =2$, in the model with potential (\protect\ref{pot}) and
$\protect\mu _{1}=\protect\mu _{2}=1$ (the first and third rows), or $%
\protect\mu _{1}=\protect\mu _{2}=0$ (the second and forth rows).}
\label{f-11}
\end{figure}

Some additional remarks are necessary here. Firstly, as mass parameters $\mu
_{1}$ and $\mu _{2}$ of potential (\ref{pot}) are vanishing, the asymmetry
between the sectors vanishes too, the configurations in both sectors getting
rotationally invariant for all values of $\kappa $, see Fig.~\ref{f-11}. In
that case, the system is stable due to the coupling between the sectors, the
coupling constant is unbounded from above, there being no critical value of $%
\kappa $ [cf. Eq. (\ref{cr})], and norms in both sectors increase equally
with the subsequent growth of $\kappa $, hence the asymmetry does not change
anymore. Furthermore, the coupling can stabilize the system even when the
mass parameters take negative values.

Secondly, for the model with ``old" potential (\ref{pot-old}), the general
evolution scenario is similar to that in the model with the double vacuum
potential (\ref{pot})\ considered above, although some peculiarities may
differ. To demonstrate that, we briefly consider model (\ref{Lag}) with
potential

\begin{figure}[th]
\refstepcounter{fig} \setlength{\unitlength}{1cm} \centering
\vspace{-0.5 cm}\hspace{-0.0cm} \includegraphics[height=5.5cm,angle=0,bb=00
90 365 755]{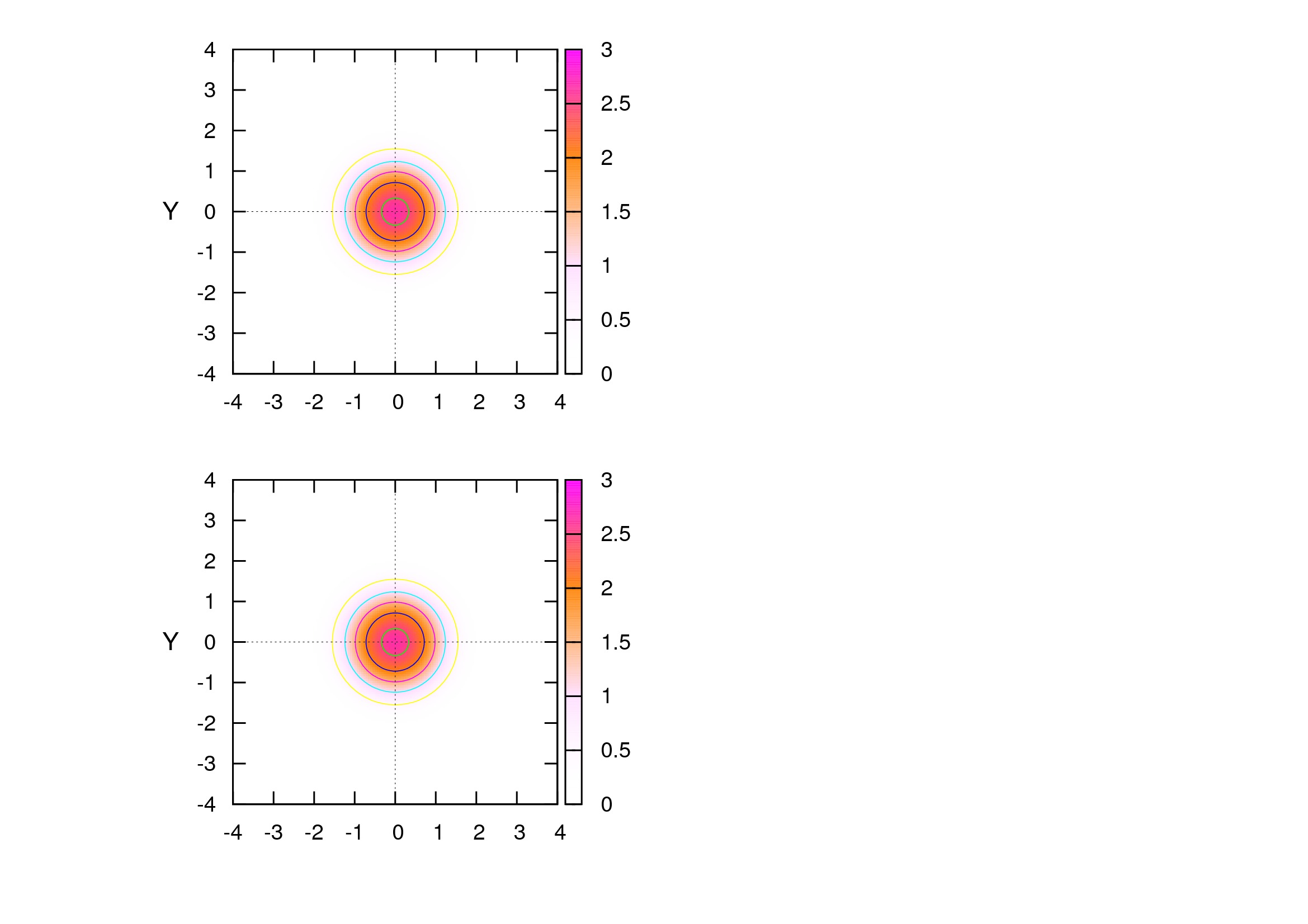}\hspace{0.2 cm} %
\includegraphics[height=5.5cm,angle=0,bb=00 90 365
755]{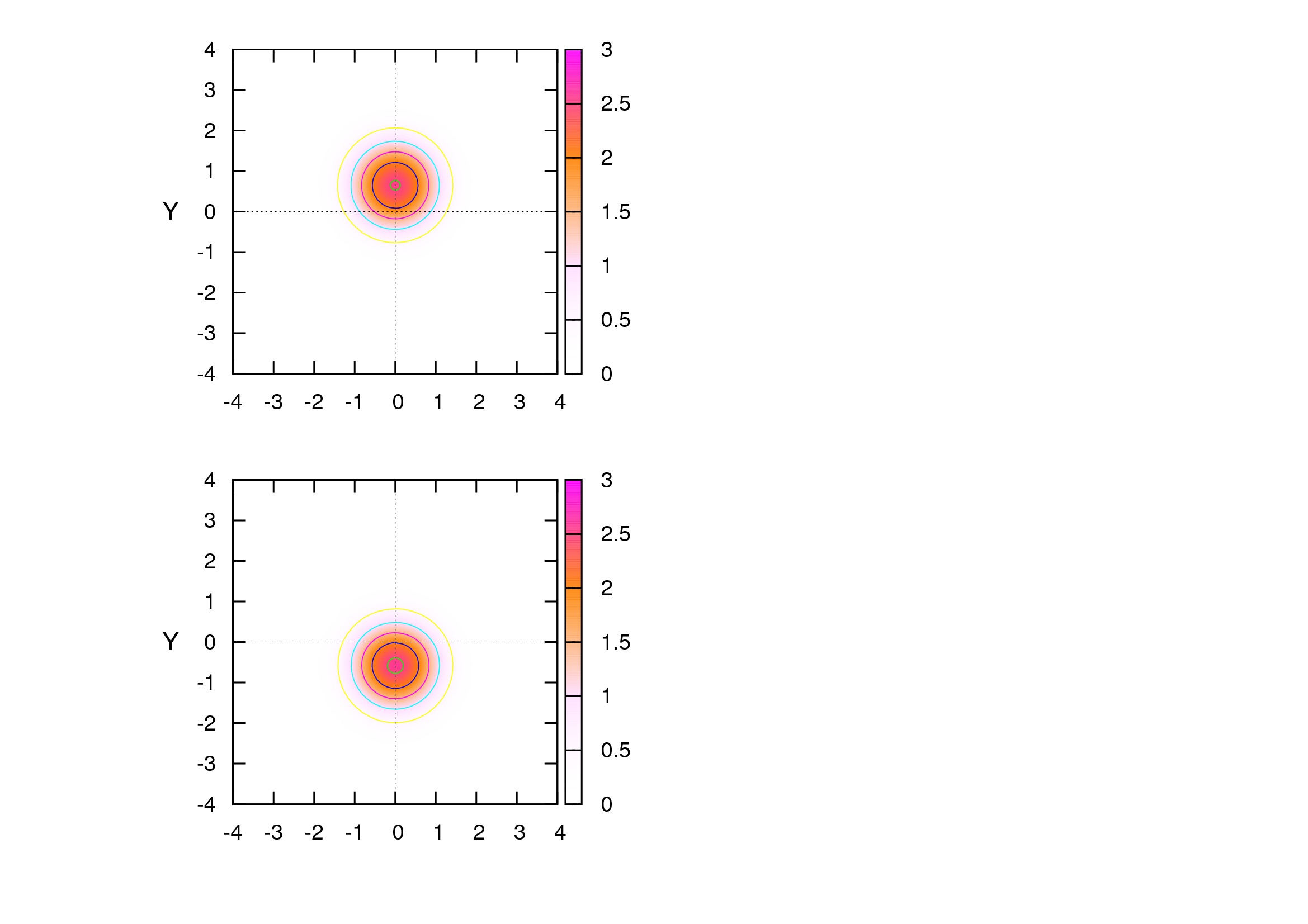}\hspace{0.2 cm} %
\includegraphics[height=5.5cm,angle=0,bb=00 90 365
755]{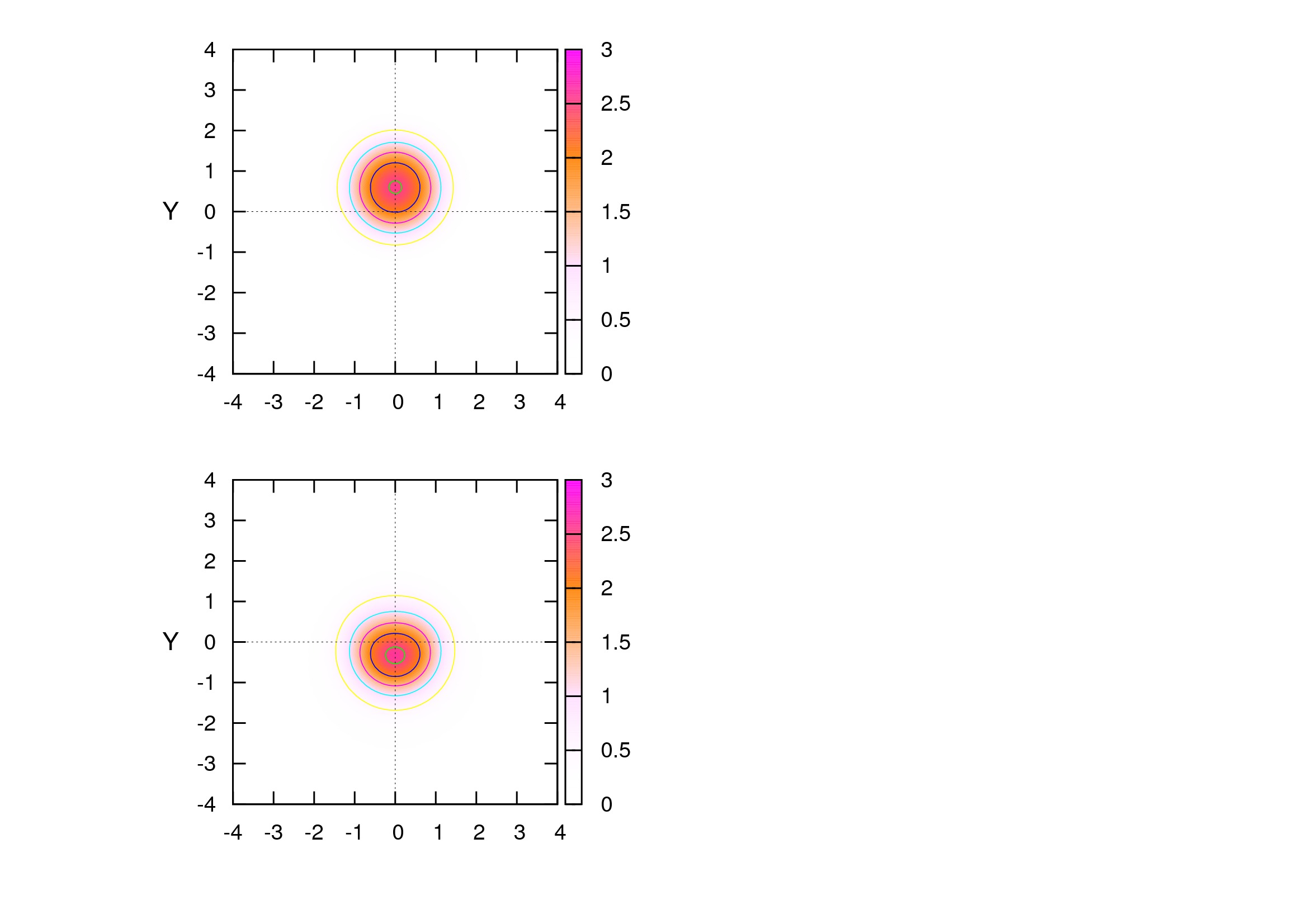}\hspace{0.2 cm} %
\includegraphics[height=5.5cm,angle=0,bb=00 90 365
755]{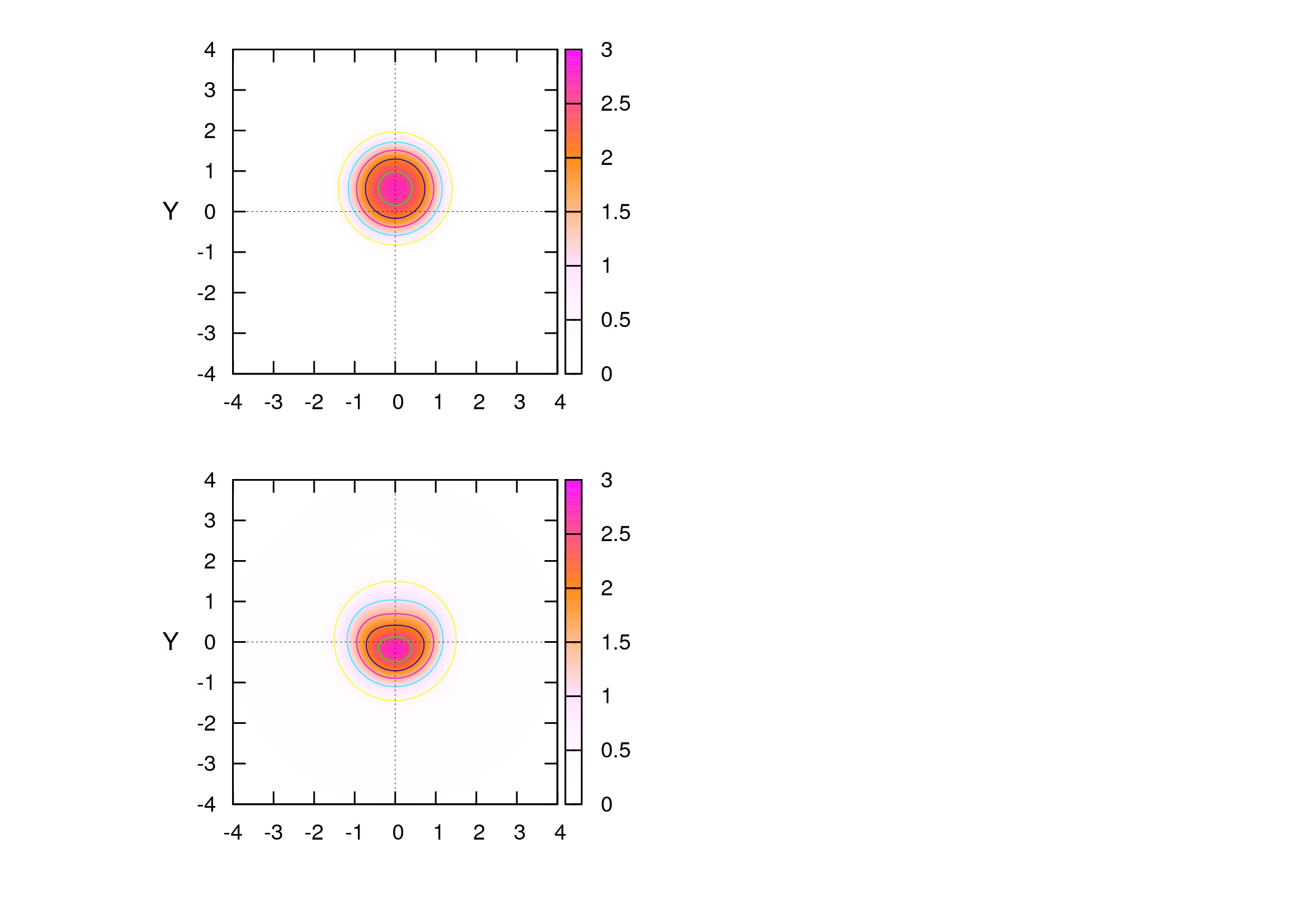}
\caption{(Color online) Contour plots of the energy density of the
components of the $(1,1)$ configuration in the model with potential (\protect
\ref{pot-old-2}), at $\protect\kappa =0,0.2,1.0,2.0$ (from left to right).}
\label{f-13}
\end{figure}
\begin{figure}[ht]
\refstepcounter{fig} \setlength{\unitlength}{1cm} \centering
\vspace{0.5 cm}\hspace{-0.0 cm} \includegraphics[height=5.5 cm,angle=0,bb=00
90 365 755]{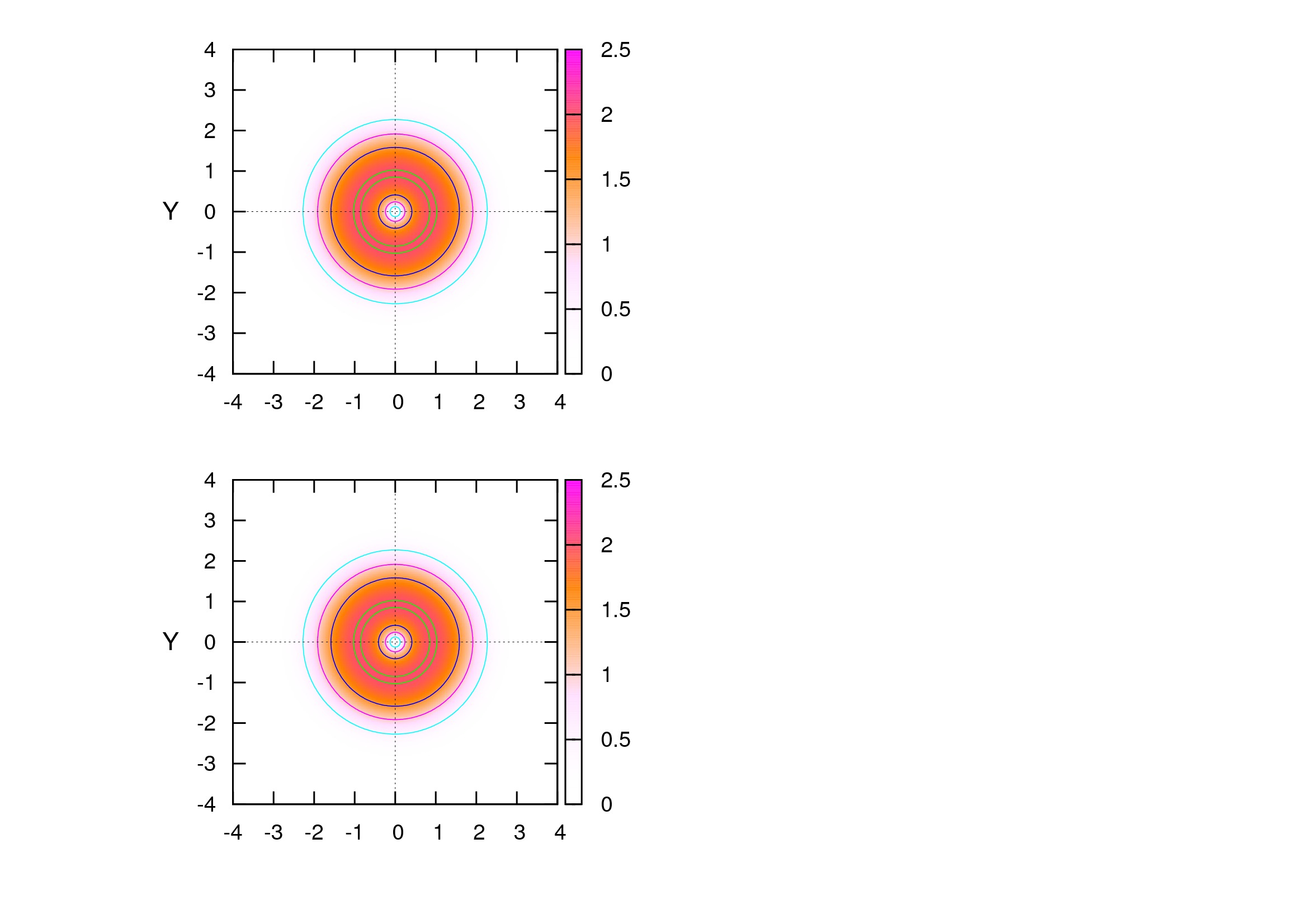}\hspace{0.2 cm} %
\includegraphics[height=5.5 cm,angle=0,bb=00 90 365
755]{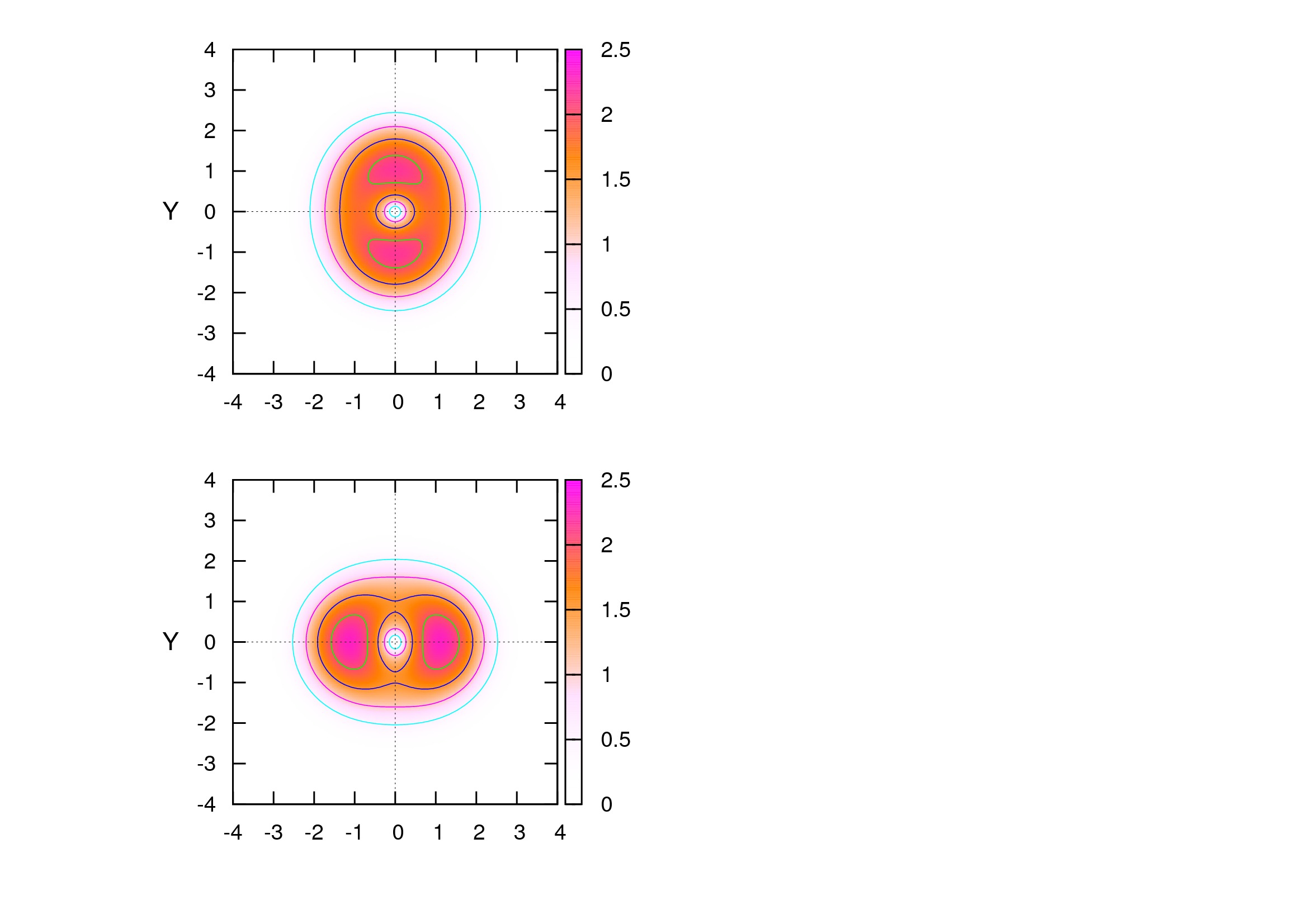}\hspace{0.2 cm} %
\includegraphics[height=5.5 cm,angle=0,bb=00 90 365
755]{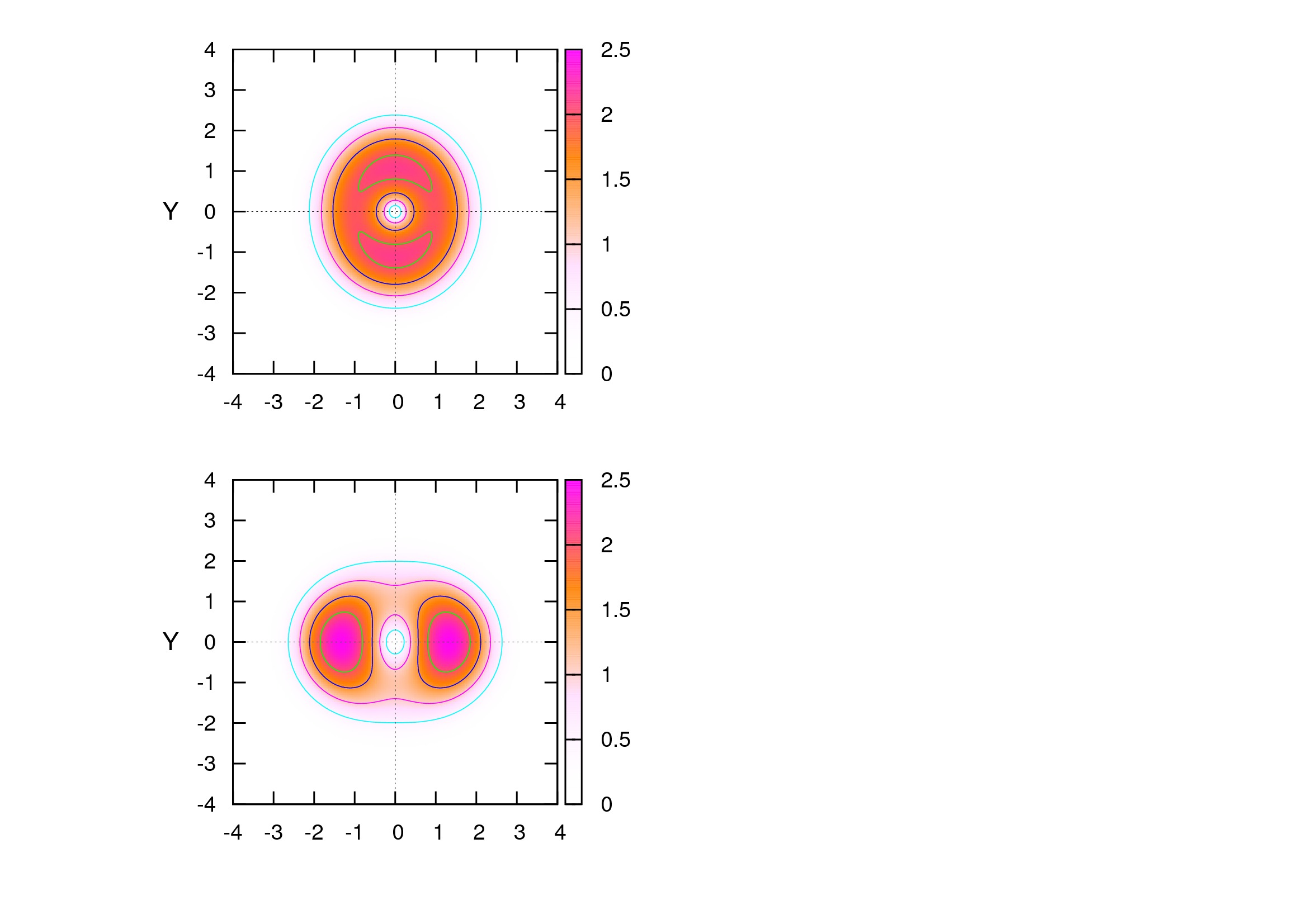}\hspace{0.2 cm} %
\includegraphics[height=5.5 cm,angle=0,bb=00 90 365
755]{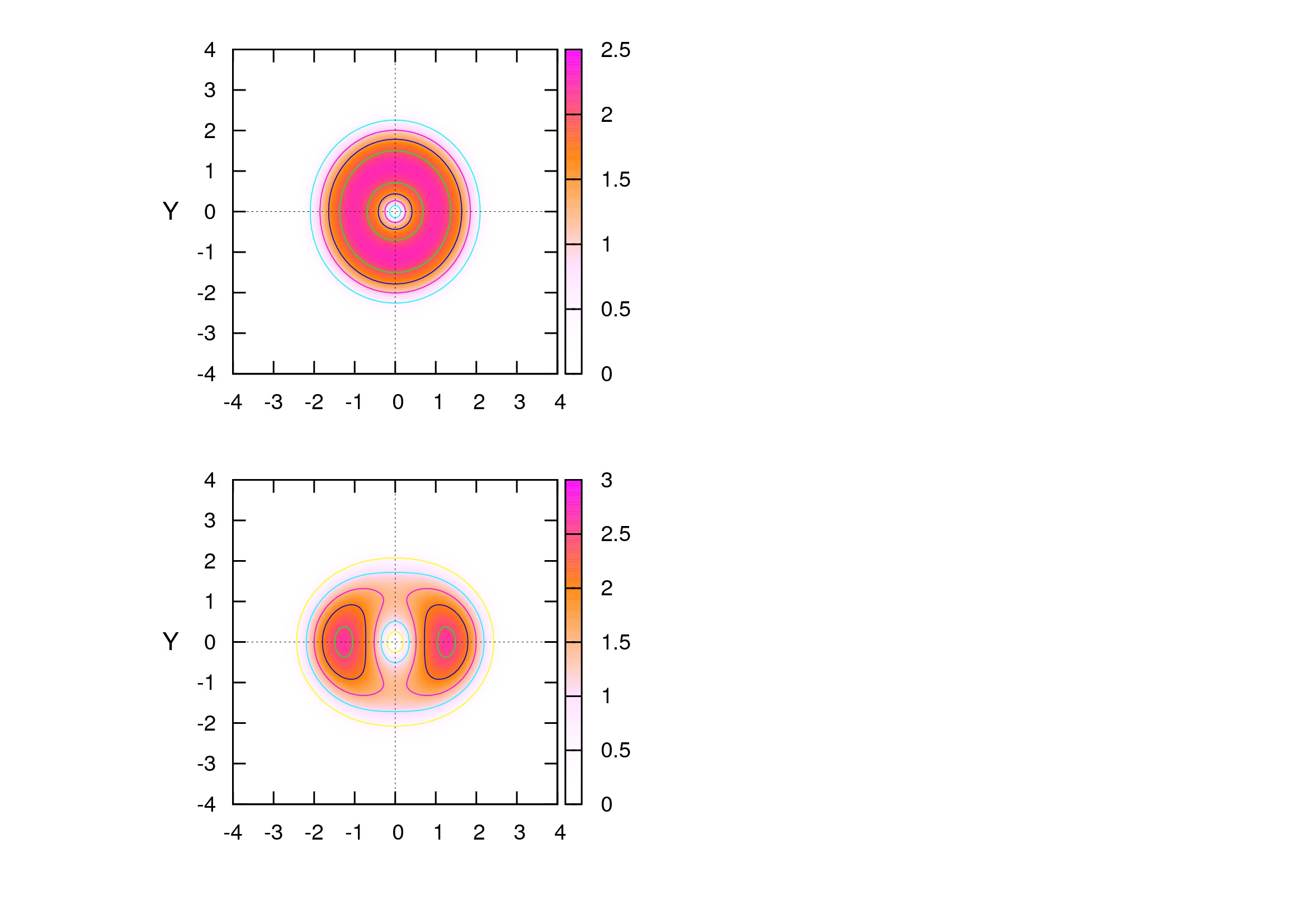}
\caption{(Color online) The energy density of components of the $(2,2)$
configuration in the model with potential (\protect\ref{pot-old-2}), at $%
\protect\kappa =0,0.2,1.0,2.0$ (from left to right).}
\label{f-14}
\end{figure}
\begin{figure}[ht]
\refstepcounter{fig} \setlength{\unitlength}{1cm} \centering
\vspace{1.0 cm}\hspace{0.0 cm} \includegraphics[height=5.5cm,angle=0,bb=00
90 365 755]{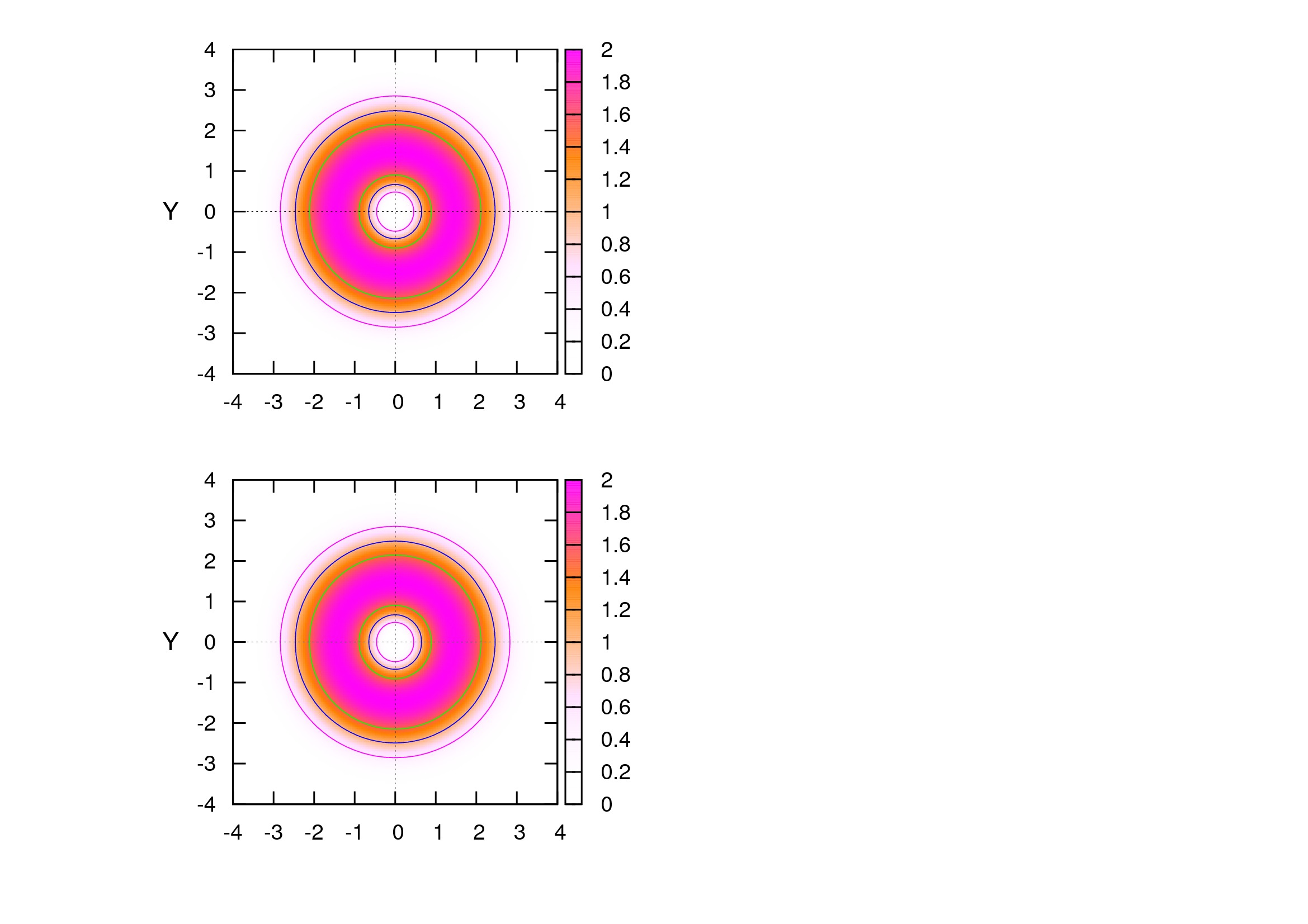}\hspace{0.2 cm} %
\includegraphics[height=5.5cm,angle=0,bb=00 90 365
755]{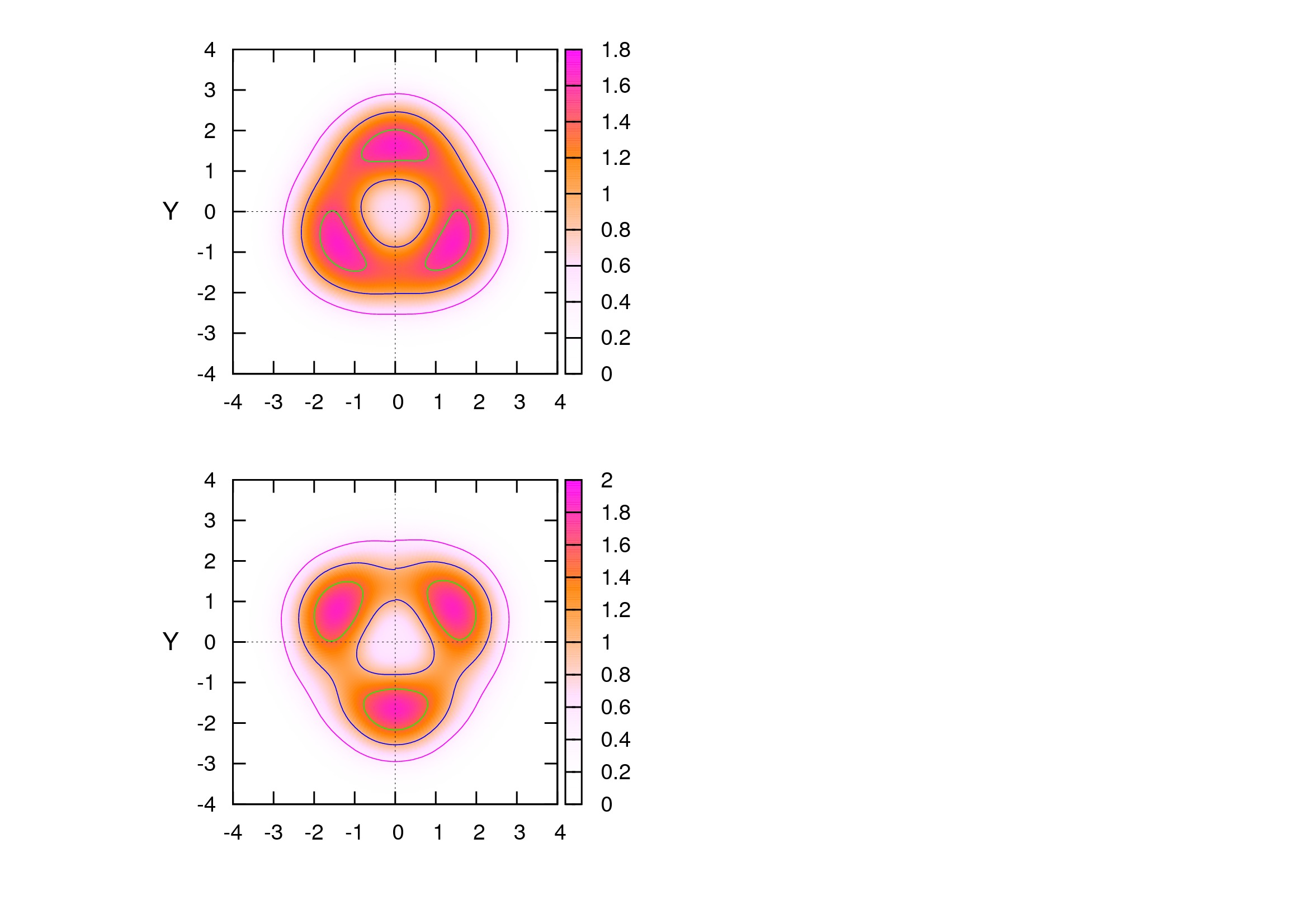}\hspace{0.2 cm} %
\includegraphics[height=5.5cm,angle=0,bb=00 90 365
755]{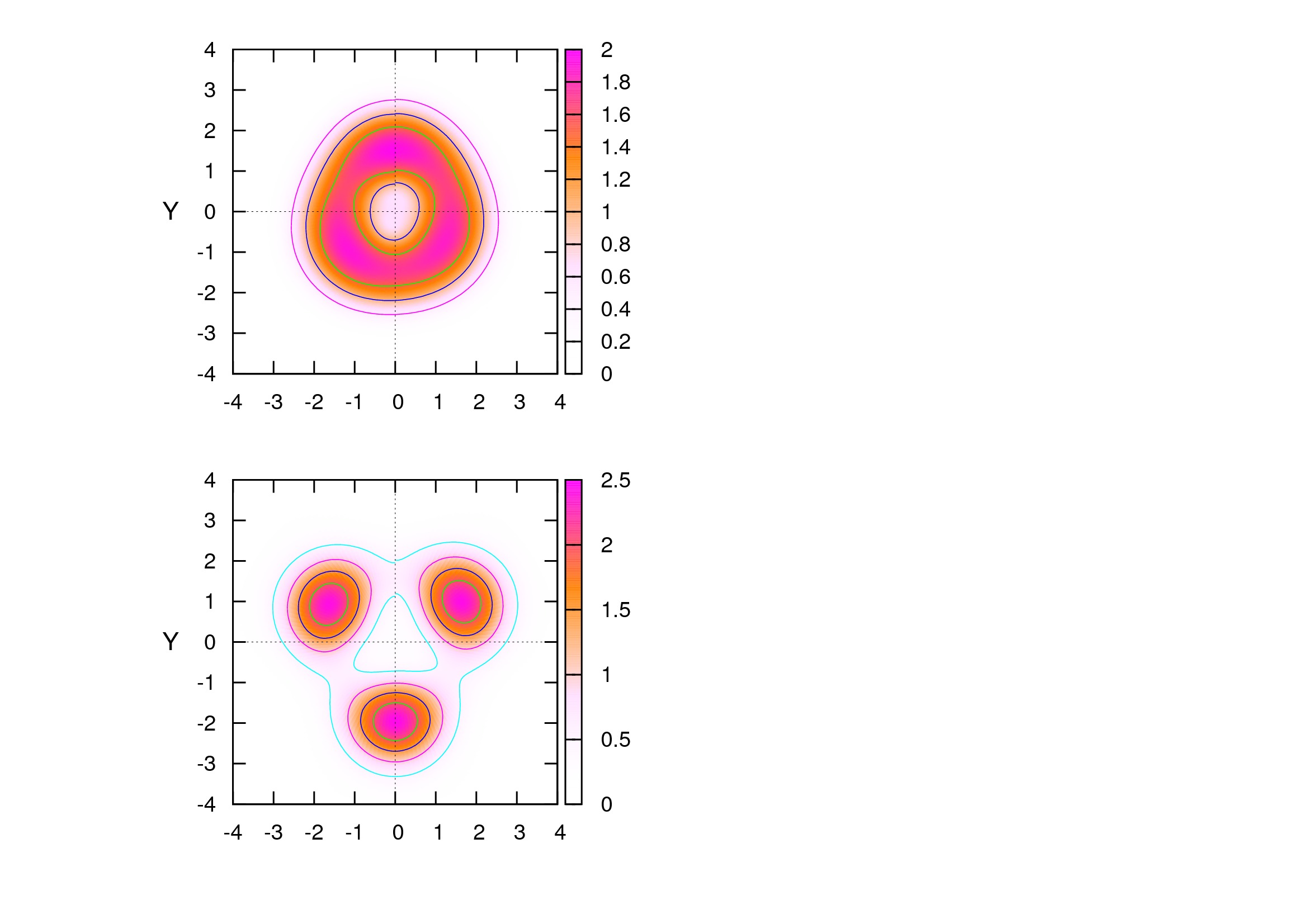}\hspace{0.2 cm} %
\includegraphics[height=5.5cm,angle=0,bb=00 90 365
755]{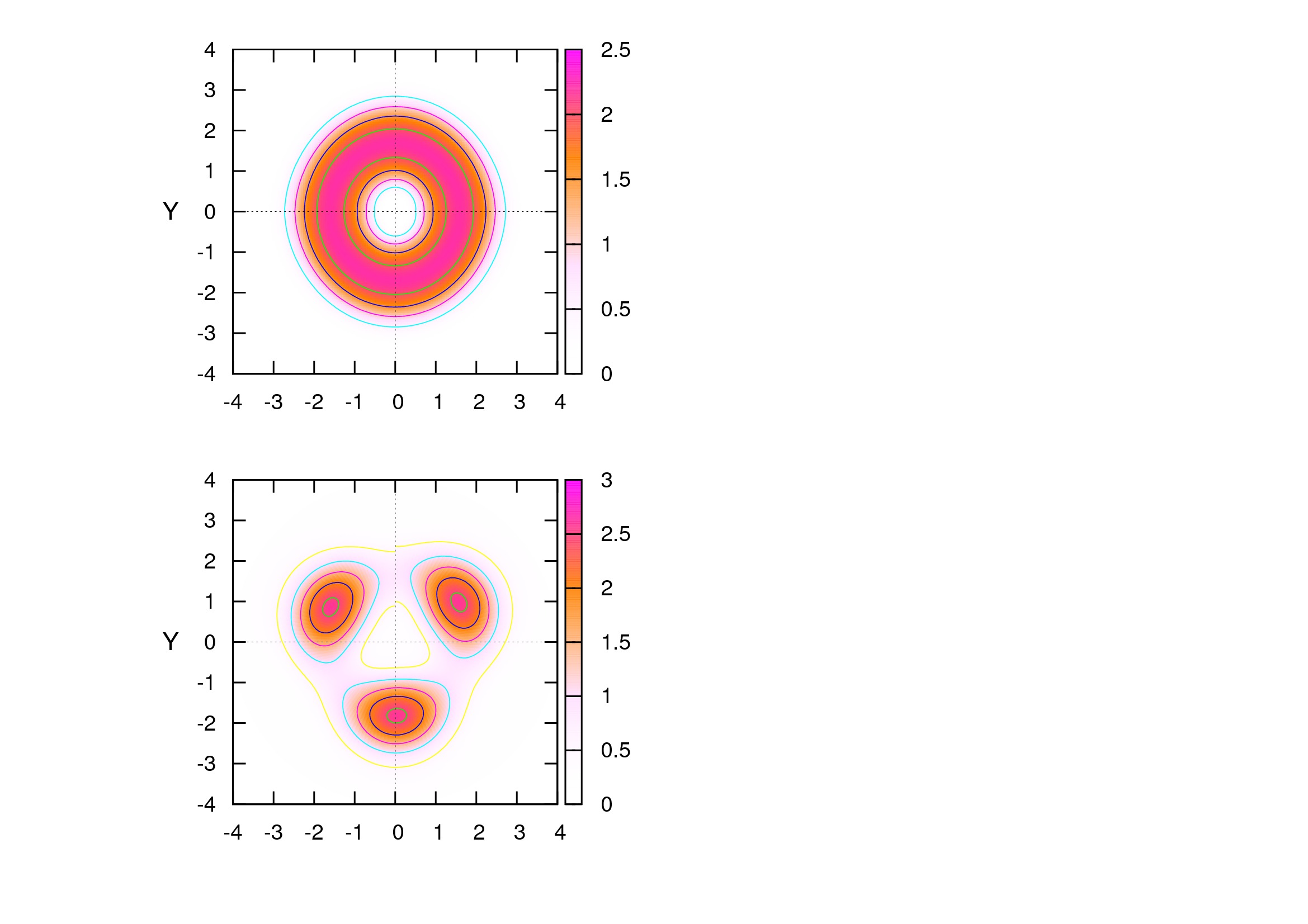}
\caption{(Color online) The energy density of components of the $(3,3)$
configuration in the model with potential (\protect\ref{pot-old-2}), at $%
\protect\kappa =0,0.3,1.0,2.0$ (from left to right).}
\label{f-15}
\end{figure}


\begin{equation}
U\left( \mathbf{\phi }^{(1)},\mathbf{\phi }^{(2)}\right) =\mu _{1}^{2}\left[
1-\left( \phi _{3}^{(1)}\right) \right] +\mu _{2}^{2}\left[ 1-\left( \phi
_{3}^{(2)}\right) \right] +\kappa \phi _{3}^{(1)}\phi _{1}^{(2)}\,.
\label{pot-old-2}
\end{equation}%
In the numerical solution, we fix the mass parameters as above, $\mu
_{1}=\mu _{2}=1$, and let the coupling constant, $\kappa $, gradually
increase from zero. Then, in Figs.~\ref{f-13}-\ref{f-15} we display contour
plots of the energy density, which illustrate a typical scenario of the
evolution of configurations $(1,1)$, $(2,2)$ and $(3,3)$ in the model with
potential (\ref{pot-old-2}), cf. similar plots in Figs.~\ref{f-1},\ref{f-16}
and \ref{f-3}. Once again, the $(1,1)$ configuration in the weak-coupling
regime exhibits separation of the components without their deformations, see
Fig.~\ref{f-13}. However as coupling becomes stronger, some novelty is
observed. The difference from the model with the double vacuum potential (%
\ref{pot}) is that, depending on values of masses $\mu _{1,2}$, the
configurations with topological charge $B^{(1,2)}\geq 2$ may not possess the
initial rotational invariance \cite{PZS}. Accordingly, the symmetry breaking
evolves differently, two distinct steps being identified in the evolution of
the coupled configuration. At first, similar to the model with potential (%
\ref{pot}), as $\kappa $ increases from zero, the energy-density
distributions of both components become symmetric with respect to dihedral
group $D_{m}$, with symmetry axes of different orders. However, as the
coupling grows stronger, at $\kappa >1$ the asymptotic forms of components $%
\phi _{1}^{(2)}$ and $\phi _{3}^{(2)}$ are not completely swapped, and the
configuration approaches another critical limit, when the first component
with topological charge $B^{(1)}=n$ again restores the rotational
invariance, while the second component with topological charge $B^{(2)}=m$
is formed as a set of $m$ individual charge-$1$ skyrmions, which is
symmetric with respect to dihedral group $D_{m}$, hence the number of the
respective segments is twice as small as in the model with the potential (%
\ref{pot}).

Thus, the above results depend on the choice of the symmetry-breaking
interaction potential, such as those given by Eq. (\ref{pot}) and (\ref%
{pot-old-2}). Other options for inducing symmetry breaking are also
possible. In particular, while the above-mentioned \textquotedblleft
straight" coupling, of the form\ of $\kappa \phi _{3}^{(1)}\phi _{3}^{(2)}$,
in Eq. (\ref{Lag}) does not, by itself, break the symmetry between the two
cores, it can be demonstrated that the linear-coupling potential chosen as $%
\kappa \phi _{1}^{(1)}\phi _{2}^{(2)}$ leads to a completely different
pattern of the symmetry breaking in the dual-core system.

Another interesting possibility is to consider the \textquotedblleft
skew-symmetrized" form of the coupling potential
\begin{equation*}
L_{\mathrm{coupling}}=\kappa \left( \phi _{3}^{(1)}\phi _{1}^{(2)}+\phi
_{1}^{(1)}\phi _{3}^{(2)}\right)
\end{equation*}%
and impose identical boundary conditions (\ref{infty2}) in both sectors to
allow the transformation of components $\phi _{1}^{(1)}$ and $\phi _{3}^{(1)}
$ too. Then, as $\kappa $ increases from zero, the initial rotational
invariance of both components of the $(m,m)$ configuration gradually becomes
broken to dihedral group $D_{2m}$ with symmetry axes of the same orders.

Finally, it can be demonstrated that the coupling of the soliton component
with a topologically trivial field in the second sector yields a
non-topological soliton (lump), whose structure precisely matches the
distribution of the coupling energy. In this configuration, only one
component of the field in the second sector, $\phi _{1}^{(2)}$, is
nontrivial at $\kappa >0$.

\section{Conclusions}

The objective of this work is to introduce a class of dual-core ($2+1$%
)-dimensional field-theory models, such as the BSM\ (baby Skyrme model).
Each skyrmion resides in its plane (core), the two parallel planes being
related by linear tunneling of the fields. This model can be implemented in
dual-layer magnetic media. As in previously studied dual-core models of
nonlinear optics and BEC, the interplay of the intra-core nonlinearity and
linear inter-core coupling ($\kappa $)\ gives rise to the SBB\
(symmetry-breaking bifurcation) of the solitons, but, on the contrary to
those models, where the SBB\ occurs with the decrease of $\kappa $, in the
present system is takes place with the increase of $\kappa $, which is
explained by the fact that cores cannot be empty even at $\kappa =0$. The
SBB\ follows the initial increase of the lateral separation between the two
components without the symmetry breaking, which is caused by the increase of
$\kappa $ from zero, and was explained by means of the effective potential
of the interaction between the two components. These evolution scenarios
were studied for different species of the two-component baby skyrmions,
categorized by values of the topological charge in the components: initially
symmetric ones, of the $\left( 1,1\right) $, $(2,2)$, $(3,3)$, and $\left(
4,4\right) $ types, and asymmetric composite states, $\left( 1,2\right) $
and $\left( 2,4\right) $.

\section*{Acknowledgments}

This work was financially supported by Alexander von Humboldt Foundation in
the framework of the Institutes linkage Programm. Ya.S. and G.Z. are
grateful to the Institute of Physics at the Carl von Ossietzky University
Oldenburg for hospitality. We thank Marek Karliner for useful discussions.

\end{document}